%% file: SSA22_WFC3.tex
\documentclass[twocolumn, times, twocolappendix]{aastex63}

\usepackage{amsmath}
\usepackage{xcolor}
\usepackage{booktabs}

\newcommand{\lya}{Ly$\alpha$}

\newcommand{\zthree}{$z\sim3$}
\newcommand{\sersic}{S\'ersic}

\newcommand{\galfit}{\texttt{GALFIT}}
\newcommand{\galfitm}{\texttt{GALFITM}}

\newcommand{\galapagostwo}{\texttt{GALAPAGOS-2}}
\newcommand{\sextractor}{\texttt{SExtractor}}
\newcommand{\adrizzle}{\texttt{AstroDrizzle}}

\newcommand{\ligh}{\texttt{Lightning}}

\newcommand{\Hubble}{\textit{Hubble}}
\newcommand{\HST}{\textit{HST}}
\newcommand{\Spitzer}{\textit{Spitzer}}
\newcommand{\Chandra}{\textit{Chandra}}

\defcitealias{steidel2003}{S03}
\defcitealias{micheva2017}{M17}

\begin{document}

\title{\large On the Nature of AGN and Star Formation Enhancement in the $z = 3.1$ SSA22 Protocluster:\\ The \textit{HST} WFC3 IR View}
\shorttitle{The \HST\ WFC3 IR View of Galaxies in the SSA22 Protocluster}
\shortauthors{Monson et al.}

\author[0000-0001-8473-5140]{Erik B. Monson}
\affiliation{Department of Physics, University of Arkansas, 226 Physics Building, 825 West Dickson Street, Fayetteville, AR 72701, USA}
\correspondingauthor{Erik B. Monson}
\email{ebmonson@uark.edu}

\author[0000-0003-2192-3296]{Bret D. Lehmer}
\affiliation{Department of Physics, University of Arkansas, 226 Physics Building, 825 West Dickson Street, Fayetteville, AR 72701, USA}

\author[0000-0001-5035-4016]{Keith Doore}
\affiliation{Department of Physics, University of Arkansas, 226 Physics Building, 825 West Dickson Street, Fayetteville, AR 72701, USA}

\author[0000-0002-2987-1796]{Rafael T. Eufrasio}
\affiliation{Department of Physics, University of Arkansas, 226 Physics Building, 825 West Dickson Street, Fayetteville, AR 72701, USA}

\author[0000-0003-0189-1805]{Brett Bonine}
\affiliation{Homer L. Dodge Department of Physics and Astronomy, The University of Oklahoma, 440 W. Brooks Street, Norman, OK 73019, USA}

\author[0000-0002-5896-6313]{David M. Alexander}
\affiliation{Centre for Extragalactic Astronomy, Department of Physics, Durham University, South Road, Durham, DH1 3LE, UK}

\author{Chris M. Harrison}
\affiliation{School of Mathematics, Statistics and Physics, Herschel Building, Newcastle University, Newcastle upon Tyne, NE1 7RU, UK}

\author[0000-0002-7598-5292]{Mariko Kubo}
\affiliation{Research Center for Space and Cosmic Evolution, Bunkyo-cho 2-5, Matsuyama, Ehime, 790-8577, Japan}

\author{Kameswara B. Mantha}
\affiliation{Department of Physics and Astronomy, University of Missouri Kansas City, 202 Flarsheim Hall, 5110 Rockhill Road, Kansas City, MO 64110, USA}

\author{Cristian Saez}
\affiliation{Departamento de Astronom\'ia, Universidad de Chile, Casilla 36-D, Santiago, Chile}

\author[0000-0002-4772-7878]{Amber Straughn}
\affiliation{NASA Goddard Space Flight Center, Code 662, Greenbelt, MD 20771, USA}

\author[0000-0003-1937-0573]{Hideki Umehata}
\affiliation{RIKEN Cluster for Pioneering Research, 2-1 Hirosawa, Wako, Saitama, 351-0198, Japan}
\affiliation{Institute of Astronomy, Graduate School of Science, The University of Tokyo, 2-21-1 Osawa, Mitaka, Tokyo 181-0015, Japan}

\begin{abstract}
We examine possible environmental sources of the enhanced star formation and active galactic nucleus (AGN) activity in the $z = 3.09$ SSA22 protocluster using \textit{Hubble} WFC3 F160W ($\sim1.6\ \micron$) observations of the SSA22 field, including new observations centered on eight X-ray selected protocluster AGN. To investigate the role of mergers in the observed AGN and star formation enhancement, we apply both quantitative (S\'ersic-fit and Gini-$M_{20}$) and visual morphological classifications to F160W images of protocluster Lyman break galaxies (LBGs) in the fields of the X-ray AGN and $z \sim 3$ field LBGs in SSA22 and GOODS-N. We find no statistically significant differences between the morphologies and merger fractions of protocluster and field LBGs, though we are limited by small number statistics in the protocluster. We also fit the UV-to-near-IR spectral energy distributions (SED) of F160W-detected protocluster and field LBGs to characterize their stellar masses and star formation histories (SFH). We find that the mean protocluster LBG is by a factor of $\sim2$ times more massive and more attenuated than the mean $z \sim 3$ field LBG. We take our results to suggest that ongoing mergers are not more common among protocluster LBGs than field LBGs, though protocluster LBGs appear to be more massive. We speculate that the larger mass of the protocluster LBGs contributes to the enhancement of SMBH mass and accretion rate in the protocluster, which in turn drives the observed protocluster AGN enhancement.
\end{abstract}

\section{Introduction}
\label{sec:intro}

Galaxy clusters, the largest virialized, gravitationally bound structures in the Universe, are currently thought to form by the dark-matter driven mergers of protoclusters, smaller groupings of galaxies in the early universe (i.e. $z\gtrsim2$; lookback times $\gtrsim$ 10 Gyr). In current cosmological models, the most luminous modern galaxies are assembled in protoclusters by mergers of smaller galaxies, where gas-rich mergers may trigger active galactic nuclei (AGN) and episodes of star formation. These protoclusters can move along dark matter filaments toward dense nodes, where they merge and collapse to form the clusters observed at lower redshifts \citep[e.g.,][]{boylan-kolchin2009}. Protoclusters have been discovered at redshifts ranging from $z \sim 2$ \citetext{e.g. \citealp{venemans2002,miley2004,capak2011}; see \citealp{overzier2016} for a review} as far as $z \sim 8$ \citep[in the BoRG58 field;][]{trenti2012}, in the epoch of galactic re-ionization, less than 1 Gyr after the Big Bang. Observations of protoclusters provide direct constraints on galaxy evolution, supermassive black hole (SMBH) growth, and the formation of large-scale structures and their galactic constituents.

\begin{figure*}
	\centering
	\includegraphics[width=0.95\textwidth]{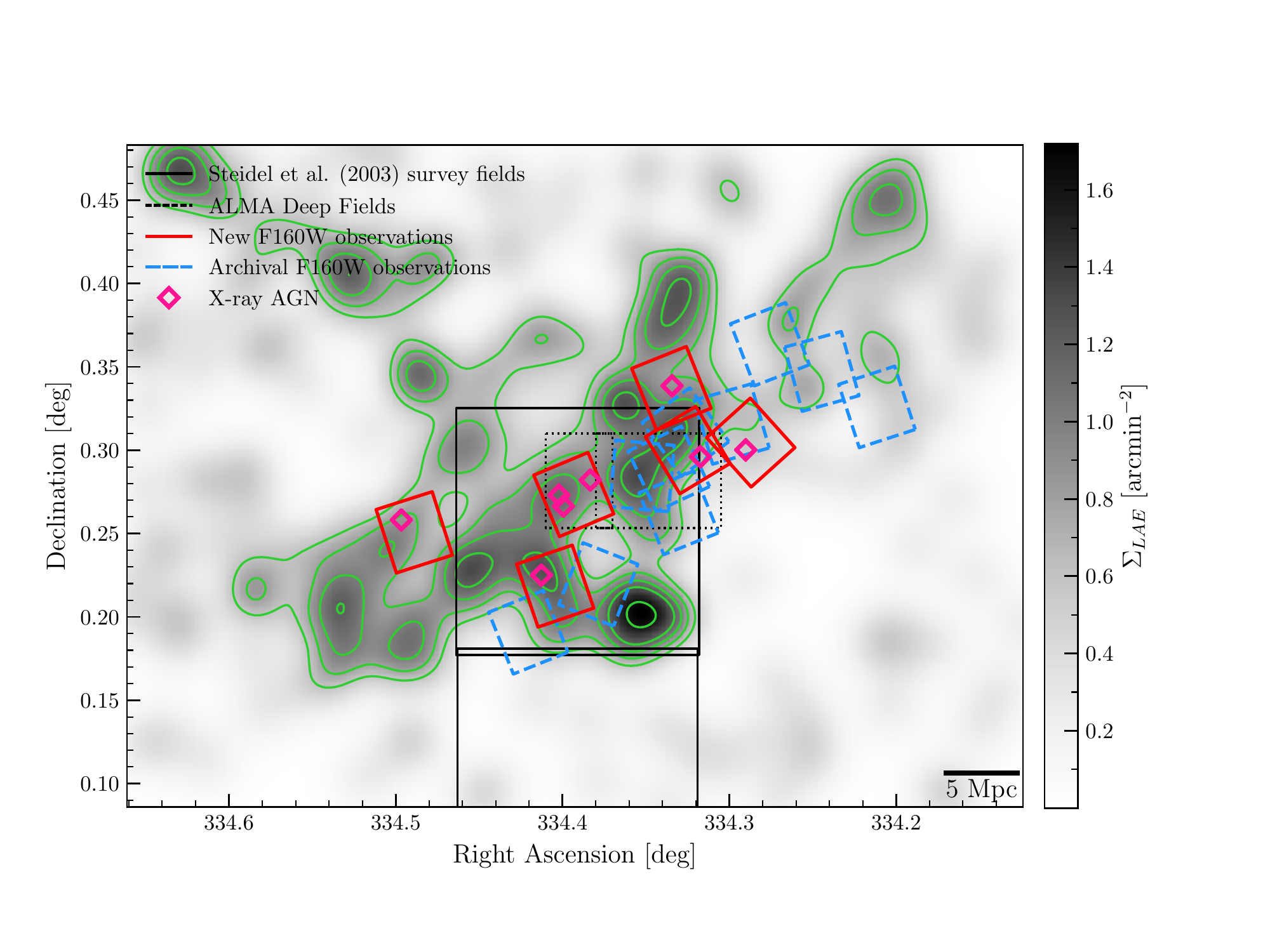}
	\caption{\label{fig:obs}Rectangles show the position and orientation of 
	our observations (thick red lines) and archival observations (dashed blue lines) 
	relative to the protocluster, which is shown as a surface density map of $z = 3.1$ 
	LAEs from \citet{hayashino2004}. To highlight the general shape of the protocluster
	we also show contours at levels $\left\{0.6,0.8,1.0,1.2,1.6\right\}\ \rm arcmin^{-2}$.
	We show the positions of the eight X-ray detected protocluster AGN from \citet{lehmer2009} and \citet{alexander2016} 
	as open magenta diamonds. For reference we also show the SSA22 fields studied 
	in \citet{steidel2003} (their SSA22a and SSA22b) in solid black lines, the approximate footprints of the 
	ALMA deep fields studied by \citet{umehata2015,umehata2017,umehata2018,umehata2019} in dotted black lines,
	and a 5 co-moving Mpc scalebar.}
\end{figure*}

The protocluster in the SSA22 survey field \citep[R.A.: $22^{\rm{h}}17^{\rm{m}}34.7^{\rm{s}}$, Dec.: $+0\degr15'7''$;][]{cowie1994} was discovered by \citet{steidel1998} as a spike in the redshift distribution of Lyman-break galaxies (LBGs) at $z=3.09$. Further observations of the same region revealed a $\sim$six-fold overdensity of LBGs consistent with a galaxy cluster in the early stages of development \citep{steidel2000, steidel2003}. Simulations suggest that the SSA22 protocluster and structures of similar scale and overdensity will evolve toward Coma-like (virialized mass $\sim10^{14}\ \rm M_\odot$) clusters at redshift $z=0$ \citep{governato1998}. Recent surveys \citep[e.g.,][]{toshikawa2016, toshikawa2018, higuchi2019} of candidate protoclusters suggest that progenitors of Coma-like clusters are not uncommon, with perhaps $76\%$ of $4\sigma$ significant overdensities projected to evolve into clusters with masses $\sim 5 \times 10^{14} \ \rm M_\odot$, though conclusive spectroscopic identifications of protoclusters remain difficult and possibly biased toward younger galaxies due to the reliance on \lya\ emission \citep{toshikawa2016, toshikawa2018}.

Follow-up narrowband observations over an area 10 times larger than that studied in \citet{steidel1998} using SUPRIME-Cam have further identified $\sim$six-fold overdense bands of Ly$\alpha$ emitting galaxies (LAEs), spectroscopically confirmed as a set of three large-scale filamentary structures \citep{hayashino2004, matsuda2005}. The co-moving scale of the largest of these filaments, shown in \autoref{fig:obs}, is on the order of 60 Mpc long and 10 Mpc wide, with a redshift range of 3.088--3.108, making it one of the largest mapped structures at $z\sim3$ \citep{matsuda2005}. Large \lya\ emitting nebulae (Ly$\alpha$ blobs; LABs) around star forming galaxies have been shown to be associated with these filaments, suggesting that they are the precursors of massive galaxies developing in the regions of greatest overdensity \citep{matsuda2005}. The filamentary structure of the protocluster has been further established by MUSE spectral-imaging observations mapping the filaments in emission around the protocluster core \citep{umehata2019}. The scale and detail at which the filamentary structures in SSA22 have been mapped remains relatively unique among high-redshift protoclusters, though this is also changing: \citet{harikane2019} mapped candidate protoclusters at $z>6$ on scales of $>100$ comoving Mpc, and \citet{daddi2021} have recently imaged filamentary structures in \lya\ emission around an overdensity of galaxies at $z=2.9$.

The SSA22 region has also been well-studied in millimeter/sub-millimeter bands \citep{tamura2009, umehata2014, umehata2015, umehata2017, umehata2018, umehata2019, alexander2016}. \citet{umehata2015} identified a concentration of 8 dusty star-forming galaxies (DSFGs) associated with the intersection of the major filamentary structures at the center of the protocluster. \textit{Herschel} SPIRE measurements suggest that DSFGs in the SSA22 protocluster account for a star formation rate density on the order of $10^3\ \rm{M_{\Sun}\ yr^{-1}\ Mpc^{-3}}$, a factor of $\sim 10^4$ increase in star formation rate density over the field at redshift $z=3.09$ \citep{kato2016}. Additionally, five of the SSA22 DSFGs are associated with X-ray luminous AGN, and two are associated with LABs. Further observations of the protocluster core with higher resolution sub-millimeter instruments have revealed more DSFGs, at least 10 of which are spectroscopically confirmed protocluster members, possibly indicating preferential formation of these galaxies in the densest region of the protocluster \citep{umehata2017, umehata2018}. Wide-spectrum ($u^*$-band to \Spitzer\ IRAC 8 \micron) photometry and spectral energy distribution (SED) fitting by \citet{kubo2013} has additionally suggested the presence of very massive galaxies in the densest regions of the protocluster. Overall, these observations support a picture of the SSA22 protocluster environment as one where massive, intensely star-forming galaxies are actively forming.

\Chandra\ X-ray observations of the region have revealed a higher rate of AGN activity compared to field galaxies at $z\sim3$ \citep{lehmer2009}, with $9.5^{+12.7}_{-6.1}\%$ and $5.1^{+6.8}_{-3.1}\%$ of protocluster LBGs and LAEs, respectively, hosting AGN with $L_{8-32~\rm{keV}} \gtrsim 3 \times 10^{43}~\rm{ergs~s^{-1}}$. These fractions are elevated by a factor of $\approx6$ compared to non-protocluster galaxies (hereafter, ``field'' galaxies) at \zthree, indicating a possible enhancement of SMBH growth in the protocluster. Similar enhancements have been observed in other $z=2$--3 protoclusters \citetext{e.g., HS1700, \citealp{digbynorth2010}; 2QZCluster, \citealp{lehmer2013}; DRC, \citealp{vito2020}}; there is also evidence that AGN fraction in overdense environments evolves with redshift \citep{martini2013}, with overdensities at higher redshifts having larger AGN fractions and modern clusters having lower AGN fractions than the field.

The AGN and highly star-forming galaxies in the protocluster are consistently found to be associated with the larger scale structure of the protocluster, embedded in the intersection of the filaments and giant \lya\ nebulae. Of the 8 X-ray detected protocluster AGN from \citet{lehmer2009}, 4 are associated with LABs, 2 of which are giant LABs larger than 100 kpc in scale \citep{alexander2016}. The implication, then, is that the enhanced AGN and star formation activity are driven by environmental factors unique to the protocluster. These enhancements may be driven by accretion episodes caused by an elevated merger rate among protocluster members, or by secular gas accretion from shared gas reservoirs \citep{narayanan2015} and filamentary structures in the intergalactic medium \citep{umehata2019}. However, the AGN enhancement may also be driven by the presence of more massive galaxies (and hence SMBH) in the protocluster, as compared to the field at $z\sim3$; it has been established that galaxies in $2 \lesssim z \lesssim 4$ protoclusters are on average more massive than their field counterparts at the same redshift \citep[e.g.,][]{steidel2005, hatch2011, cooke2014}.

Motivated by the elevated AGN fraction observed in the protocluster, \citet{hine2015} used archival \Hubble\ ACS F814W optical observations of SSA22 (probing rest-frame UV emission) to visually classify LBGs in the protocluster, finding a marginally enhanced merger fraction among protocluster galaxies ($48 \pm10\%$) as compared to field galaxies ($30 \pm 6\%$). However these results are limited by the small-number statistics of the protocluster and suffer from ambiguous interpretation due to the patchiness of rest-frame UV observations, which are highly influenced by star formation activity and attenuation. High-resolution near-infrared (i.e., 1--2 \micron) observations, which probe rest-frame optical wavelengths at $z=3.09$, can better trace the stellar mass content of the protocluster galaxies, less influenced by individual bursts of star formation and more sensitive to merger activity. 

\input{table1.tex}

In this work, we investigate the possible contributions of mergers to the increased AGN and SF activity in the protocluster by applying three separate morphological analysis techniques to galaxies detected in \Hubble\ WFC3 infrared (F160W, $\lambda_p = 1.537\ {\rm \mu m}$) observations in SSA22, targeting the environment around the X-ray detected protocluster AGN studied in \citet{alexander2016}. We use parametric model fitting techniques to extract \sersic\ model parameters from F160W-detected LBGs, compare the measured morphologies of protocluster LBGs to a sample of field LBGs in GOODS-N, and analyze the residuals after \sersic\ model subtraction for indications of merger activity. We additionally use nonparametric morphological measures (the Gini coefficient $G$, moment of light $M_{20}$, and concentration $C$) to compare the morphologies of protocluster and field galaxies and attempt to classify mergers. Lastly, we apply a similar visual analysis as \citet{hine2015} to our sample of F160W-detected LBGs to compare the rest-frame optical merger fraction for the SSA22 protocluster to the field at \zthree.

We also investigate the possibility that more massive galaxies are the driver of AGN and SF enhancement by fitting the SEDs of a subset of our F160W-detected SSA22 LBG sample. We compare the distribution of stellar mass and the mean star formation history of protocluster LBGs to a sample of \zthree\ field LBGs in GOODS-N.

The paper is organized as follows: \autoref{sec:data_analysis} describes the observations, data reduction, catalog generation, and sample selection; \autoref{sec:morph_analysis} describes our analysis of the morphological properties of protocluster galaxies; \autoref{sec:phys_analysis} describes our analysis of the SEDs and physical properties of protocluster galaxies; in \autoref{sec:discussion} we discuss our results and attempt to connect the morphologies and physical properties of protocluster galaxies to the protocluster environment; lastly, in \autoref{sec:summary} we summarize our results and their implications for understanding the galaxy assembly process at \zthree.

Coordinates in this work are J2000, magnitudes are given in the AB system, and we adopt a \citet{kroupa2001} IMF. We adopt a cosmology with $H_0 = 70\ \mathrm{km\ s^{-1}\ Mpc^{-1}}$, $\Omega_M = 0.3$, and $\Omega_\Lambda = 0.7$, yielding a lookback time of 11.42 Gyr, a 2.04 Gyr universe age, and a physical scale of 7.63 proper ${\rm kpc\ arcsec^{-1}}$ at $z = 3.09$. We use proper scales when discussing lengths on the scale of galaxies, and co-moving scales when discussing lengths on the scale of the protocluster itself. Hereafter, we define SSA22 protocluster galaxies as those galaxies in our F160W images with $3.06 \le z \le 3.12$. Over this range of redshift, the rest-frame wavelength probed by our WFC3 IR observations ranges from $3740 - 3790\ {\rm \AA}$.

\section{Data Analysis}
\label{sec:data_analysis}

\subsection{Data Reduction}
\label{sec:data_reduction}

We use sixteen \HST\ WFC3 F160W images of the SSA22 field for our morphological analyses (\HST\ proposals 13844, 11735, 11636, 14747). We summarize the locations and exposure times of these fields in \autoref{table:obs} and show their footprints superimposed on the protocluster structure in \autoref{fig:obs}. Six of these fields (\HST\ proposal 13844) were new observations obtained to cover the eight protocluster AGN detected in \Chandra\ observations by \citet{lehmer2009} that were studied by \citet{alexander2016}. These six observations were taken at two-orbit depth for ease of comparison to the CANDELS-Wide fields \citep{koekemoer2011, grogin2011}.

The STScI \adrizzle\ package\footnote{\url{http://drizzlepac.stsci.edu}} was used for image recombination and data reduction. We re-bin our images to a scale of $0.065~\rm{arcsec}\ \rm{pixel}^{-1}$; at this scale the PSF FWHM of the images is $\approx2.6\ \rm{pixels}$. We adopt an inverse variance weighting scheme when combining the exposures. We use the inverse variance weight maps generated by \adrizzle\ to create a map (a ``sigma image'') of the estimated total standard deviation (in electrons) for each pixel in the sky-subtracted science image as 
\begin{equation*}
	\sigma = \sqrt{p_{>0} + \frac{1}{\rm{weight}}},
\end{equation*}
where $p_{>0}$ denotes the greater of the science image pixel value (in electrons) and zero. 

We adopt a model PSF based on the median of 33 isolated stars in our two-orbit depth WFC3 F160W images, identified using \sextractor\ \citep{bertin1996} by making a box selection in the \verb|MAG_AUTO|--\verb|FLUX_RADIUS| plane; we select sources with $0.13'' < \verb|FLUX_RADIUS| < 0.15''$ and $21.1 < \verb|MAG_AUTO| < 22.8$ as stars.


\subsection{Catalog Generation}
\label{sec:catalog_generation}
We use \galapagostwo\ \citep{haussler2013, barden2012} to carry out source detection (using \sextractor), background determination, model fitting (using \galfitm, an updated multi-band version of \galfit; see \citealp{vika2013}), and catalog compilation. We briefly describe the methods adopted for the catalog generation tasks here, since they are relevant to our methodology \citep[for a more thorough description, see][]{barden2012}.

\sextractor\ is designed to de-blend sources so that nearby sources are detected and cataloged separately. To prevent over-de-blending of bright sources while still detecting faint sources in deep images, \galapagostwo\ runs \sextractor\ twice, once with a low threshold for detection and once with a high threshold. Detections from the high-threshold run are accepted and cataloged automatically; detections from the low-threshold run are rejected if they fall inside the isophotal ellipse of a high-run source, and accepted otherwise. Science image and sigma image cutouts are then constructed for each source.

\sextractor\ is known to overestimate the sky level \citep{haussler2007}, so \galapagostwo\ determines a local sky value of its own and holds the sky fixed during the later model-fitting step. The procedure is described in detail in \citet{barden2012}. Briefly, \galapagostwo\ identifies the brightest secondary sources in each cutout, enlarges their \sextractor\ Kron isophotes by a pre-determined factor (default 2.5 times), masks out every pixel inside the resulting ellipses, and takes the median of the remaining pixels as the local background.

To generate our preliminary source catalog, the detection criteria for the ``hot" low-threshold run was 15 pixels $\ge 2.5\sigma$ above the background after convolution with a Gaussian kernel with a FWHM of 2 pixels (i.e., the default 5x5 pixel Gaussian kernel supplied with \sextractor). For the ``cold," high-threshold run the detection criteria were strengthened to 30 pixels $\ge 3.5\sigma$ above the background; no convolution filter was applied on the cold run. Sources were de-blended using 64 thresholds and a minimum contrast of 0.001. The \sextractor\ Kron isophotes were enlarged by a factor of 2.5 when combining the hot and cold catalogs. 

To estimate the completeness of our preliminary catalog under these \sextractor\ parameters, we added a total of 3000 simulated stars and 3000 simulated galaxies (generated using \galfit, with morphological parameters uniformly selected from ranges taken from the \citet{vanderwel2012} fits to \zthree\ GOODS-N galaxies) to our two-orbit images and re-detected them with the same procedures. We find that our preliminary catalog is 90 (50) percent complete to a F160W magnitude of 24.8 (25.7) for galaxies and 25.1 (26.2) for point sources detected in two-orbit images. We repeated the process for the single-orbit images, finding 90 (50) percent completeness limit magnitudes of 24.7 (25.4) for galaxies and 24.8 (26.0) for point sources.

To clean spurious or un-physical \sextractor\ detections from the catalog, we require detections to have $S/N \ge 5$ in a $1''$ diameter aperture, $\verb|FWHM_IMAGE| > 2.6\rm~pix$, and $50\%$ $\verb|FLUX_RADIUS| > 0$. Following the above procedures results in a ``main catalog'' of 7538 F160W-detected sources. 

\input{table2.tex}

We cross-match detections in our catalog with sources in reference catalogs, provided they are within $1''$ of the position returned by \sextractor. To identify LBGs we use the catalogs produced by \citet{steidel2003} (hereafter S03) and \citet{micheva2017} (hereafter M17). The \citetalias{micheva2017} LBG sample is an expanded version of an LBG sample from \citet{iwata2009}, containing U-dropouts with VLT/VIMOS followups that confirm redshifts $z \ge 3.06$; while this sample overlaps in part with the \citet{steidel2003} LBG sample we keep \citetalias{steidel2003} and \citetalias{micheva2017} LBGs separate in subsequent figures and analysis out of an abundance of caution concerning possible differences in, e.g., color selection criteria. To identify LAEs we use the catalog from \citet{yamada2012} and the \citetalias{micheva2017} catalog. We identify AGN based on the X-ray point source catalog from \citet{lehmer2009} and Lyman-continuum (LyC) emitters based on the \citetalias{micheva2017} catalog. For spectroscopic redshifts, if not available in one of the aforementioned catalogs, we have taken redshifts from the spectroscopic SSA22 surveys by \citet{saez2015} and \citet{kubo2015}, and the VLT-VIMOS Deep Survey (VVDS) \citep{lefevre2013}. We find spectroscopic redshifts from the above references for 216 of the sources in our catalog. For the subsequent analysis, we require galaxies to have spectroscopic redshifts in order to conclusively identify them as protocluster members or \zthree\ field galaxies.  We find that of these sources, 91 have redshifts $2.9 \leq z \leq 3.3$, with 72 galaxies in the protocluster redshift range ($3.06 \le z \le 3.12$), and thus 19 galaxies in the field redshift range. By construction, the \citetalias{micheva2017} LBGs have a spectroscopic redshift; for \citetalias{steidel2003} LBGs we estimate that our requirement of spectroscopic redshifts may exclude as many as four protocluster LBGs from our subsequent analysis. For additional photometry (covering $u^*$-band to \Spitzer\ IRAC 8 \micron; see  \autoref{table:sedfilter} for the full list of filters), we have used the photometric catalog of \citet{kubo2013}. We also include narrowband magnitudes at 4972 \r{A} (the observed wavelength of the Lyman $\alpha$ line at $z=3.1$) from \citet{yamada2012}, and we report $\log L_{X}$ (measured from 2--8 keV, approximately 8--32 rest-frame keV) for the X-ray sources in \citet{lehmer2009}. We estimate that for the majority of the reference catalogs the number of possible mismatches with our catalog is on the order of a few galaxies. For the larger \citet{kubo2013} photometric catalog, the number of mismatches could be as large as 200, though this is still $\sim 11\%$ of the overall number of matches we find with the \citet{kubo2013} catalog. We note that these numbers of false matches are likely overestimated, since the angular separations of the matches are typically much less than an arcsecond, and there is a large number of sources in our main catalog. Excerpts of the main catalog are provided in \autoref{table:morphcatalog} and \autoref{table:photcatalog} for protocluster LBGs with acceptable fits from \galfitm\ (as defined in \autoref{sec:models}).

In \autoref{fig:sample_flow} we show how we divide our main catalog into sub-samples (based on requirements for spectroscopic redshift, etc.) for the analyses presented in the following sections. Due to the availability of an additional \citetalias{steidel2003} LBG sample in GOODS-N, and concerns about how different LBG color selection criteria might harm any protocluster-field comparisons, we focus the majority of the following analysis on the sample of \citetalias{steidel2003} LBGs in our catalog. Our main catalog contains 26 \citetalias{steidel2003} LBGs with $2.9 \le z \le 3.3$, to a maximum $R-$band magnitude of 25.4. These 26 \zthree\ \citetalias{steidel2003} LBGs amount to 13 protocluster LBGs and 13 field LBGs. Our main catalog also contains 13 LBGs in the same redshift range which are unique to the \citetalias{micheva2017} catalog; these have a maximum $R-$band magnitude of 25.5, and amount to 11 protocluster LBGs and 2 field LBGs.

\subsection{Comparison LBG Sample in GOODS-N}
\label{sec:comparison_sample}

We constructed an additional comparison sample of LBGs in the GOODS-N field based on the the \citetalias{steidel2003} catalog. The \citetalias{steidel2003} GOODS-N catalog contains 40 LBGs with $2.9 \le z \le 3.3$, to a maximum $R-$band magnitude of 25.6. \sersic\ parameters for these galaxies were retrieved by cross-matching with the \citet{vanderwel2012} single-\sersic\ fitting catalog. We find that a subset of 33 \zthree\ LBGs have acceptable single-\sersic\ fits from \citet{vanderwel2012}. We use these \sersic\ parameters to increase the size of our field galaxy comparison sample in the analysis described in \autoref{sec:models}.

We use UV-to-mid-IR ($U$-band to \textit{Spitzer} MIPS 70 \micron) photometry from the \citet{barro2019} catalog in the CANDELS survey areas for our comparison LBG sample in GOODS-N, retrieved from the \textit{Rainbow} database\footnote{\url{http://rainbowx.fis.ucm.es/Rainbow_navigator_public/}}. We searched the catalogs for the closest match within $1''$ to each LBG in the \citet{steidel2003} GOODS-N sample. We estimate that $\lesssim 10$ galaxies could be mismatched with the CANDELS photometric catalog, though this could amount to a significant fraction of the 40 \citetalias{steidel2003} LBGs in GOODS-N. However, we again expect that this number is overestimated based on the small angular separations between matches. We list the filters used for this photometry in \autoref{table:sedfilter}.

\section{Morphological Analysis}
\label{sec:morph_analysis}

\subsection{Parametric Morphology Fitting}
\label{sec:models}
To analyze the morphologies of our detected galaxies in the SSA22 field, we began by fitting 2D parametric models to the data. The surface brightness profile of an elliptical or spheroidal galaxy without a well-resolved disk is well described by the S\'ersic law, a symmetric profile specified by two parameters, the \sersic\ index $n$ and effective radius $r_e$:
\begin{equation}
	\label{eq:sersic}
	I(r) = I_e \exp \left[ -b_n \left( \left( \frac{r}{r_e} \right)^{\frac{1}{n}} - 1 \right) \right],
\end{equation}
where $I_e$ is the surface brightness at $r = r_e$ and $b_n$ satisfies
\begin{equation}
	\Gamma \left( 2n \right)  = 2 \gamma \left( 2n, b_n \right);
\end{equation}
$\Gamma$ and $\gamma$ are the complete and lower incomplete gamma functions, respectively \citep{graham2005}. In the general case of an elliptical profile, the $r$ in the equation above is a function of the profile's center and elliptical axis ratio $q = b/a$.

Multiple studies have demonstrated that single \sersic\ model fitting with \galfit\ \citep{peng2002} can be used to extract galaxy morphologies from large \HST\ datasets \citetext{e.g. GEMS, \citealp{haussler2007}, and CANDELS, \citealp{vanderwel2012}}. \galfit\ has the ability to de-blend nearby sources by simultaneous fitting, allowing accurate photometric measurements in crowded images and the examination of galaxies with close projected companions for evidence of mergers. \galapagostwo\ uses \galfitm\ \citep{vika2013} for single \sersic\ model fitting. \galfitm\ is a modified version of \galfit, which retains all of the same functionality and runs on the same Levenberg-Marquardt algorithm.

The \galapagostwo\ fitting procedure distinguishes between the ``primary" source in a cutout, i.e. the main source currently being fit, and ``secondary" sources, nearby objects bright enough to bias the photometry of the primary. For accurate fitting to the primary source, secondary sources must be fit simultaneously. Sources are sorted and fit in order of decreasing brightness, and every source gets a turn as the primary. If a secondary source is present and was already fit (i.e., if the secondary source is brighter than the current primary), the parameters from that fit are reused and held fixed \citep[provided the secondary is in the same image as the primary source; see Figure 9 in][]{barden2012}. While secondary sources are fit simultaneously, sources that are present in the cutout but are faint enough compared to the primary are not fit, and instead masked out such that \galfitm\ ignores any pixels corresponding to these sources. The fit results presented in \autoref{table:morphcatalog} and in what follows represent the primary fit to each source.

Following the generation of the source catalog, \galfitm\ was used to fit a single \sersic\ profile to each F160W detection. We focus on single component fits for individual galaxies, reasoning that for the $z \sim 3$ we are primarily interested in bulge-like and disk-like components were unlikely to be resolved separately. The initial guesses and constraints for our fits come from \sextractor\ parameters. For a given object, the initial guess for the \sersic\ model magnitude is the \sextractor\ \verb|MAG_BEST|; the initial value for $r_e$ is the $50\%$ \verb|FLUX_RADIUS| raised to the power of 1.4. The initial value for the \sersic\ index is 2.5 for all galaxies, and the initial position of the \sersic\ model is the position determined by \sextractor.

\galfit\ (and by extension, \galfitm) allows parameter value ranges to be limited and coupled. The following bounds on $r_e$, $n$, and the \sersic\ profile magnitude $m$ were adopted for this work:
\begin{align*}
	0.002\arcsec < r_e < 26\arcsec&;\\
	                  0.2 < n < 10&;\\
	              |m_{SE} - m |< 5&,
\end{align*}
where $m_{SE}$ is the magnitude reported by \sextractor. In pixel units the constraint on S\'ersic radius is $0.3 < r_e < 400$, where the lower limit is hard-coded into \galapagostwo. These constraints are a slight relaxation of the \galapagostwo\ defaults, which are themselves selected to do a good job of keeping the fit from wandering into unphysical regions of the parameter space without being overly restrictive. The center of the model is constrained so that it can only move within $10\%$ of the cutout size from the initial position.

\input{table3.tex}

Following \citet{vanderwel2012}, we flagged fits where the primary fit is unlikely to represent the galaxy well as unacceptable, based on the following criteria. We flagged fits where any of the final parameter values were equal to one of the bounds listed above and fits where the final S\'ersic index was equal to the initial value of 2.5 as unacceptable. The \galfit\ algorithm occasionally converges on an arbitrarily small axis ratio for low-$S/N$ objects with small apparent sizes. For this reason, we also flagged fits with axis ratios less than 0.125. Apparently well-converged fits that do not represent the data well also have exceptionally large errors in the resulting total \sersic\ magnitude and effective radius, so we also flagged fits where the magnitude error estimated by \galfitm\ is greater than 5 mag (more typical errors are on the order of 0.08 mag for acceptable fits and 1.4 mag for unacceptable fits), and flagged fits where the recovered effective radius was consistent with 0 within $1\sigma$. 

\begin{figure*}
	\centering
	\includegraphics[width=0.90\textwidth]{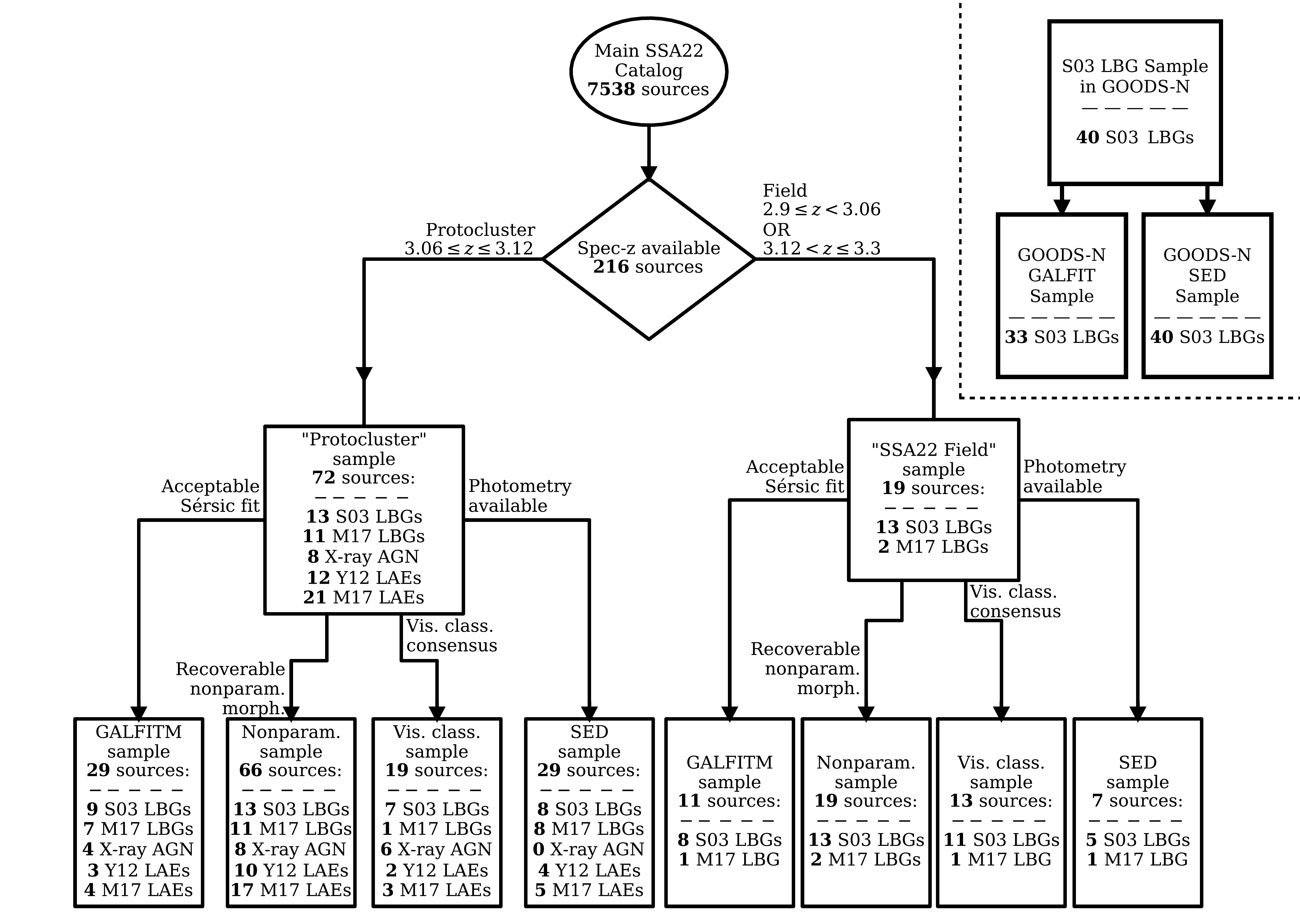}
	\caption{\label{fig:sample_flow}We illustrate how each of the protocluster and field subsamples used throughout this paper stem from the main SSA22 catalog. We require galaxies to have spectroscopic redshifts from the literature to distinguish between protocluster and field galaxies. To be included in the analysis of our \textit{\galfitm\ sample}, we require sources to have acceptable \sersic\ fits, as defined in \autoref{sec:models}. In our analysis of the \textit{non-parametric morphology sample}, we require sources to have a recoverable nonparametric morphology using the procedures described in \autoref{sec:npmorphologies}. In our \textit{visual classification sample}, we include only sources for which our classifiers reached a consensus about whether the system was merging or an isolated galaxy, as described in \autoref{sec:visual-classification}. Our \textit{SED sample} includes only galaxies for which we could retrieve photometry from the \citet{kubo2013} photometric catalog. For each subsample, we show the total number of sources as well as the number in different classes and the reference to the classification in the literature. While we focus on the LBG and AGN populations in subsequent sections we also list the number of LAEs in each sample. We abbreviate the references as: S03 -- \citet{steidel2003}; Y12 -- \citet{yamada2012}; M17 -- \citet{micheva2017}. The X-ray AGN are those studied in \citet{alexander2016}; they are all found in the protocluster, and our SED fitting sample excludes them by construction. Due to the nature of the narrowband LAE selection, there are no field LAEs in our catalog. In the upper right corner we show the breakdown of our \textit{comparison LBG sample in GOODS-N}, made up of 40 GOODS-N LBGs from the \citetalias{steidel2003} catalog, 33 of which have acceptable single-\sersic\ fits from \citet{vanderwel2012}.}
\end{figure*}

Of our main catalog, 2833 detections ($37.6\%$ of detections) have acceptable single \sersic\ fits according to the criteria above. Sources with acceptable fits tend to be brighter and have larger effective sizes than sources with bad fits; foreground stars and other point-like sources tend to have bad fits, as do faint, compact galaxies like LAEs. In terms of \sextractor\ parameters, poorly fit sources tend to have $23.5 \leq \verb|MAG_BEST| \leq 26.1$ and (in pixel units) $4.41 \leq \verb|FWHM_IMAGE| \leq 9.84$; sources with acceptable fits tend to have $21.9 \leq \verb|MAG_BEST| \leq 24.7$ and $5.71 \leq \verb|FWHM_IMAGE| \leq 13.25$. The fraction of acceptable fits increases slightly among sources with $2.9 \leq z \leq 3.3$: 40 ($44.0\%$) have acceptable \sersic\ fits. In \autoref{fig:sample_flow} we show how the sources with acceptable fits break down between the protocluster and field, and the numbers of LBGs and other categories of galaxies with acceptable fits. Among the sources with acceptable fits, we identify 29 spectroscopically-confirmed protocluster members, 9 of which are LBGs from the \citetalias{steidel2003} catalog and 7 of which are LBGs exclusive to the \citetalias{micheva2017} catalog. Only 4 of the 8 X-ray detected protocluster AGN from \citet{lehmer2009} have acceptable \sersic\ fits under the criteria above; the fits to the rest-frame optically bright AGN (e.g., the quasar SSA22a-D12) do not return especially meaningful or well-constrained \sersic\ model parameters due to their point source-like profiles.

\begin{figure*}
	\centering
	\includegraphics[width=0.9\textwidth]{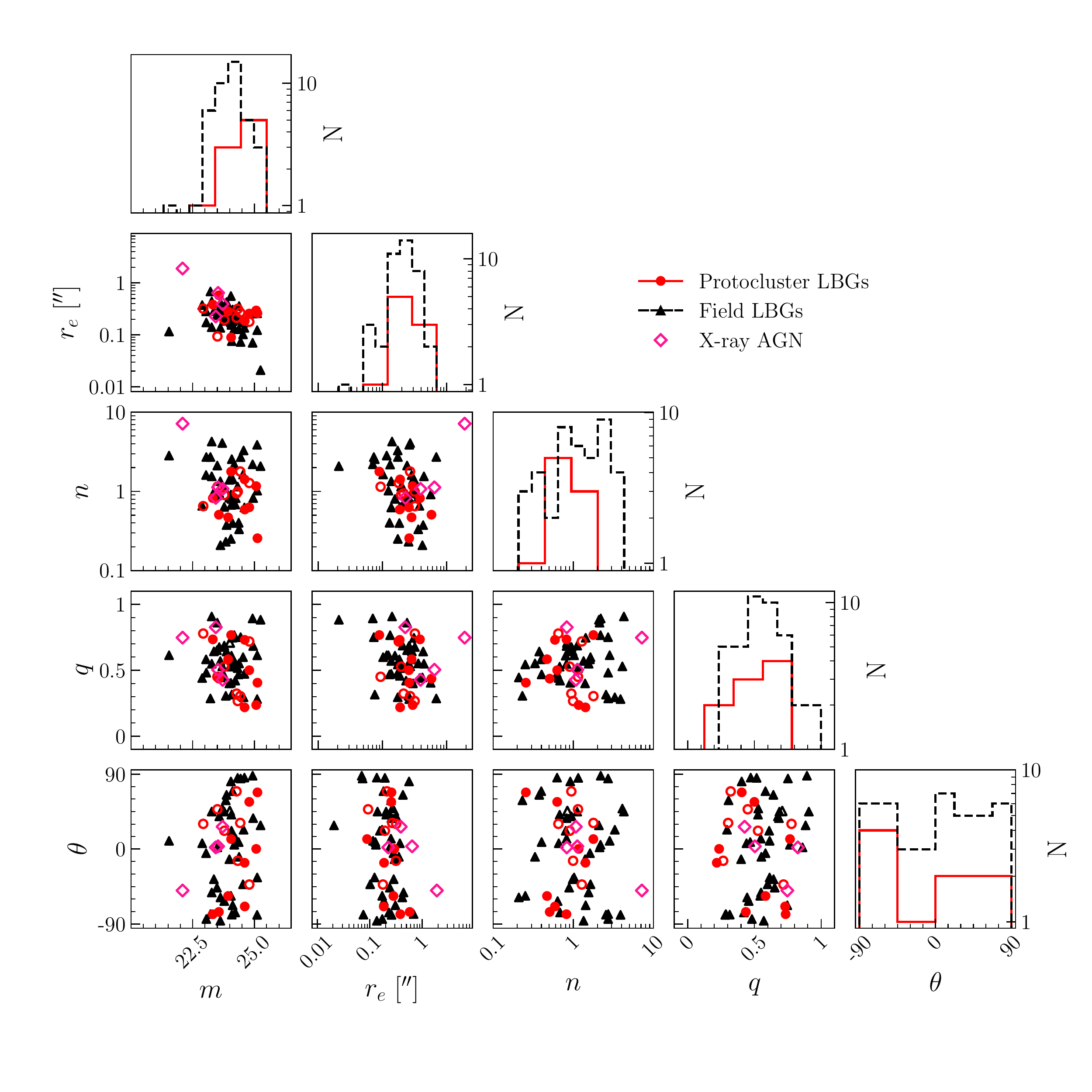}
	\caption{\label{fig:paramplot}On the diagonal, we show the distributions of 
	\sersic\ fit parameters -- integrated F160W magnitude $m$, effective radius $r_e$, 
	\sersic\ index $n$, axis ratio $q$ and position angle $\theta$ -- for the \citetalias{steidel2003} LBGs with 
	acceptable fits in the final SSA22 catalog and the comparison
	GOODS-N sample with fits from \citet{vanderwel2012}. Off-diagonal, we show correlations between
	the parameters for \citetalias{steidel2003} LBGs, \citetalias{micheva2017} LBGs, and protocluster AGN with acceptable fits. 
	We show LBGs associated with the protocluster ($3.06 \le z \le 3.12$) in red 
	(circles, solid lines), and field LBGs in black (triangles, dashed lines). While the histograms on the diagonal are limited to only \citetalias{steidel2003} LBGs, on the off-diagonal scatter plots we show both \citetalias{steidel2003}
	LBGs (filled symbols) and \citetalias{micheva2017} LBGs (open symbols). The X-ray 
	detected protocluster AGN are shown on the scatter plots as open magenta diamonds.}
\end{figure*}

We present the parameters derived from model fitting in \autoref{fig:paramplot}; to broaden the field galaxy comparison sample, we include single-\sersic-fit parameters from the 33 galaxies in our GOODS-N comparison sample with good fits from \citet{vanderwel2012} (see \autoref{sec:comparison_sample}) in the figure.

To assess whether or not the morphologies of the protocluster and field samples are drawn from the same underlying population, we use 1-D and 2-D two-sample Kolmogorov-Smirnov (hereafter, KS) tests, under the null hypothesis that the results for the protocluster and field samples are drawn from the same distribution. For a fair comparison between LBGs selected by the same color criteria, we initially limited the tests to \citetalias{steidel2003} LBGs. The tests are consistent with the null hypothesis that the \citetalias{steidel2003} protocluster and \citetalias{steidel2003} field LBGs are drawn from the same morphological population. We then performed the same KS tests with the addition of the \citetalias{micheva2017} LBGs, finding again that all the tests on the parametric morphologies are consistent with the null hypothesis. We show the results for both sets of 1-D and 2-D KS tests in \autoref{table:ksall}.

We note the one-to-two orbit depth of our images is not ideal for parametric model fitting; however, in \autoref{app:morph_sim} we use simulated galaxies similar to our LBG sample to investigate how decreasing signal-to-noise affects the reliability of our fits. We find that the fits are generally reliable for low-$n$ galaxies with $S/N \gtrsim 100$. Noting that only two galaxies in \autoref{table:morphcatalog} fall below this rough threshold and that their \sersic\ model parameters appear well constrained according to the error estimates from \galfitm, we are confident in the reliability of the fits to the LBG samples we use above.

In hierarchical models, galaxies with bulge-dominated early-type morphologies are assembled by past mergers. More evolved galaxies that have experienced a number of past mergers should then have larger \sersic\ indices. Since we probe a young population by relying on a Lyman break selected sample, larger-than-expected \sersic\ indices could be an indication of more rapid, merger-driven morphological evolution. We note that the majority of our LBGs have more disk-like morphologies with $n<2.5$, and all of the protocluster LBGs fall below this line. The protocluster LBGs do not tend to have larger $n$ than field galaxies, nor do they tend to be larger, suggesting that the morphologies of LBGs in the protocluster are not evolving faster than their field counterparts at \zthree. We note that in \autoref{app:morph_sim} we find that our \sersic\ model fits typically recover $n$ smaller than the ``true'' value of $n$ due to the broadening effects of the PSF on the \sersic\ model. For the 4 protocluster LBGs with recovered \sersic\ indices $1.25 < n < 2$, we thus expect that the true \sersic\ index could be as much as a factor of 2 larger in the case of $S/N \gtrsim 100$, possibly indicating an underlying bulge dominated morphology smeared out by the effects of the PSF. However, we note that the PSF also affects the field LBGs, and given again that the protocluster and field LBGs cluster together strongly in the space of the \sersic\ model parameters we are confident in assessing that the protocluster LBGs are not more morphologically evolved than the field LBGs.

\input{table4.tex}

\subsection{Residual Image Analysis}
\label{sec:resid}

\begin{figure*}[p]
	\centering
	\includegraphics[width=0.815\textwidth]{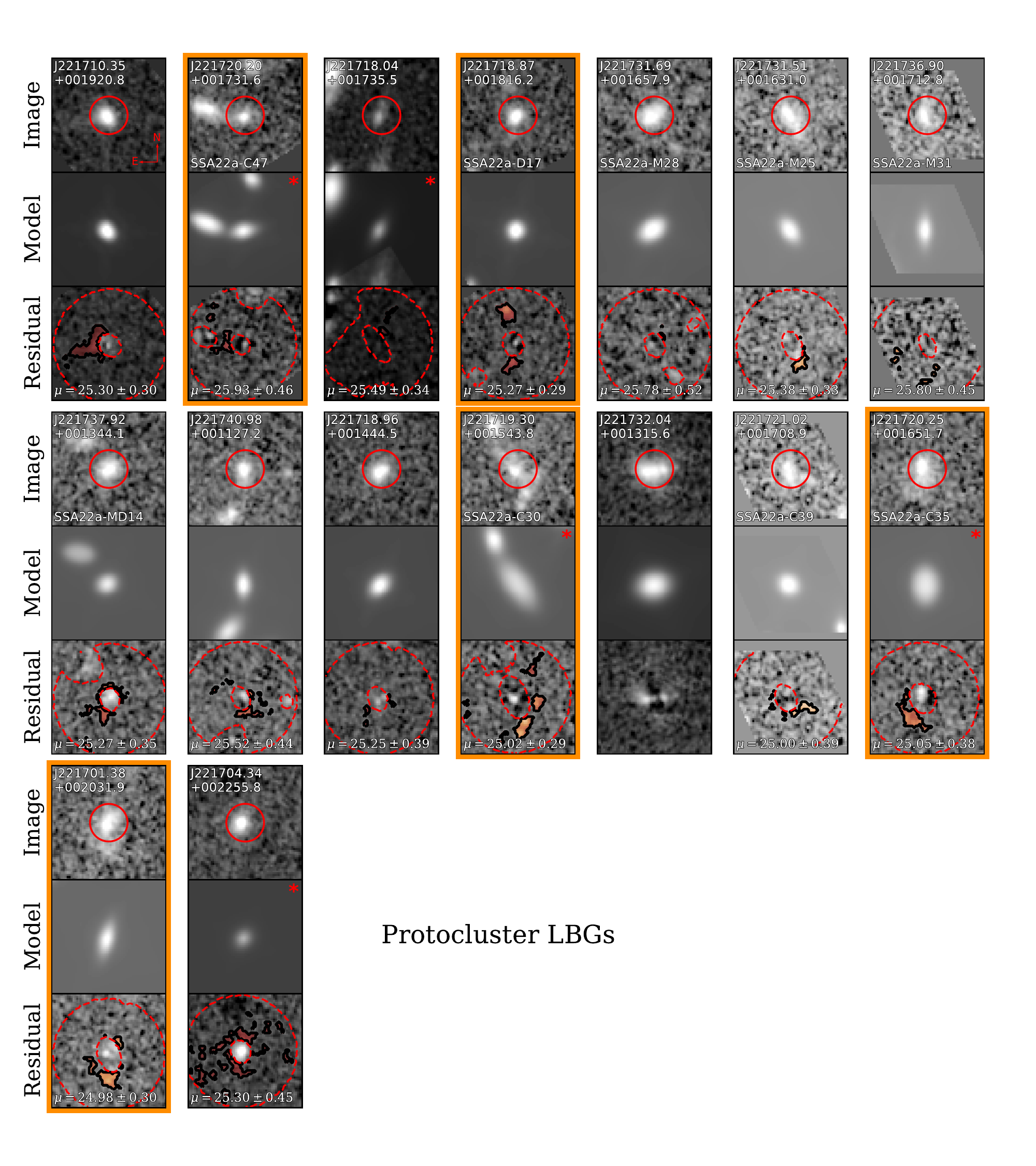}
	\caption{\label{fig:lbgmontageproto}SSA22 protocluster LBGs in our sample 
	with acceptable single \sersic\ fits. We display the original F160W
	image, model, and residual at the same normalization and stretch for each galaxy.
	In four cases we have re-fit the model 
	with adjustments to the position of the models to allow for cleaner tidal feature
	extraction; we mark these cases with a red star in the model panel. 
	The red circle shows a $10$ proper kpc diameter aperture at the position of the primary source.
	In the top-left corner of each F160W cutout, we print the galaxy's J2000 positional identifier;
	for LBGs in the \citetalias{steidel2003} catalog we print their catalog designation in the lower-left corner.
	In the residual panel, we highlight features extracted by the tidal feature pipeline in an orange colormap 
	and print the surface brightness of the extracted features in $\rm mag\ arcsec^{-2}$. The regions 
	inside the inner dashed red line and outside the outer dashed red line were excluded from feature extraction.
	Image blocks for galaxies where we have found a plausible tidal feature are outlined with a thick orange border.}
\end{figure*}

\begin{figure*}[t]
	\centering
	\includegraphics[width=0.815\textwidth]{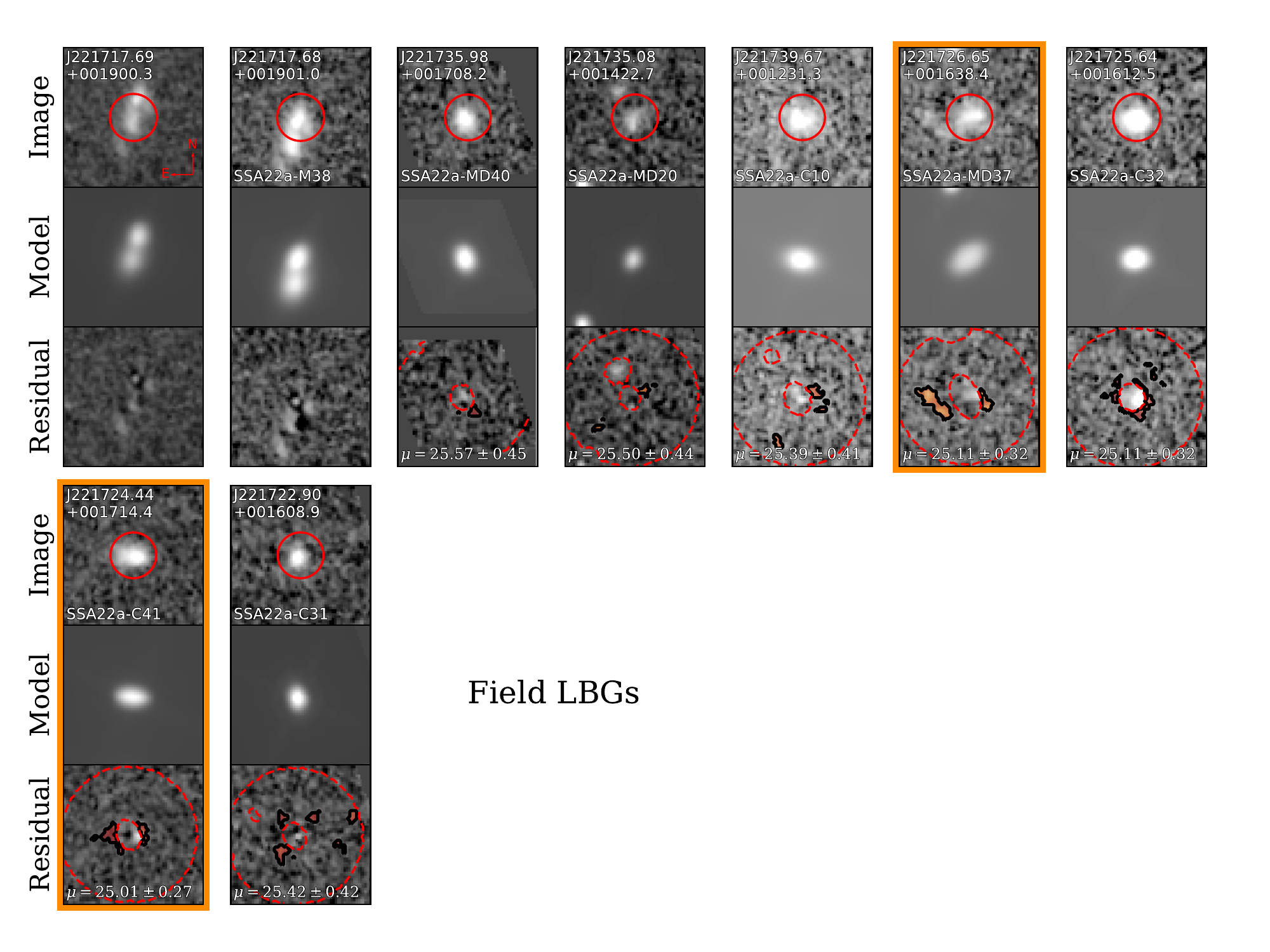}
	\caption{\label{fig:lbgmontagefield}SSA22 field LBGs in our sample 
	with acceptable single \sersic\ fits. All annotations are the same as \autoref{fig:lbgmontageproto}.}
\end{figure*}

For the subset of 25 SSA22 LBGs (16 protocluster and 9 field) in our sample with acceptable fits, we show the residuals after model subtraction in \autoref{fig:lbgmontageproto} and \autoref{fig:lbgmontagefield} for protocluster and field galaxies, respectively. We used the \texttt{Python}-based tidal feature extraction pipeline\footnote{\url{https://github.com/AgentM-GEG/residual_feature_extraction}} from \citet{mantha2019} to examine our single-\sersic\ fit residuals for evidence of potential tidal features related to recent mergers. Briefly, the \citet{mantha2019} method identifies flux- and area-wise significant contiguous pixel regions in residual images. We set the pipeline to search for connected regions of pixels $\ge 2\sigma$ above the background after convolution with a boxcar filter 3 pixels wide. Based on the average galaxy size at \zthree, we searched for residual features within 15 kpc of the main galaxy. To ensure that under-subtracted regions at the center of a galaxy were not erroneously detected as possible tidal features, we also set the pipeline to mask out an ellipse centered on the source position, which has major and minor axes scaled up 2.5 times from the \sextractor\ detection ellipse. Additionally, since the goal of the pipeline is to extract low surface brightness features associated with mergers, bright features are masked in the image before extraction, in order to exclude bright companion galaxies from being identified as tidal features.

Three galaxies in our LBG sample are deliberately excluded from this analysis: J221732.04+001315.6, a protocluster LBG, which appears to be 2--3 sources blended within 5 kpc, and both J221717.69+001900.3 and J221717.68+001901.0 (SSA22a-M38), a pair of field LBGs which are also blended together within 5 kpc, though they were detected and fit separately by our \galapagostwo\ pipeline. We excluded these blended galaxies even though they may be merging systems that are physically associated. In such close associates, the blending makes it difficult to reliably fit and mask the images in a way that prevents the pipeline from extracting under-fit components of the blended system as tidal features. 

To ensure the cleanest possible residuals, we re-fit the ``original'' models from the \galapagostwo\ pipeline for five galaxies with small adjustments to improve their positioning. For instance, the fitting cutout for J221718.04+001735.5 contains a bright, unrelated point source which was originally poorly fit with a \sersic\ model and adversely affected the positioning of the other models in the image. We fit this cutout again, with the point source properly modeled by a PSF, and allow the positions of the other models to vary. We also found that secondary galaxies in the cutouts containing J221720.20+001731.6 (SSA22a-C47) and J221719.30+001543.8 (SSA22a-C30) were under-subtracted due to offsets in the positions of their models; we re-fit them with the magnitude and shape parameters fixed, and the position parameters allowed to vary.

In the case of both J221720.25+001651.7 (SSA22a-C35) and J221704.34+002255.8, the main concentration of the galaxy appears offset from the center of the original \galapagostwo\ pipeline model, which is fit to both the concentrated component and an apparent fainter, diffuse component that extends asymmetrically to the southeast of the main concentration. To recenter the fit on the dominant, concentrated feature, we computed the centroid of the pixels $\ge 4\sigma$ above the background and re-fit the cutout with the primary model fixed to that position. The best fit model parameters of the primary galaxy do not change significantly in any of these cases, except for the position of the fits to SSA22a-C35 and J221704.34+002255.8, by construction.

We show the features extracted by the tidal feature pipeline in the residual panels of \autoref{fig:lbgmontageproto} and \autoref{fig:lbgmontagefield}, along with their surface brightness in $\rm mag\ arcsec^{-2}$ and the unmasked area in which they were extracted.

The extracted features are all of low surface brightness. The range of $2\sigma$ limiting surface brightness in our two-orbit images is 25.0--25.4 ${\rm\ mag\ arcsec^{-2}}$; in the single-orbit images the range is 24.6--25.1 ${\rm\ mag\ arcsec^{-2}}$. In terms of surface brightness alone, none of the extracted features are unambiguous; only four galaxies have extracted features with surface brightness brighter than the $2\sigma$ limit: J221710.35+001920.8, J221718.87+001816.2 (SSA22a-D17), J221719.30+001543.8 (SSA22a-C30), and J221701.38+002031.9. While the surface brightness of the features in the other galaxies is on the order of the limiting surface brightness, the sizes of the features in most cutouts indicate that they are unlikely to be due to noise alone. In one case, J221731.69+001657.9 (SSA22a-M28), the area of the largest residual feature is small enough to be consistent with noise, and we thus exclude the extracted features in this galaxy cutout from consideration as plausible tidal features; in the remainder of cases we estimate a probability $p < 0.02$ that the largest feature is due to noise, based on simulations of the image background. In classifying residual features as plausible tidal features resulting from merger activity, we focus on three additional criteria: (1) asymmetry with respect to the primary galaxy, (2) extension, and (3) plausible physical association with the primary galaxy. As a rule-of-thumb, we consider features that reach within 5 kpc of the primary galaxy's center to plausibly be physically associated with the galaxy.

While the residual feature in the J221710.35+001920.8 cutout is extended, asymmetric, and plausibly associated with the main galaxy, we note that it is positioned near the expected location of a diffraction spike from the WFC3 PSF. Though the extracted feature is low surface brightness, the galaxy is very concentrated in appearance, and the \sersic\ model fit is concentrated and visibly PSF-like in appearance. For these reasons, we do not consider the feature extracted from the J221710.35+001920.8 cutout to be plausibly tidal.

Two features are extracted in the SSA22a-D17 cutout, associated with clumpy features to the north and south of the main galaxy in the original image. The features are asymmetrical in size and shape, and both are within 5 to 10 kpc of the model barycenter. The residual features in the SSA22a-C30 cutout are also associated with clumpy structures which are apparent in the original image, to the southwest of the main galaxy. These clumpy features are also asymmetrical with respect to the main galaxy, and extend between 5 and 10 kpc away from the primary model's barycenter. Similar clumpy residual features have been observed by \citet{mantha2019} in mock two-orbit F160W observations of merging galaxies in the VELA cosmological simulations \citep{ceverino2014, zolotov2015}. The simulated observations suggest that similar features may be associated with the late stages of a major merger (i.e. 0.15--0.80 Gyr after the galaxies coalesce), when multiple nuclei are no longer apparent \citetext{see Figure 9 in \citealp{mantha2019}}. The feature extracted in the J221701.38+002031.9 cutout is brighter than the limiting surface brightness in the image, and appears to be asymmetrical and plausibly physically associated with the galaxy. Its shape and offset from the main concentration of the galaxy suggest that it may be tidal in origin, though this galaxy is in a single-orbit image. We are thus less confident in assessing this as a plausible tidal feature.

The apparent association of the small residual feature in the SSA22a-C47 cutout with both galaxies may suggest interaction; residual features bridging the two galaxies seem to be common in mergers \citetext{see, e.g., Figures 7 and 9 in \citealp{mantha2019}}. If these two galaxies are at the same redshift, they might then be a pre-coalescence merging pair, based on the residual feature and the apparent bridge between the galaxies in the original image. However, we have not found a spectroscopic redshift in the literature for the projected companion galaxy, nor do we have independent photometric redshifts for both galaxies, so we are unable to establish whether the apparent companion is physically close to the primary galaxy.

The irregularly shaped galaxy SSA22a-C35 has a diffuse feature offset to the southeast from the main concentration of the galaxy, with a surface brightness of $\approx 25\ {\rm mag\ arcsec^{-2}}$. This feature, based on its shape and plausible physical association with the galaxy, may also be associated with the late, post-coalescence stages of a merger. However, the data we used to fit this galaxy was taken at single-orbit depth, and the quality of the fit is poor. We are thus less confident in assessing this as a plausible tidal feature.

Based on the above, we find 2--5 residual features that may plausibly be tidal in our protocluster LBGs. The clumpy features associated with SSA22a-D17 and SSA22a-C30 are the most plausible, and the diffuse features associated with SSA22a-C35 and J221701.38+002031.9 are the least plausible. If we extrapolate this to a merger fraction based on the number of \citetalias{steidel2003} LBGs we applied the pipeline to, we find a protocluster LBG merger fraction of 0.22--0.44, comparable to the merger fraction we derive by na\"ive visual classification of \citetalias{steidel2003} LBGs in \autoref{sec:visual-classification}. If we include \citetalias{micheva2017} LBGs we find a protocluster merger fraction of 0.13--0.33.

Of the 7 field LBGs we have applied the pipeline to, J221735.08+001422.7 (SSA22a-MD20), J221726.65+001638.4 (SSA22a-MD37), and J221724.44+001714.4 (SSA22a-C41) have offset features apparent in the original image, though the feature near SSA22a-MD20 is bright enough to be masked by the residual extraction pipeline. For consistency with the above we consider only the features near SSA22a-MD37 and SSA22a-C41. These features are both consistent with the $2\sigma$ limiting surface brightness in their respective images. Both features are asymmetric and plausibly physically associated with the galaxy, though the feature in the SSA22a-C41 cutout is smaller and closer to the main concentration of the galaxy. We note the superficial similarities of the extracted features in the SSA22a-MD37 cutout to the clumpy features associated with SSA22a-D17 and SSA22a-C30, and conclude that they may be tidal features. The feature in the SSA22a-C41 cutout is more ambiguous; in the original image it appears that the galaxy is asymmetric, with the eastern side of the galaxy being fainter and more diffuse. The extracted feature is apparently associated with this diffuse region of the main galaxy, similar to SSA22a-C35 above, but smaller in size. We thus find this to be a low-confidence tidal feature, as with SSA22a-C35. Based on the 7 field LBGs we applied the pipeline to, we find a field merger fraction of 0.14--0.28, comparable to the protocluster merger fraction we found above, and the merger fractions derived by visual classification in \autoref{sec:visual-classification}. 

We note that conclusive or completely quantitative identification of residual features as being due to mergers is beyond the scope of this work. The relationship between the observed strength and shape of tidal features and the different stages of a merger is not yet fully explored, and will require comprehensive simulations to establish. Thus, rather than attempting to conclusively identify mergers with this technique, we have classified features only as ``plausibly'' tidal above, and we treat this method as a supplement to the more established methods of morphological analysis we use in the other sections of this paper.

To roughly estimate the mass of the plausible tidal features above, we used the protocluster and field SED models described in \autoref{sec:phys_analysis} to calculate mass-to-light ratios in the F160W band. We find $M/L = 0.45\ {\rm M_{\odot}/L_{\odot}}$ for the protocluster model and $M/L = 0.22\ {\rm M_{\odot}/L_{\odot}}$ for the field model, both assuming $z=3.1$. These are in agreement with rest-frame $B-$band mass-to-light ratios calculated from the models using the $B-R$ color relationship in \citet[][see their Appendix B]{zibetti2009}. Using the F160W mass-to-light ratios we derived, we find that the plausible tidal features associated with the protocluster galaxies have masses on the order of $10^9\ {\rm M_{\odot}}$, ranging from $\log_{10}M_{\star}/ \rm M_{\odot} = 9.10$ to $9.78$, suggesting that the largest of the clumps have masses comparable to the Small Magellanic Cloud. Assuming a typical stellar mass of $10^{10}\ \rm M_{\odot}$ for protocluster LBGs (see \autoref{fig:mainseq}) we find feature mass to total stellar mass ratios ranging from 0.13--0.60. These feature mass ratios, along with the compact or clumpy nature of some of the residuals (e.g., SSA22a-C30), suggest that this technique may be sensitive to minor mergers (mass ratio $< 0.25$), and that some of the features we extract may be infalling satellite galaxies.

\subsection{Non - Parametric Morphological Analysis}
\label{sec:npmorphologies}

To mitigate the limits imposed by requiring well-converged \sersic\ fits to our data, we also pursued non-parametric morphological analyses. We applied this analysis to the \zthree\ SSA22 \citetalias{steidel2003} and \citetalias{micheva2017} LBGs in our sample, excluding LBGs from the GOODS-N comparison sample. The Gini coefficient $G$ of a galaxy's flux \citep{abraham2003} and the second order moment of light statistic $M_{20}$ for the brightest $20\%$ of light from a galaxy \citep{lotz2004} can be used in concert to identify merger candidates. We used the definition of $G$ from \citet{lotz2004}:
\begin{equation}
	G = \frac{1}{\overline{\lvert f \rvert} n \left( n - 1 \right)} \sum_i^n \left( 2i - n - 1 \right)\lvert f_i \rvert,
\end{equation}
where $\{\lvert f_i \rvert\}$ contains the absolute values of the source flux, sorted from smallest to largest, and $\overline{\lvert f \rvert}$ denotes the mean of these values. The Gini coefficient describes the equality of the distribution of light in a galaxy, with values close to 0 indicating an egalitarian distribution of flux and values approaching 1 indicating an unequal distribution.

The total second order moment of light is defined as 
\begin{equation}
	M_{tot} = \sum_i^n f_i r_i^2 = \sum_i^n f_i \left[ \left( x_i - x_c \right)^2 + \left( y_i - y_c \right)^2 \right],
\end{equation}
where $r_i$ is the distance from the pixel containing flux $f_i$ to the center $(x_c,y_c)$ of the galaxy; in this context, the center is defined as the location that minimizes $M_{tot}$. $M_{20}$ is then computed as 
\begin{equation}
	M_{20} = \log_{10} \left( \frac{\sum_j^m f_j r_j^2}{M_{tot}} \right),
\end{equation}
where $m$ is the largest index satisfying
\begin{equation}
	\sum_j^m f_j \le 0.2 f_{tot}
\end{equation}
when $\{f_j\}$ contains the flux of the pixels sorted from largest to smallest. Values of $M_{20}$ close to 0 indicate excesses of flux further from the galactic centers, which may indicate star forming knots, multiple nuclei, or otherwise disturbed morphologies. Smaller (more negative) values indicate concentration of light at the center of the galaxy; local quiescent elliptical galaxies have average $M_{20} \sim -2.5$, for example \citep{lotz2004}.

Locally, the difference between mergers and ``normal'' galaxies is well-established in the Gini-$M_{20}$ plane; disturbed morphologies create an unequal, off-center distribution of light, so mergers tend to fall above the Gini-$M_{20}$ trend -- canonically, the line $G = -0.14 M_{20} + 0.33$ \citep{lotz2008} -- with larger values of $M_{20}$, while normal galaxies fall below. The Gini coefficient is expected to remain relatively unbiased at $z \gtrsim 2$ given high signal-to-noise and resolution better than $500\ \rm proper\ pc\ pixel^{-1}$; $M_{20}$ may be biased by the flattening of the angular size of features at high redshift, though its large dynamic range may still prove useful in distinguishing between disturbed and undisturbed morphologies \citep{lotz2004}. The physical resolution of our images is $496\ \rm proper\ pc\ pixel^{-1}$ at $z=3.09$, approaching the limit of what \citet{lotz2004} recommend. In \autoref{app:morph_sim}, we used simulations of galaxies similar to our LBG sample to investigate how $G$, $M_{20}$, and $C$ are biased by decreasing signal-to-noise, finding that they are relatively stable over the range of $S/N$ in our sample.

As a non-parametric analog to the S\'ersic index, we also calculated the concentration parameter $C$, defined as
\begin{equation}
	C = 5 \log\left(\frac{r_{80}}{r_{20}}\right),
\end{equation}
where $r_{80}$ and $r_{20}$ are the radii enclosing $80\%$ and $20\%$ of the total flux, respectively \citep{conselice2003}. Here, the total flux is defined as the flux contained within $1.5$ Petrosian radii. The value of $C$ tends to increase for more concentrated, bulge-dominated morphologies (i.e., with increasing $n$).

We adopted a similar method as \citet{lotz2004} to compute $G$, $M_{20}$, and $C$. We defined a new segmentation map for each detection by first computing its elliptical Petrosian radius $r_p$. We then smoothed the image of the galaxy with a Gaussian with $\sigma = r_p / 5$, measured the mean flux $\mu(r_p)$ of the pixels at $r_p$, and assigned any pixels with flux greater than $\mu(r_p)$ to the new segmentation map. Finally, we computed $G$ and $M_{20}$ for each detection using all the pixels included in this new segmentation map. It is possible for this process to fail if, for example, the Petrosian radius cannot be computed or the new segmentation map contains multiple disjoint features. In these cases (6 of the 72 galaxies with protocluster redshifts, all of which are LAEs) we do not report $G$, $M_{20}$, or $C$ in the catalog. However, we were able to recover all three quantities for all of the protocluster and field LBGs in our F160W images, as well as the 8 protocluster AGN in our images.

\begin{figure*}
	\centering
	\plotone{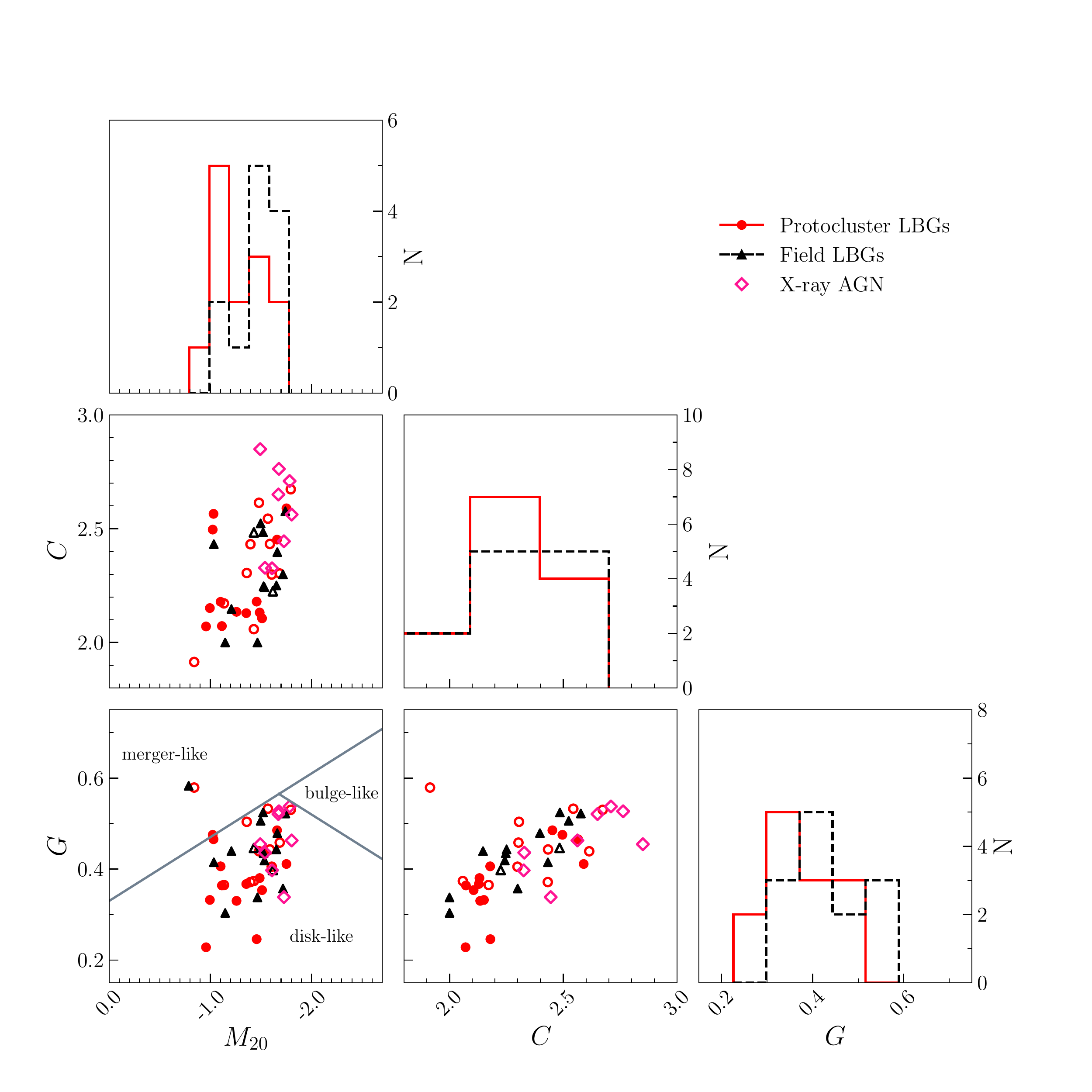}
	\caption{\label{fig:ginim20cplot}We plot the projections in $G-M_{20}-C$ space of the 13
	\citetalias{steidel2003} protocluster LBGs (filled circles), 11 
	\citetalias{micheva2017} (open circles) protocluster LBGs,
	13 \citetalias{steidel2003} field LBGs (filled triangles), 
	2 \citetalias{micheva2017} field LBG (open triangles), and 8 X-ray detected
	protocluster AGN (open diamonds), where $G$ is the Gini coefficient, $M_{20}$ is the second-order moment of light, 
	and $C$ is the concentration statistic. 
	For reference we also plot the $G-M_{20}$ classification regions of \citet{lotz2008}.
	While we show both \citetalias{steidel2003} and \citetalias{micheva2017} LBGs 
	in the scatter plots, we restrict the histograms
	to the \citetalias{steidel2003} LBG sample.}
\end{figure*}

We show the derived values of $G$, $M_{20}$, and $C$ for protocluster and SSA22 field LBGs in \autoref{fig:ginim20cplot}. Both \citetalias{steidel2003} protocluster and SSA22 field LBGs cluster together with median $(G, M_{20}, C)$ of $(0.37, -1.26, 2.15)$ and $(0.44, -1.52, 2.25)$, respectively. While most of the X-ray AGN fall in the same area of $G-M_{20}$ space as our LBG samples, they tend to be more concentrated than normal galaxies, falling mainly to the upper right of the trend in $G-C$ space, as expected in cases where the AGN dominates the rest-frame optical emission from the galaxy. 

In \autoref{fig:ginim20cplot}, we adopt the $G-M_{20}$ classifications from \citet{lotz2008}, which show that galaxies with $G > -0.14 M_{20} + 0.33$ are merger-like and galaxies with $G \le -0.14 M_{20} + 0.33$ and $G > 0.14 M_{20} + 0.80$ are bulge-dominated (i.e, Hubble classes E, S0, and Sa). The majority of our galaxies occupy the third region defined by these two lines, where galaxies with irregular and disk-dominated morphologies (i.e Hubble classes Sb, Sc, and Ir) fall at low redshift. However, there is significant concern about the use of cuts in the $G-M_{20}$ plane to classify mergers at high redshift. Artificial redshifting of simulated merging systems suggests that the typical $G-M_{20}$ criteria may miss a significant number of mergers at high redshift, and that any apparent trend toward the merger-like region of the $G-M_{20}$ plane may only be a function of mass \citep{abruzzo2018}. \citet{snyder2015} also find that the joint distribution of $G$ and $M_{20}$ is narrow at \zthree, even for diverse morphological types, due to the $G-M_{20}$ segmentation algorithm excluding the low surface brightness outer regions of the galaxy. As such mergers without clear-cut cases of multiple nuclei may not separate from normal galaxies in the $G-M_{20}$ plane.

It is also apparent from our analysis that the $G-M_{20}$ criteria might miss mergers: the systems in which we see plausible tidal features in \autoref{sec:resid} do not all fall above the merger-like/disk-like dividing line in \autoref{fig:ginim20cplot}, nor do the galaxies visually classified as mergers in \autoref{sec:visual-classification}. Only two galaxies fall in the merger region: J221740.98+001127.2, a \citetalias{micheva2017} protocluster LBG, and J221726.65+001638.4 (SSA22a-MD37) a \citetalias{steidel2003} field LBG. Neither of these galaxies has obvious multi-nuclear structure; rather, their elevated $G$ and $M_{20}$ values appear to be due to diffuse, asymmetric features that extend away from the main concentration of the galaxy. As we have noted in \autoref{sec:resid}, such features could be tidal structures associated with the late stages of a merger. In \autoref{sec:resid} we identified clumpy structures associated with SSA22a-MD37, though we found no significant residual features associated with J221740.98+001127.2. Given this inconsistency, and the fact that other galaxies in which we have identified plausible tidal features do not have $G$ and $M_{20}$ values consistent with the \citet{lotz2008} merger classification, we hesitate to draw any conclusions about the SSA22 protocluster merger fraction from the $G-M_{20}$ classification.

Protocluster LBGs also do not appear to be more morphologically evolved, or bulge-like, by the \citet{lotz2008} criteria, though given that the merger criterion evidently misses mergers we cannot conclusively apply the bulge criterion here. Additionally, the protocluster and field LBGs cluster along the same locus in the $G-C$ plane, with neither set of galaxies conclusively being more concentrated or bulge-dominated than the other.

Beyond using $G$ and $M_{20}$ to attempt to classify mergers, we also attempted to use them to distinguish between the protocluster and field populations. We show 1-D and 2-D KS test results for the individual and joint distributions of non-parametric morphologies in \autoref{table:ksall}. We again limited these tests to \citetalias{steidel2003} LBGs, but here we are restricted to the protocluster and the SSA22 field. The KS test results for the \citetalias{steidel2003} LBGs alone suggest that the the protocluster LBGs are flatter than their field counterparts: the test for $G$ admits the rejection of the null hypothesis, and the tests on the $G-M_{20}$ and $G-C$ joint distributions also suggest differences in the distributions of light of protocluster and field galaxies. However, the protocluster and field galaxies do not cleanly separate in any of the projections of $G$--$M_{20}$--$C$ space. Additionally, if we include the \citetalias{micheva2017} LBGs, all of the KS tests on the nonparametric morphological measures are consistent with the null hypothesis. Thus we are unable to draw any conclusions about morphological differences between the protocluster and field LBGs.

\subsection{Visual Classification}
\label{sec:visual-classification}

For a direct comparison of our rest-frame optical morphologies with \citet{hine2015}, we also pursued a similar visual classification scheme for protocluster candidate LBGs using F160W images. Cutouts of SSA22 LBGs were mixed with cutouts of LBGs from the GOODS-N field and blindly distributed to seven voters, who placed each galaxy in one of six categories, as defined in \citet{hine2015}:

\begin{itemize}
	\item{\textbf{C1}: One clearly distinct, compact nucleus}
	\item{\textbf{C2}: Single nucleus, but less compact or with minor asymmetry}
	\item{\textbf{M1}: Clear evidence of a second nucleus; all flux falls in a $1\arcsec$ 
	diameter aperture centered at the source position}
	\item{\textbf{M2}: Clear evidence of a second nucleus; some flux falls outside the 
	$1\arcsec$ aperture}
	\item{\textbf{M3}: More than two nuclei or more complex clumpy structure; some flux 
	falls outside the $1\arcsec$ aperture}
	\item{\textbf{M4}: More than two nuclei or more complex clumpy structure; all flux 
	falls inside the $1\arcsec$ aperture}
\end{itemize}

\begin{figure}
	\centering
	\plotone{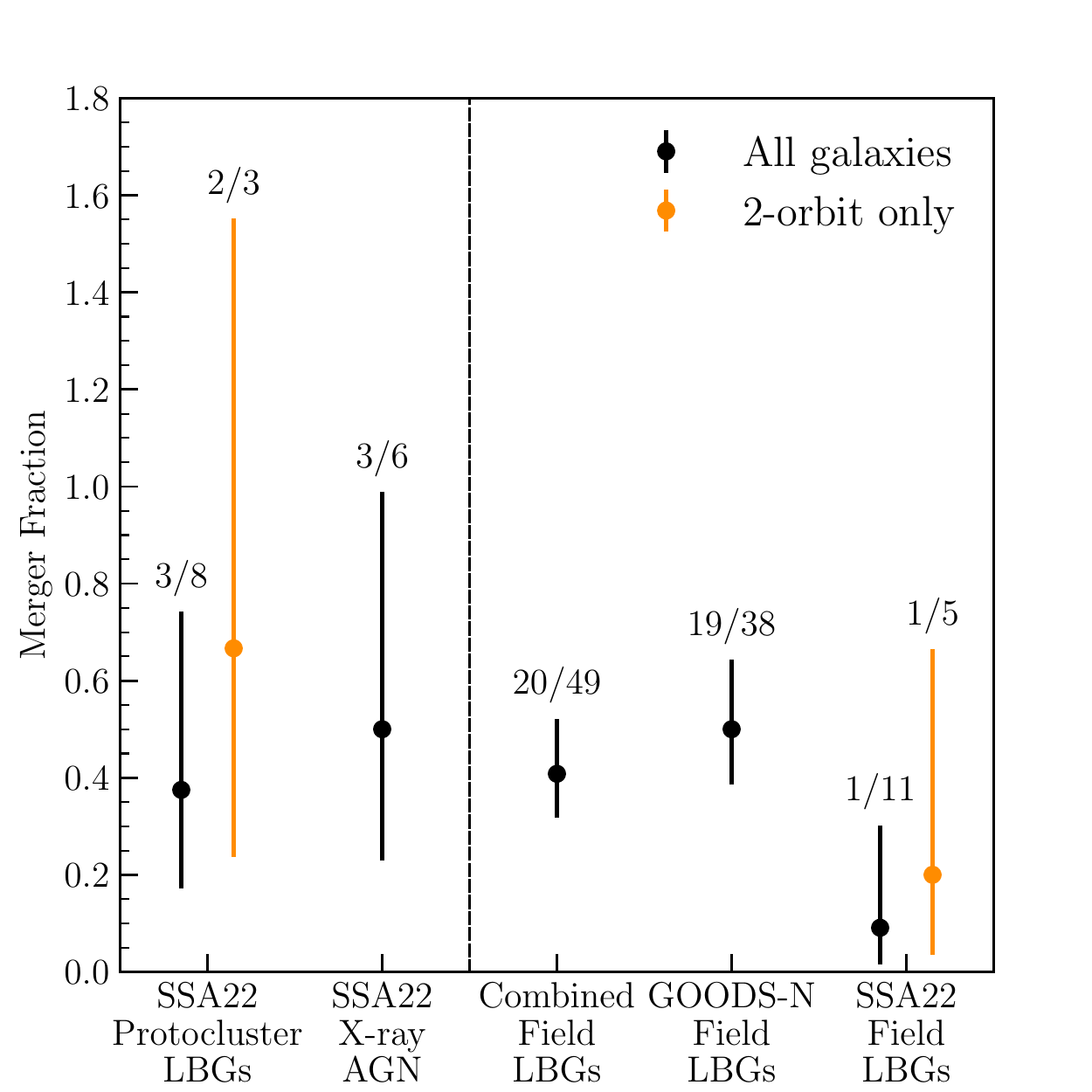}
	\caption{\label{fig:visual-classification}We show the fraction of \citetalias{steidel2003} LBGs in classes 
	$\rm M1-4$ for the protocluster, combined field, GOODS-N field, and SSA22 field, along with the fraction of X-ray selected
	protocluster AGN in merger classes. In orange, we show the merger fractions for the protocluster and SSA22 field 
	if we consider only galaxies in 2-orbit images. The uncertainty on the merger fraction is derived from Poisson statistics. 
	The vertical dashed line separates protocluster and field merger fractions.}
\end{figure}

Voters were asked to assign a confidence level from 1 (low) to 5 (high) to their classification, which we use to weight the votes. To artificially enforce consensus on our classifications, we summed the confidence scores of the voters for each galaxy and set a confidence threshold of 65\%, classifying galaxies with $\ge 65$\% of the total confidence in categories M1, M2, M3, and M4 as mergers, while galaxies with a $\ge65$\% of the total confidence in categories C1 and C2 were classified as isolated. This threshold was chosen to exclude cases where voters were split 4-to-3 between merger and isolated categories. This method excludes 2 protocluster LBGs and 1 X-ray AGN for which the confidence threshold is not met (i.e., the voters were not confident in classifying the galaxy as a merger or isolated). We classify these galaxies as ``ambiguous." A plurality of votes for these ambiguous galaxies place them in category C2, which allows for diffuse morphology and asymmetry about a single nucleus. This is consistent with the voting scheme in \citet{hine2015}, where many ambiguous galaxies were classified as C2. We also used the confidence scores to assign each galaxy a final category from the above list by weighting each vote by the voter's assigned confidence. We show the results of this analysis in \autoref{tab:visual-classification} and \autoref{fig:visual-classification}.

\input{table5_gehrels.tex}

The outcome of our visual classification of LBGs supports the results of the parametric and non-parametric morphological analysis, finding no significant difference in the observed fraction of galaxies in mergers between the SSA22 protocluster and the combined \zthree\ field. Specifically, we find merger fractions of $0.38^{+0.37}_{-0.20}$ for protocluster LBGs, $0.50^{+0.49}_{-0.27}$ for the X-ray selected protocluster AGN, and $0.41^{+0.11}_{-0.09}$ for the combined SSA22 and GOODS-N field samples. We note also that if we consider the case where all of the ambiguous cases mentioned above were, in reality, mergers, we would have a protocluster LBG merger fraction of $0.50^{+0.34}_{-0.22}$, which remains consistent with the GOODS-N and combined field merger fractions. However, our reported merger fraction of $0.09^{+0.21}_{-0.08}$ for the SSA22 field is based on one identified merger and is considerably lower than the merger fraction of $0.50^{+0.14}_{-0.11}$ for the GOODS-N field. Additionally, \citet{hine2015} report a merger fraction of $0.33 \pm 0.18$ for the SSA22 field based on F814W ACS imaging data. We note that we use a different redshift range to define \zthree\ field galaxies: \citet{hine2015} used galaxies from $2.5 \le z \le 3.5$ in their field samples, while we include only galaxies with $2.9 \le z \le 3.3$. However, this does not appear to be a significant driver of the very low merger fraction we observe in the SSA22 field: if we widen our criteria to include galaxies with $2.5 \le z \le 3.5$, the field merger fractions remain consistent with the values reported above.

Regardless of the redshift range we adopt, if we assume that the true merger fraction in the SSA22 field is equal to the GOODS-N merger fraction, we find a Poisson probability of $p < 0.03$ that we would observe one or fewer mergers among the SSA22 field LBGs classified here by chance alone. If we compare the SSA22 protocluster and field directly by assuming that the protocluster merger fraction is correct, we find that the elevation seen in \autoref{tab:visual-classification} and \autoref{fig:visual-classification} is apparently marginal: there is a Poisson probability $p < 0.09$ of observing a merger fraction less than or equal to the SSA22 field merger fraction, regardless of how we define the field.

Varying WFC3 IR imaging depth across the SSA22 field of view appears to play a role in the classification of galaxies as merging or isolated, and may be the primary driver of the protocluster-over-field elevation of the merger fraction that we observe in SSA22. If we divide the galaxies by depth, we find two-orbit merger fractions $0.67^{+0.89}_{-0.43}$ (2/3) for protocluster LBGs and $0.20^{+0.73}_{-0.06}$ (1/5) for SSA22 field LBGs, and one-orbit merger fractions $0.20^{+0.73}_{-0.06}$ (1/5) and $0.00^{+0.31}_{-0.00}$ (0/6) for the protocluster and SSA22 field LBGs, respectively. Again assuming that the merger fraction calculated from the GOODS-N LBG sample represents the true field galaxy merger fraction at \zthree, we calculate the Poisson probability of finding a merger fraction less than or equal to the observed SSA22 field merger fraction in the two-orbit (one-orbit) images to be 0.29 (0.05). If we evaluate the significance of the protocluster-over-field elevation after separating the galaxies by depth, we find that the elevation is no longer significant: if we assume the merger fraction from the two-orbit (one-orbit) protocluster galaxies, we find a Poisson probability of 0.16 (0.30) of observing a merger fraction less than or equal to the SSA22 field merger fraction at the same depth.

We take this to mean that our classification misses some mergers at single-orbit depth, consequently underestimating the SSA22 field merger fraction, since the sample of SSA22 field galaxies we classified coincidentally contains a larger proportion of galaxies observed at single-orbit depth. Our merger classifications require that the irregular morphological features induced by a merger be discernible by eye; it is possible that the lower signal-to-noise in the single-orbit images obscures the low surface brightness features associated with the late stages of a merger. Given these issues, the small number of sources in our samples, and the small number of sources observed at two-orbit depth, our constraints on any enhancements on merger fractions in the protocluster compared to the field are weak and not statistically significant at present.

The wavelength range of our observations is also expected to affect how many galaxies are classified as mergers. \citet{hine2015} used rest-frame UV ACS F814W observations to classify their galaxies. Over their adopted redshift ranges, this probes wavelengths (based on the F814W reference wavelength) $\sim1960-1990\ \rm\AA$ in the protocluster and $\sim1790-2310\ \rm\AA$ throughout the range $2.5 \le z \le 3.5$. High angular resolution and the ability to trace star formation have made rest-frame UV observations a typical choice for merger classification \citep[e.g.][]{lotz2004,lotz2006}, though we note that the patchiness of UV observations due to individual star forming clumps may make otherwise ``ordinary'' star-forming galaxies look irregular. In our case, our F160W observations probe (again based on the filter reference wavelength) $\sim3740-3790\ \rm\AA$ in the protocluster and $\sim3580-3950\ \rm\AA$ throughout the range $2.9 \le z \le 3.3$. Over this range of redshift the $4000\ \rm\AA$ break moves through the F160W bandpass. By construction and the cosmological constraints of observing at \zthree, the galaxies in our LBG sample are dominated by young stellar populations and consequently have weak $4000\ \rm\AA$ breaks (see, e.g. the model SEDs in \autoref{fig:meansfhplot}). Since there is significant continuum emission on either side of the break in our galaxies, we do not expect that observing across the $4000\ \rm\AA$ break should impact our morphological classifications in any significant way.

\section{SED Fitting and Physical Property Analysis}
\label{sec:phys_analysis}

To quantify the stellar mass distribution of our sample, we fit the spectral energy distributions (SED) of LBGs in the redshift range $2.9 \le z \le 3.3$. To avoid biasing our measurements of stellar masses and star formation rates, we exclude galaxies known from the literature to be hosting AGN and galaxies with rest-frame 8-32 keV luminosities $\ge 10^{43.5} {\rm\ erg\ s^{-1}}$ \citep{lehmer2009b}. We require the SSA22 galaxies we fit to have optical-to-NIR photometry available from \citet{kubo2013}, and we extract F160W fluxes from our own images. For consistency with the \citet{kubo2013} photometry, we deconvolve our F160W images from the PSF described in \autoref{sec:data_reduction}, smooth the result to a Gaussian PSF with a $1\arcsec$ FWHM, and extract photometry from a $2\arcsec$ diameter circular aperture. We use all of the filters listed in the top half of \autoref{table:sedfilter}, where available. Some of our SSA22 galaxies are not detected in the $u^*$ band due to their strong Lyman breaks, and some galaxies do not have photometry available in all 4 IRAC bands. We exclude missing and non-detected bands from the fits.

\input{table6.tex}

For the GOODS-N sample, we used the photometric uncertainties derived by \citep[][accepted in ApJ]{doore2021}. These uncertainties were recalibrated to include systematic uncertainties beyond single-instrument calibration, including the use of different photometric methods and systems in the observations, uncertainty and variation in Galactic extinction, blending of sources, and systematic effects created by the assumptions of our SED model.

We performed SED fitting using \ligh\ \citep{eufrasio2017}, which fits non-parametric star-formation histories (SFH) in discrete, variable or fixed-width stellar age bins. We made use of the most recent update to \ligh, which uses an adaptive Markov Chain Monte Carlo (MCMC) algorithm \citep{doore2021}. We chose SFH bins of $0-10\ \rm{Myr}$, $10-100\ \rm{Myr}$, $100\ \rm{Myr} - 1\ \rm{Gyr}$, and $1 - 2\ \rm{Gyr}$, where the upper age limit of the final bin is allowed to vary based on the age of the Universe at the redshift of the galaxy being fit. We assumed a Kroupa IMF \citep{kroupa2001} and fit using two metallicities: $Z_{\Sun}$ and $0.655 Z_{\Sun}$, corresponding to the average metallicity of the Universe at $z=3.1$ as given by the best-fit model in \citet{madau2017}. To generate the stellar population models, we used \texttt{P\'EGASE} \citep{fioc1997, fioc1999}, running it once for each metallicity. For intrinsic attenuation, we adopted a \citet{calzetti2000} extinction law, modified as in \citet{noll2009} to include a UV bump at 2175 \AA\ and a parameter $\delta$ to control the slope of the attenuation curve. We further modified the attenuation curve by including a birth cloud component, which is applied to the emission from the stars in only the youngest age bin. Our SED model then has a total of 7 free parameters: 4 SFH coefficients and 3 attenuation parameters. For a more thorough description of the stellar emission and attenuation prescriptions available in \ligh, we refer the reader to \citet{eufrasio2017} and \citet{doore2021}. To account for Galactic reddening, we used the standard \citet{fitzpatrick1999} curve. The Galactic $A_V$ varies with the position of each galaxy, based on the Galactic dust extinction estimates of \citet{schlafly2011}, which we retrieved using the IRSA \texttt{DUST} web application\footnote{\url{https://irsa.ipac.caltech.edu/applications/DUST/}}. 

We find the quality of our fits to both SSA22 LBGs and GOODS-N LBGs acceptable based on the distributions of $\chi^2$ for each sample. For the fits with $Z=0.655Z_{\odot}$ the median and 16th to 84th percentile range of the $\chi_{min}^2$ distribution is $5.79_{-2.98}^{+3.69}$ for the SSA22 LBGs we discuss below, with a median of 6 degrees of freedom, and $11.03_{-4.58}^{+9.45}$ for the GOODS-N LBGs, from a median of 10 degrees of freedom (note that the number of degrees of freedom is larger for the GOODS-N galaxies due to the larger number of available bands; see \autoref{table:sedfilter}). This corresponds to probabilities $p_{null}=0.40_{-0.29}^{+0.39}$ for the SSA22 LBGs and $p_{null}=0.24_{-0.20}^{+0.47}$ for the GOODS-N LBGs; here we define $p_{null}$ as the probability of accepting the hypothesis that the data are generated by the model. The majority of fits are thus not ruled out by a $\chi^2$ test. The quality of the fits does not change significantly for the fits with $Z = Z_{\odot}$.

We show example SED and SFH fit results in \autoref{app:SED}. The SFH parameters are sampled from the last 1000 steps of the MCMC chains. We use the sampled SFH to calculate the stellar mass of each galaxy, and compute the recent star formation rate (SFR) over the last 100 Myr as the age-bin weighted average of the most recent two bins of the SFH. The mass-weighted age is computed by weighting the average age of the stars in each bin by the mass in the bin. 

\begin{figure*}
	\centering
	\plotone{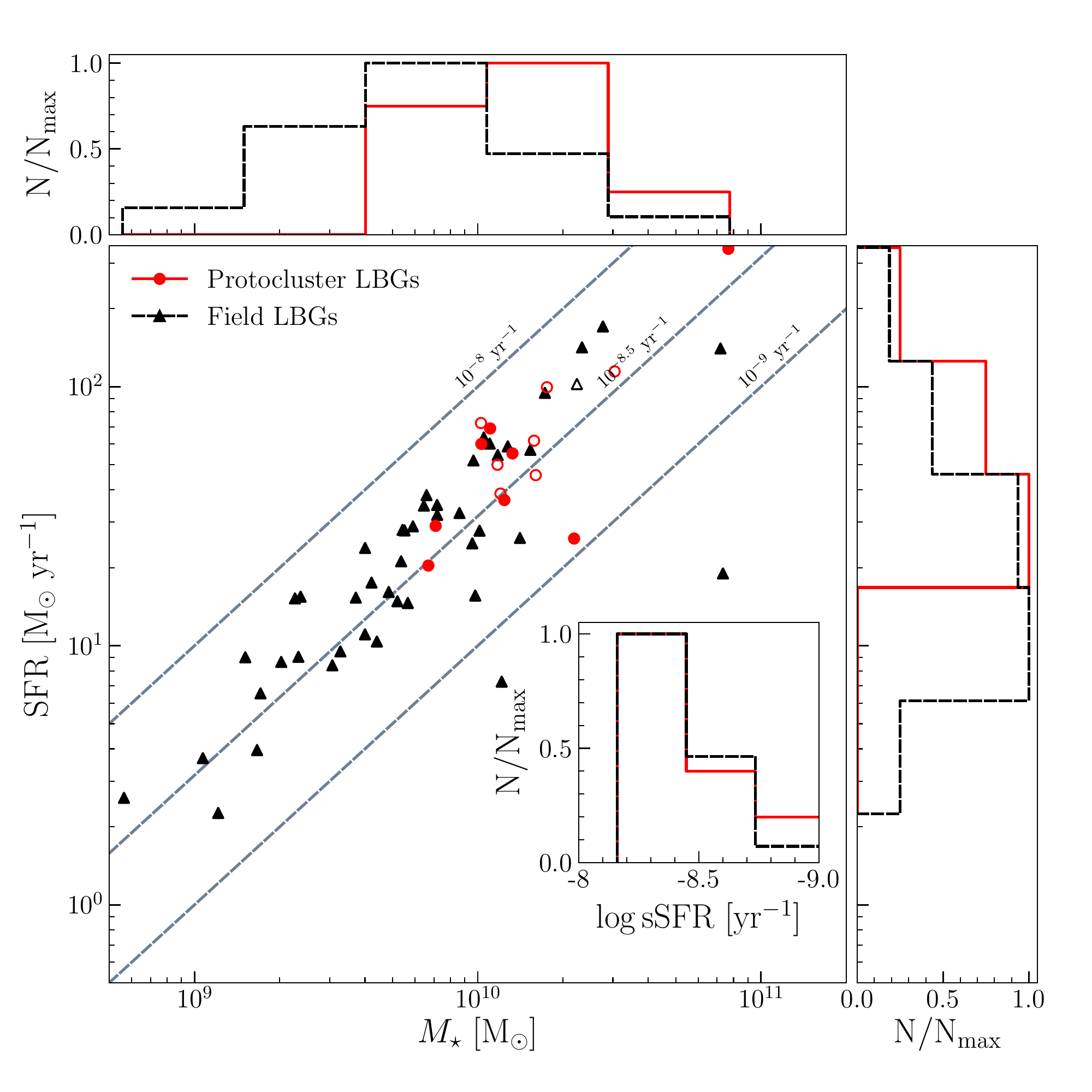}
	\caption{\label{fig:mainseq}Recent SFR versus stellar mass $M_{*}$ for protocluster and field 
	LBGs from the SED fits with $Z = 0.655 Z_{\odot}$. LBGs from the \citetalias{steidel2003} catalog 
	are shown as filled symbols, and LBGs from the \citetalias{micheva2017} catalog as open symbols. 
	For reference we show dashed lines of constant specific star formation rate (sSFR) in gray, 
	covering $10^{-9}\ {\rm yr^{-1}}$ to $10^{-8}\ {\rm yr^{-1}}$ at increments of 0.5 dex. In the 
	histograms in the margins we show the distributions of SFR and $M_{*}$, and in the inset histogram 
	we show the distribution of sSFR. For the histograms we again include only LBGs from the 
	\citetalias{steidel2003} catalog. The protocluster LBGs appear to trend toward larger masses 
	and SFR compared to the field LBGs.}
\end{figure*}

We show the distributions of stellar mass, SFR, and specific star formation rate (sSFR) in \autoref{fig:mainseq}. The SSA22 protocluster LBGs largely appear to follow the same star-forming main sequence as the field LBGs, though they populate the upper end. The typical star formation rates of our protocluster LBGs, 20--200 ${\rm M_{\odot}\ yr^{-1}}$, are significantly smaller than the IR-dervied SFRs for the DSFGs in the core region of the protocluster, which typically range from $\sim10^2$--$10^3\ {\rm M_{\odot}\ yr^{-1}}$ \citep{umehata2015,kato2016}; that is, we are not probing the most intensely star-forming population of the protocluster. Three of the eight galaxies hosting X-ray detected AGN, which we excluded from our SED fitting, have ALMA derived SFR $\approx 220$--$410\ {\rm M_{\odot}\ yr^{-1}}$, and the remaining five, which are not ALMA-detected, have upper limits $<130$--$210\ {\rm M_{\odot}\ yr^{-1}}$. These upper limits are consistent with the LBGs in our sample with the largest SFRs.

Two-sample KS tests comparing the SFH-derived properties of \citetalias{steidel2003} protocluster and field LBGs (see \autoref{table:ksall}) indicate a significant difference between the protocluster and field distributions of stellar mass; in \autoref{fig:mainseq} the protocluster galaxies cluster at higher masses than field galaxies. One of the \citetalias{micheva2017} LBGs, J221718.04+001735.5, is best fit by an extremely high SFR on the order of $10^3\ {\rm M_{\odot}\ yr^{-1}}$. Visual inspection of this galaxy (see \autoref{fig:lbgmontageproto}) shows that there is a bright unrelated point source nearby, which may be blended with the galaxy in near-IR photometry, producing an IR-heavy, high-attenuation, high-SFR best-fit SED with $p_{null}=0.99$. We therefore exclude it when we perform KS tests on the combined \citetalias{steidel2003} and \citetalias{micheva2017} LBG samples. The KS tests on the combined LBG samples indicate significant differences between the protocluster and field distributions of stellar mass and SFR, with protocluster galaxies having, on average, larger masses and larger SFR. There does not appear to be a significant difference between the sSFR distributions of the protocluster and field galaxies (as visible in \autoref{fig:mainseq}) or the mass-weighted age distributions of the protocluster and field galaxies.

\begin{figure*}
	\centering
	\includegraphics[width=\textwidth]{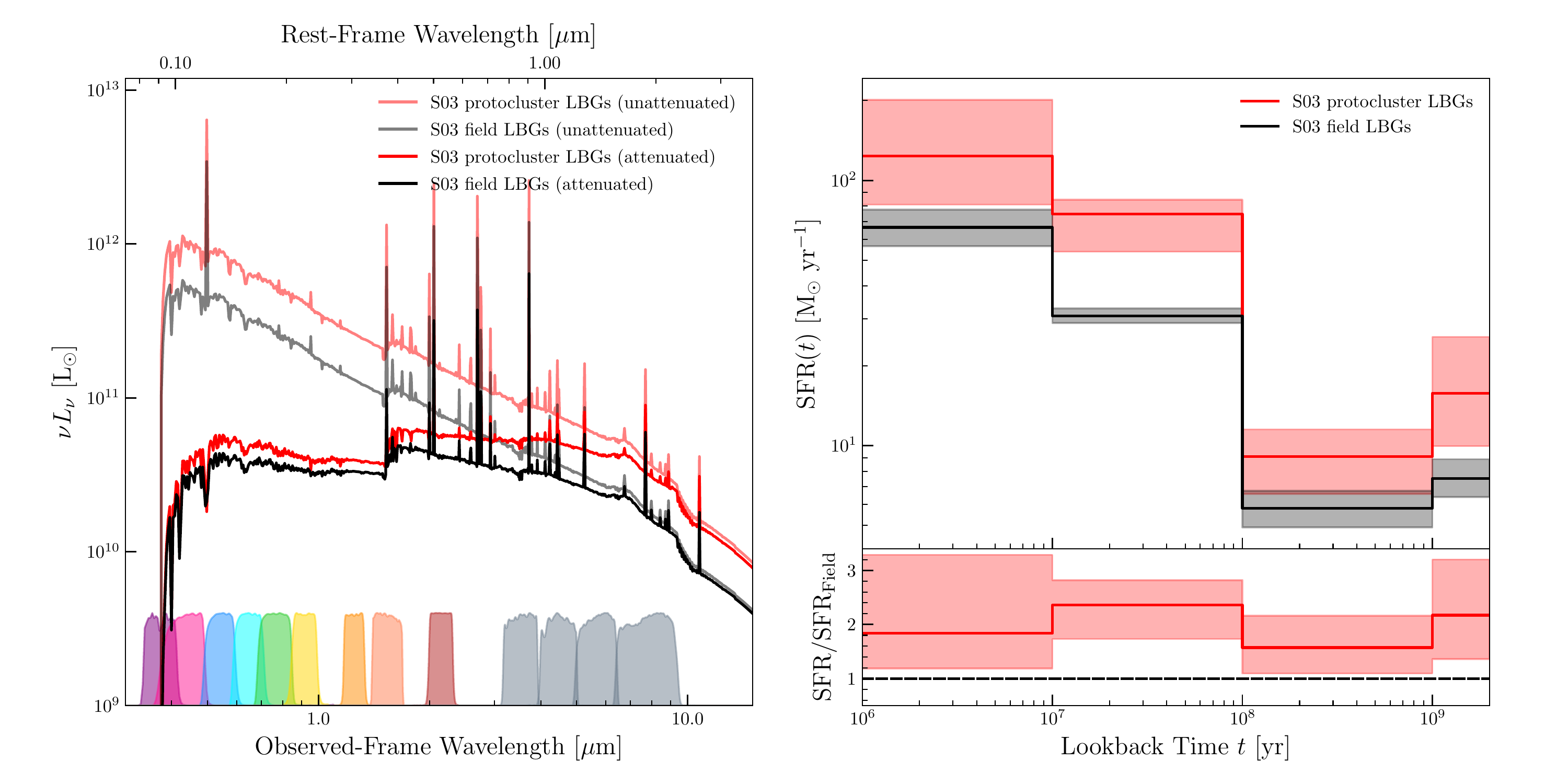}
	\caption{\label{fig:meansfhplot}\textit{Left panel}: We show the sample average model SED 
	for the 8 \citetalias{steidel2003} protocluster (red) and 45 \citetalias{steidel2003} field 
	(black) LBGs for which we have SED fits, both before (faded) and after (solid) application of 
	the sample average attenuation model. Underneath the SED we show the filter curves for the 
	bands we use for fitting for the SSA22 sample (see \autoref{table:sedfilter}), where coverage 
	could be improved throughout the mid-IR to better constrain the attenuation in protocluster 
	galaxies. The protocluster model is more attenuated than the field model. \textit{Right panel}: 
	We show the sample average SFH of the same protocluster (red) and combined field (black) samples. 
	Below, we show the ratio of the protocluster SFH to the field SFH, with a dashed line at unity 
	for reference. The shaded regions show the 16\textsuperscript{th} to 84\textsuperscript{th} 
	percentile range. In both panels we show only the results from fitting with $Z = 0.655 Z_\odot$. 
	For this plot we have set the redshift for both models at $z=3.1$.}
\end{figure*}

We took advantage of our non-parametric SFH fitting technique to investigate the average SFH of SSA22 protocluster LBGs. The SFR $\psi_i$ in each of our four stellar age bins is fit as a free parameter. We constructed a sample average SFH chain for the protocluster and field samples by averaging the $\psi_i$ values across each sample's chains. For this exercise we used the last 1000 values of $\psi_i$ in the MCMC chains, thus yielding sample average SFH chains with 1000 entries. We then sampled the sample average SFH chains to construct the average SFH and model SED of both samples, which we show in \autoref{fig:meansfhplot}. Regardless of the assumed metallicity, we find that the SFH of the protocluster sample is significantly elevated over the combined field SFH; for the fits with sub-solar metallicity we find that the SFH is more significantly elevated at the earlier times, while for solar metallicity we find that the elevation is more significant for the most recent age bin. For an assumed metallicity of $Z = 0.655 Z_\odot$ ($Z = Z_\odot$) the maximum SFR enhancement for \citetalias{steidel2003} protocluster LBGs is $2.36_{-0.63}^{+0.46}$ ($2.02_{-0.70}^{+0.82}$)\footnote{SED-fit derived parameters and their uncertainties are reported as the median and 16th to 84th percentile range of the last 1000 steps in the MCMC chain.} in the $10-100$ Gyr ($0-10$ Myr) stellar age bin. We list the protocluster-over-field SFH ratio for both metallicities and each stellar age bin in \autoref{tab:sfr_enhancement}.

Due to the elevation of the mean protocluster SFH over the mean field SFH, the mass of the mean protocluster LBG as determined from the mean SFH is greater by a factor of 1.99 than the mean field LBG: the mean \citetalias{steidel2003} protocluster LBG has $\log M_*/{\rm M_\Sun} = 10.31_{-0.07}^{+0.07}$ while the mean \citetalias{steidel2003} field LBG has $\log M_*/{\rm M_\Sun} = 10.01_{-0.02}^{+0.02}$. We find that the mean protocluster LBG has a mass-weighted age consistent with the mean field LBG: $\log t_{AGE}/{\rm yr} = 8.85_{-0.04}^{+0.04}$ for the protocluster LBGs, and $\log t_{AGE}/{\rm yr} = 8.84_{-0.02}^{+0.02}$ for the field LBGs.

If we construct sample average attenuation curves by treating the attenuation parameters in the same way as the SFH, we find that the protocluster LBGs are more attenuated than their field counterparts. For the fits with $Z=0.0655 Z_\odot$ the optical depth in the rest-frame $V-$band, $\tau_V$, is $0.49^{+0.06}_{-0.07}$ for the mean \citetalias{steidel2003} protocluster LBG and $0.19^{+0.02}_{-0.02}$ for the mean \citetalias{steidel2003} field LBG.

The larger contribution to the SED from the older stellar population and the increased attenuation together produce a mean protocluster LBG SED slightly redder than the mean field LBG SED. We computed IR colors from the model SEDs in \autoref{fig:meansfhplot}, finding that protocluster and field LBGs may be distinguished by future observations with JWST bands: $J-{\rm F444W}=1.68_{-0.43}^{+0.46}$ for the protocluster model, while $J-{\rm F444W}=1.31_{-0.17}^{+0.17}$ for the field model. However, both SED models are still dominated by young stars. Given the uncertainties on the model parameters, these colors remain uncertain, and color differences between protocluster and field LBGs remain speculative. Observations of the protocluster with JWST, which, with NIRCam, could reach F444W $S/N \approx 60$ in 1900 s exposures, will vastly improve the constraints of our SED models across the 4000 \AA\  break, allowing to reduce the uncertainties on our models and determine whether the color differences we have extrapolated from the models are real.

\input{table7.tex}

\section{Discussion}
\label{sec:discussion}

\subsection{Star Formation and AGN Enhancement in the SSA22 Protocluster}
\label{sec:mainpoints}

\citet{lehmer2009} suggested two plausible explanations for how the SSA22 protocluster environment might lead to the observed AGN fraction excess: SMBH accretion activity may be increased through (1) more frequent accretion episodes - possibly triggered by major mergers - and higher SMBH accretion rates in the dense regions of the protocluster, or (2) an increase in the X-ray luminosity of protocluster SMBHs due to the presence of more massive galaxies (and hence SMBHs) in the protocluster.

We have tested scenario 1 by searching for evidence of major mergers in protocluster LBGs detected in the fields of the X-ray detected protocluster AGN. We note that we have not focused directly on the AGN; at the wavelengths we probe it is difficult to extract morphological information from the AGN, as the AGN contributes significantly to the observed light, resulting in a point-source-like profile superimposed on the host galaxy's light profile (see, e.g., \autoref{fig:ginim20cplot}). We have instead focused on the inactive LBG population, to attempt to discern how mergers contribute to the overall growth of galaxies in the protocluster. If major mergers are a significant factor in the growth of galaxies in the protocluster, we would expect to observe differences in the morphologies of protocluster and field LBGs. Our results from quantitative and visual morphological analyses suggest that this is not the case: our samples of protocluster and field LBGs appear to be drawn from the same morphological population. We find a marginal result that the protocluster LBGs from the \citetalias{steidel2003} catalog appear to have larger values of $G$ (indicating flatter distributions of light) than their field counterparts, but when we take other Lyman-break selected galaxies from the \citetalias{micheva2017} catalog into account, we find that the \sersic\ parameters and nonparametric morphologies of protocluster and field LBGs are consistent with each other. Our results from model-fitting suggest that the majority of protocluster LBGs have \sersic\ indices $n < 2.5$, consistent with disk-dominated light profiles, even accounting for the effects of noise and the PSF (see \autoref{app:morph_sim}) which may cause $n$ to be underestimated by up to $50\%$.

While previous work by \citet{hine2015} found a marginally elevated merger fraction in the protocluster, we find that the merger fraction among X-ray AGN in the protocluster is consistent with both the merger fraction among \citetalias{steidel2003} protocluster and field LBGs, though we are limited by the small number of protocluster LBGs we are able to use. We have attempted to go beyond the typical methods for counting mergers by examining the residuals after subtracting \sersic\ models for evidence of mergers, finding a merger fraction in rough agreement with the one derived by visual classification, though this is also limited by small numbers.

The increased merger fractions in \citet{hine2015} may be due to the influence of star formation on their F814W observations, probing the rest-frame UV. Individual UV-luminous clumps of star formation may be difficult to discern by eye from multiple nuclei: in 6 out of 10 cases, the classifications they assigned to protocluster LBGs often indicated ``more than two nuclei or complex clumpy structure" rather than clear-cut cases of a double nucleus. In addition, \citet{hine2015} found that the rest-frame UV asymmetry in protocluster LBGs, often used as a merger diagnostic, indicates fewer mergers than their visual classification; however, at high redshift, calculation of the asymmetry statistic $A$ suffers from the same limitations on resolution and signal-to-noise per pixel as we have encountered in computing our own nonparametric morphologies. Comparisons to protocluster merger fractions from the literature outside of SSA22 are limited; \citet{lotz2013} found a merger fraction of $0.57_{-0.14}^{+0.13}$ in a $z=1.62$ protocluster (XMM-LSS J02182-05102; also called IRC-0218A). By comparison they measure a field merger fraction of $0.11\pm0.03$, indicating significant enhancement of merger activity in the protocluster. The enhancement of merger activity in an overdense environment is also in line with theory: in studies of Millennium simulation merger trees, \citet{fakhouri2009} found that overdensity enhanced merger rate relative to the mean field environment at all redshifts $z\le2$. However, their results also suggest that this enhancement grows weaker with increasing redshift, with overdensities at $z\sim2$ seeing less merger rate enhancement relative to overdensities at lower redshift \citetext{see e.g., Figure 5 in \citealp{fakhouri2009}}.

Our discussion thus far has been focused on major mergers (mass ratio $\ge 0.25$), which are expected to cause the largest and clearest morphological disturbances and possibly trigger an AGN phase. However, if the overall merger rate is enhanced in overdense environments, minor mergers ($0.10 \le$ mass ratio $< 0.25$) should also be more abundant. Our techniques are not particularly well suited for detection of minor mergers, largely due to surface brightness limits making detection of low-mass satellite galaxies difficult. In particular, the algorithm used to create the $G-M_{20}$ segmentation map tends to exclude low surface brightness satellite galaxies, including only the bright core region of the primary galaxy and thus making the $G-M_{20}$ merger diagnostic insensitive to galaxies accreting lower-mass satellites. In cases where low-mass satellites are segmented properly, their presence still may not move the system in the merger region of the $G-M_{20}$ diagram. \citet{lotz2010} plotted tracks in $G-M_{20}$ space for the course of mergers at a variety of mass ratios and viewing angles, finding that the $G-M_{20}$ diagnostic is not very sensitive to the early stages of low mass ratio mergers. They also find that flyby cases with low mass ratios are not cleanly diagnosed by the $G-M_{20}$ diagnostic and do not trigger long-lasting asymmetries that might be visible by eye. For the visual classifications we performed, the categories we asked the voters to use were designed with major mergers in mind. However, our residual analysis may be more sensitive to minor mergers. Estimating the feature masses and feature mass to total mass ratios of the features we have extracted from the residual images, we found that the masses of some features are consistent with dwarf galaxies, and that the mass ratios approach 0.10 at the low end. Some of the clumpy features we see in the residuals may thus be infalling satellites. While it is possible that low-mass ratio or early stage mergers our techniques are less sensitive to could contribute to the enhancement of star formation in the protocluster, studies vary on whether mergers do \citep[e.g.,][]{zamojski2011, kartaltepe2012} or do not \citep[e.g.,][]{targett2011} play a significant role in triggering bursts of star formation at $z \sim 2$, suggesting that the influence of mergers on star formation may vary with redshift and among galaxies selected by different methods.

At $z>1$, the highest luminosity (bolometric luminosity $\gtrsim 10^{46}\ {\rm erg\ s^{-1}}$) AGN are preferentially found in disturbed systems, believed to be evidence of recent mergers, though AGN luminosity and merger fraction both scale with mass at fixed redshift. \citet{mcalpine2020} find, on the basis of EAGLE simulations, that while high-luminosity AGN are more likely to be found in mergers, major mergers are only an effective driver of AGN fraction enhancement for lower-mass ($<10^{11}\ {\rm M_{\odot}}$) host galaxies. Five of the eight protocluster X-ray AGN in our F160W images (those without broad optical emission lines) have stellar mass estimates from optical SED fitting by \citet{kubo2015}, finding stellar masses in the range 0.3--$2\times10^{11}\ {\rm M_{\odot}}$, on the edge of where major mergers are expected to be effective triggers for AGN activity from EAGLE simulations.

We note, however, that our reliance on a Lyman-break selected sample excludes some high mass galaxies, which are more likely to be found in mergers. Indeed, \citet{kubo2017} found some evidence of merger-driven evolution in a group of massive quiescent galaxies at the protocluster core. It may be then that while our results suggest only that the protocluster LBG population (or equivalently, galaxies in an LBG phase of their evolution) are not any more likely to be found in major mergers than their field counterparts and that the protocluster overall may not be a more merger-rich environment than the field, mergers may still play a role in the evolution of the most massive protocluster galaxies.

While our results suggest that accretion due to major mergers is likely not a major environmental difference between the protocluster and field, and hence not the primary source of the X-ray enhancement in the fields we have studied here, galaxy-scale accretion is not ruled out. Recent simulations \citep[e.g.][]{romanodiaz2014} taken together with IFU observations also suggest that major mergers may not be the dominant mode for the accumulation of stellar mass in the SSA22 protocluster. Rather, smooth accretion of gas flows along filaments of the cosmic web between galaxies may power in-situ star formation, SMBH growth, and AGN activity. These filaments have been imaged in emission in the SSA22 protocluster, and rough calculations estimate that they may contain $\sim10^{12}\ M_\Sun$ of gas available for accretion \citep{umehata2019}. However, it is difficult to establish whether inflows along these filaments exist, or whether they could carry enough gas into a galaxy to power an AGN.

The nodes of the imaged web notably coincide with the massive, star-forming submillimeter galaxies observed in \citet{umehata2018}. Four out of the eight X-ray AGN studied in \citet{alexander2016} are co-located with large \lya\ nebulae. We find that two of the \citetalias{steidel2003} protocluster LBGs in our \galfitm\ sample, SSA22a-C47 and SSA22a-M28, are associated with \lya\ nebulae from the \citet{matsuda2004} \lya\ blob survey: LAB 11 and LAB 12, respectively. The \sersic\ model parameters and nonparametric morphologies for these galaxies are comparable to those for the other protocluster and field LBGs in our sample, and SSA22a-M28 is classified as isolated but slightly irregular by our visual classification scheme, while SSA22a-C47 was not classified by our voters. However, we find some evidence of a possible tidal bridge between SSA22a-C47 and its projected companion when examining its residual after \sersic\ model subtraction (see \autoref{fig:lbgmontageproto}). With such a small sample we are unable to establish whether the LABs, which appear to be associated with AGN activity, are also associated with major mergers. We attempt to draw more detailed connections between the morphologies and physical properties of our LBG sample and their local, Mpc-scale environment in \autoref{sec:physcorr}.

We investigated scenario 2 \citetext{which already has significant evidence from prior SED fitting and sub-mm studies, e.g. \citealp{kubo2013, umehata2018}} by fitting the SEDs and non-parametric SFHs of our samples of protocluster and field LBGs.  Typical studies of high-redshift AGN focus on hard X-ray detected AGN with large bolometric luminosities, making studies of the physical causes of AGN fraction enhancement difficult to control for mass. The result of \citet{yang2018} establishes that, at $z>2$, AGN fraction does not depend on environment when galaxies with similar stellar masses are compared. Galaxies are expected to be more massive in protoclusters at intermediate redshifts (with $2 \lesssim z \lesssim 4$) when compared to the field \citep[e.g.,][]{steidel2005, hatch2011, cooke2014}. Since SMBH mass should scale with stellar mass \citep[e.g.,][]{ferraresse2000, ding2020} we should thus expect to have enhanced AGN activity in protoclusters with respect to the field at intermediate redshifts \citep[e.g.,][]{lehmer2009, lehmer2013, digbynorth2010, vito2020}. Using KS tests we found that the protocluster and field samples appear to be drawn from the same distributions of sSFR and mass-weighted stellar age, while the protocluster galaxies have stellar mass and SFR distributions significantly weighted toward higher mass and higher SFR. We have also found that the mean SFH of our sample of protocluster LBGs is elevated by a factor of about 2 over the mean field LBG SFH in the earliest stellar age bins, from 10 Myr to approximately 2 Gyr. This elevated star formation rate in the oldest stellar population leads to a mean protocluster LBG about 2.2 times more massive than the mean field LBG. Thus, the observed AGN fraction enhancement in the SSA22 protocluster may largely be an effect of the enhanced mass of the typical protocluster galaxy.


\subsection{Correlations Between Local Environment and Galaxy Properties}
\label{sec:physcorr}

To assess the variance in morphology and star formation properties throughout the protocluster, we use a Gaussian kernel density estimation (KDE) to estimate the LAE surface density, $\Sigma_{LAE}$. At the location of each $z \approx 3.1$ LAE from \citep{hayashino2004} we place a circular 2D Gaussian with a FWHM of $2'$, corresponding to 3.75 co-moving Mpc. The resulting distribution is re-normalized as a surface number density, and sampled at the positions of our protocluster LBG sample. \citet{hayashino2004} place the threshold of the ``high-density region'' of the SSA22 protocluster at a LAE surface density of $0.26\ {\rm arcmin^{-2}}$. We find that due to the construction of our fields, all of our LBGs are in the high-density region, and they all have local $\Sigma_{LAE} > 0.5\ {\rm arcmin^{-2}}$, which is $>5$ times the average density of the control field in \citet{hayashino2004}.

\begin{figure}
	\centering
	\plotone{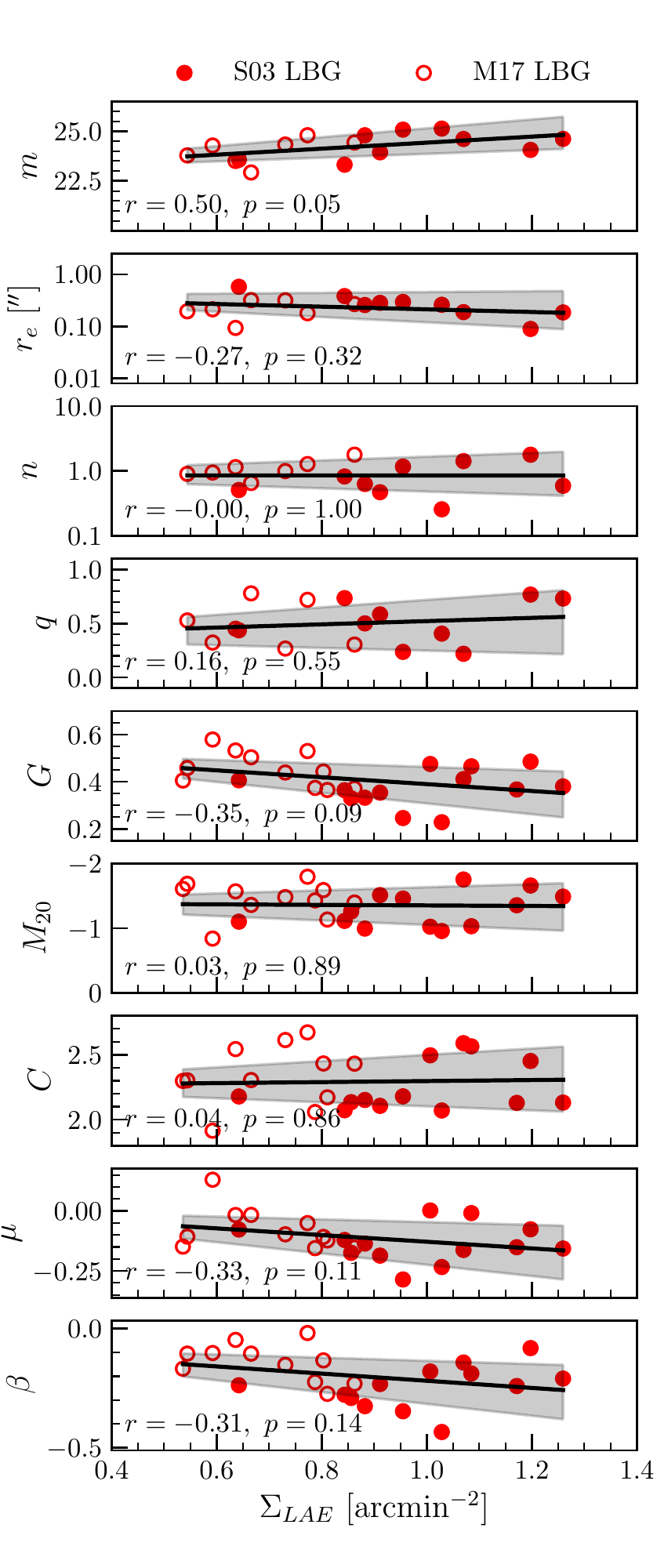}
	\caption{\label{fig:laedensmorph}We show all of our morphology 
	measurements as a function of LAE surface density for our sample
	of protocluster LBGs. From the top panel down: \sersic\ model
	F160W magnitude $m$, \sersic\ fit effective radius $r_e$, \sersic\
	index $n$, \sersic\ fit axis ratio $q$, Gini coefficient $G$, second
	order moment of light $M_{20}$, concentration index $C$, and the $G-M_{20}$
	merger and bulge statistics $\mu$ and $\beta$. We show LBGs from 
	the \citetalias{steidel2003} catalog as filled symbols, and LBGs from the 
	\citetalias{micheva2017} catalog as open symbols.
	For each property we plot a linear regression to both sets of LBGs
	with the 16th to 84th percentile interval for 
	the slope (as computed from bootstrap resampling). We print the Pearson 
	test statistic and probability for each measurement \textit{when both \citetalias{steidel2003} 
	and \citetalias{micheva2017} LBGs are included} in the corner of each panel.}
\end{figure}

\begin{figure}
	\centering
	\plotone{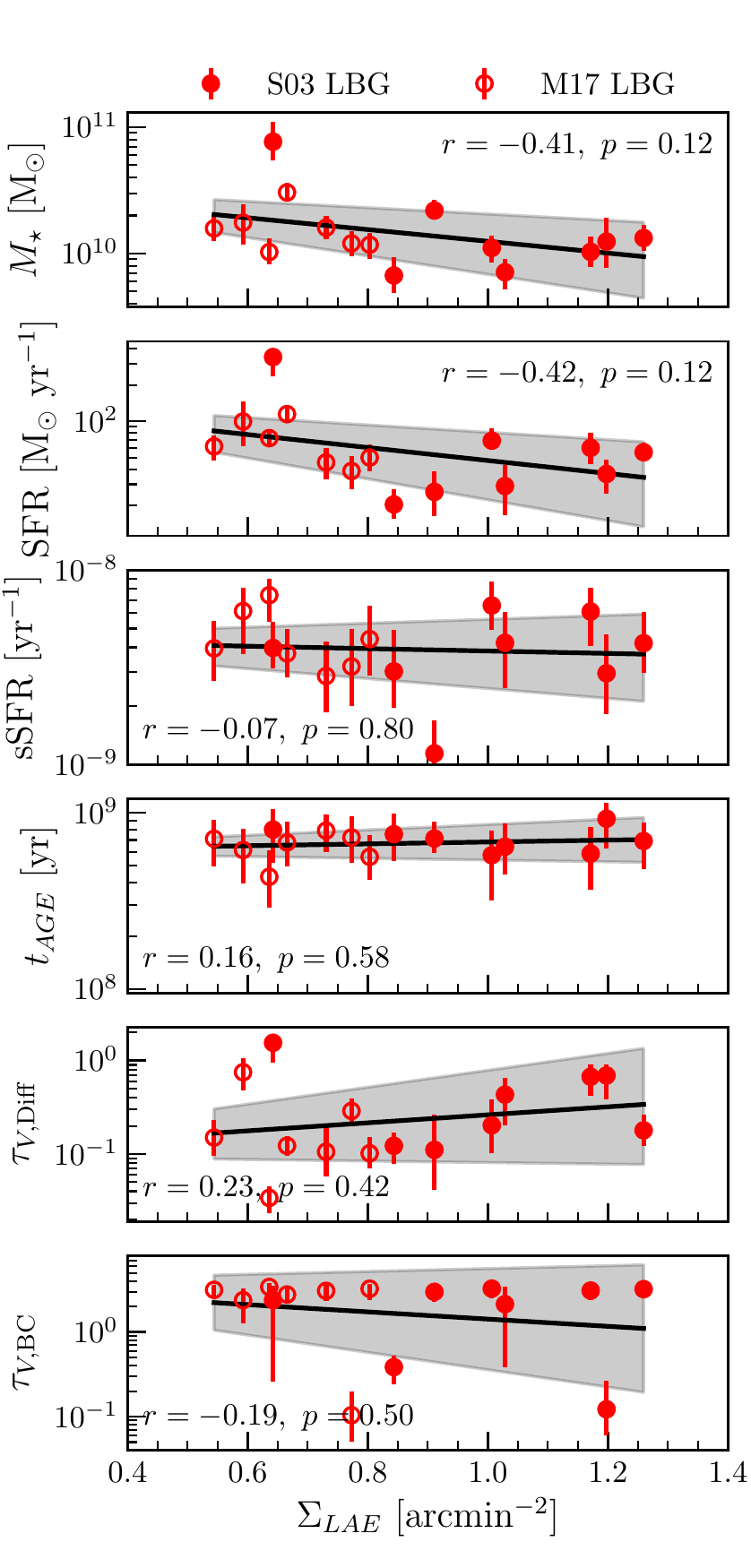}
	\caption{\label{fig:laedensphys}We show all of our SFH-derived
	measurements as a function of LAE surface density for the $Z = 0.655 Z_\odot$
	SED fits to our sample of protocluster LBGs. From the top panel down:
	stellar mass $M_\star$, star formation rate over the last 100 Myr, specific 
	star formation rate over the last 100 Myr, mass-weighted age $t_{AGE}$, 
	diffuse optical depth at the rest-frame $V-$band $\tau_{V,\rm Diff}$, and 
	birth cloud optical depth at the rest-frame $V-$band $\tau_{V,\rm BC}$. We 
	show LBGs from the \citetalias{steidel2003} catalog as filled symbols, and LBGs from the 
	\citetalias{micheva2017} catalog as open symbols. For each property we plot 
	a linear regression to both sets of LBGs with the 16th to 84th percentile interval for the slope 
	(as computed from bootstrap resampling). We print the Pearson 
	test statistic and probability for each measurement \textit{when both \citetalias{steidel2003} 
	and \citetalias{micheva2017} LBGs are included} in the corner of each panel.}
\end{figure}

We plot the morphological and physical properties of our samples of protocluster LBGs in \autoref{fig:laedensmorph} and \autoref{fig:laedensphys}, respectively. To examine whether galaxies move through $G-M_{20}$ space as a function of $\Sigma_{LAE}$, we define two additional morphological measures, the merger statistic $\mu$ and the bulge statistic $\beta$, as the signed perpendicular distance from the lines $G = -0.14 M_{20} + 0.33$ and $G = 0.14 M_{20} + 0.80$, respectively:
\begin{align}
	\mu = &0.14 M_{20} + 0.99 G - 0.33 \\
	\beta = -&0.14 M_{20} + 0.99 G - 0.79
\end{align}
Galaxies with more merger-like morphologies \citetext{in the sense of \citealp{lotz2008}; see \autoref{fig:ginim20cplot}} thus have larger values of $\mu$ and galaxies with more bulge-like morphologies have larger values of $\beta$.

If we suppose that mergers are more common in the denser regions of the protocluster, we should expect the morphologies of galaxies to have identifiable trends with the projected LAE density. We find no strong correlations between the parametric or nonparametric morphological measurements and $\Sigma_{LAE}$. There is an apparent downward trend in $G$, but a Pearson test shows that it is marginal with $p = 0.09$, and $n$ and $C$, which also probe the concentration of the galaxies, exhibit no such trend. The only other parameter that shows any marginal correlation with $\Sigma_{LAE}$ is the \sersic\ model magnitude $m$ ($r=0.50$, $p=0.05$), growing fainter with increasing density. We find that this is likely due to biases in our images: the two-orbit depth images we use are targeted on the denser regions, so in the less dense regions we preferentially select brighter galaxies, and only brighter galaxies have acceptable \sersic\ fits. If we repeat the Pearson test for all \citetalias{steidel2003} and \citetalias{micheva2017} protocluster LBGs regardless of \sersic\ fit quality, using the F160W aperture magnitudes we measured for SED fitting, we find that in this larger sample there is no significant correlation between magnitude and $\Sigma_{LAE}$. \citet{kubo2013} and \citet{kubo2017} found that several massive galaxies in the densest regions of the protocluster have compact sizes and $n > 2.5$ \sersic\ profiles similar to local early-type galaxies (ETGs). Our results do not reflect this, though we note that the maximum recovered \sersic\ index among our LBGs is $<2$ and our F160W imaging and SSA22 LBG samples do not cover the entirety of the ``core'' region of the protocluster targeted by the ALMA deep fields (ADF). In particular, the AzTEC14 group at the protocluster core \citep{kubo2015b}, where \citet{kubo2017} found massive galaxies similar to local ETGs, is not covered by our F160W images. In this region, where the proto-brightest cluster galaxy (BCG) is predicted to be evolving, mergers may be a driving factor behind evolution. However, \citet{kubo2013} and \citet{kubo2017} studied only a subset of the massive galaxies in the protocluster and \citet{kubo2017} noted a possible deficiency in low-mass galaxies in the AzTEC14 group, implying that a more sensitive mass complete sampling of the protocluster core may be necessary to establish whether mergers are ongoing or in the past. 

We find that none of the SED-fit derived physical properties are correlated with $\Sigma_{LAE}$.
It is well established that there are intensely star forming galaxies (many of which also host AGN) in the densest regions of the protocluster with IR-derived star formation rates on the order of $10^2$--$10^3\ {\rm M_{\odot}\ yr^{-1}}$ \citep[e.g.]{umehata2015, alexander2016, kato2016}. By the placement of our fields and the construction of our SED fitting sample, which excludes known AGN, we have excluded these DSFGs. Sub-millimeter observations and previous SED fitting have also shown that there are massive galaxies in the most dense region of the protocluster \citep[e.g.][]{kubo2013, kubo2015b}. Our results, which finds no strong trend between the SFR or mass of protocluster LBGs and $\Sigma_{LAE}$ may not conflict with these established results: it may be that the general star-forming galaxy population is not more massive or more intensely star forming in the denser regions of the protocluster, but rather that some galaxies in the densest regions (i.e., AzTEC14) are exceptionally massive, the possible predecessors of what will become the BCG as the protocluster evolves. In addition to our results that find that the general LBG population of the protocluster does not appear to be rapidly, currently evolving, we thus also find that it appears that LBGs in denser areas of the protocluster are evolving no more rapidly than elsewhere in the protocluster.

\subsection{JWST Prospects for SSA22}
\label{sec:future}

While the primary limits on this study are imposed by the small numbers of protocluster and field galaxies, morphological studies at \zthree\ are also limited by the sensitivity and resolution of current near-IR telescopes. Our two-orbit depth F160W images are insufficient to map the full extent of plausible tidal features, and the angular resolution in our images (slightly less than 500 $\rm kpc\ pix^{-1}$) is at the limit of what \citet{lotz2004} recommend for nonparametric morphological measurements. Next-generation space telescopes such as the James Webb Space Telescope (JWST) will present significant improvements in both regards. The search for low-surface brightness tidal features will benefit from improved sensitivity, while quantitative morphological measurements like \sersic\ model fitting will also benefit from improved angular resolution throughout the rest-frame optical.

\begin{figure}
    \plotone{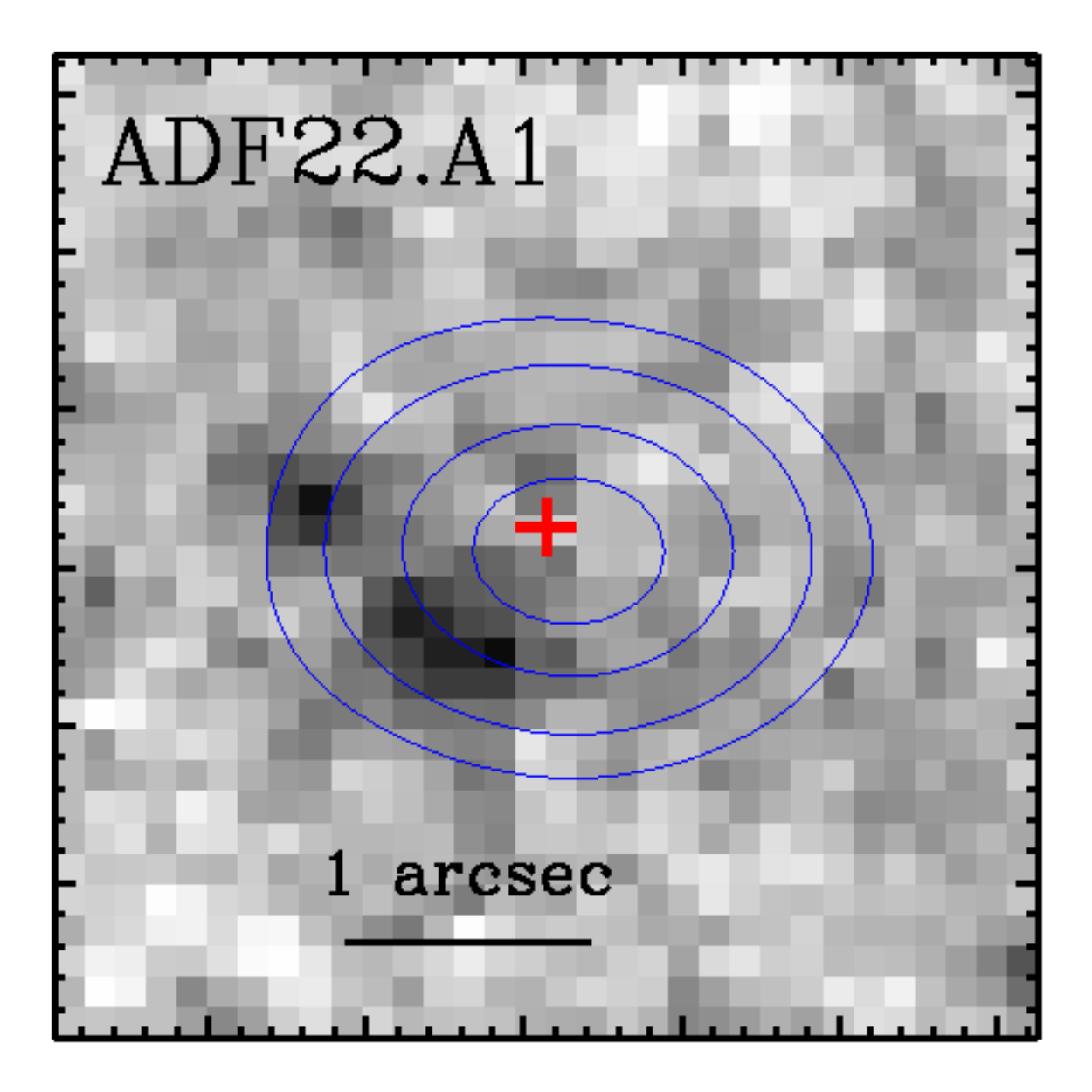}
    \caption{\label{fig:aztec1}The AzTEC1/ADF22.A1 system, as seen by WFC3/F160W. The system consists 
    of at least two closely associated ($<5$ kpc) F160W-visible galaxies, with offset Chandra (red cross) 
    and ALMA 1.1 mm (blue contours) detections. The offset source is not detected in ACS bands, and appears 
    very faintly in F160W.}
\end{figure}

For example, the field of view of our images contains at least one interesting system believed to be a complex merger: the AzTEC1/ADF22.A1 system \citep{tamura2010, umehata2014, umehata2015, umehata2017}, shown in \autoref{fig:aztec1}. The primary component of the system is an ALMA 1.1 detection with a coincident \textit{Chandra} detection, with faint companion galaxies that are not visible at wavelengths $<1$ \micron. The companions, of which there appear to be at least two, are visible in F160W, though they are not included in our catalog due to the $S/N$ cuts we impose, and they are not Lyman-break selected due to their non-detection blueward of 1 \micron. The SED of the system appears to be consistent with an AGN buried in a dusty, highly star forming galaxy. The system is thus speculated to be a protoquasar, possibly fueled by the major merger of the F160W-detected galaxies nearby. JWST ETC simulations using the currently observed SED suggest that NIRCam and MIRI observations will produce near- and mid-IR detections of all three components this system at higher resolution than even the currently available \HST\ data, allowing us to search for merger features throughout the IR and to treat this system as a case study in whether mergers fuel AGN in the protocluster environment.

While we have been unable to establish whether major mergers are a dominant mode of growth among galaxies throughout the protocluster, we have not examined the very core of the protocluster where the system that will become the BCG may be evolving. The BCGs in Coma-like clusters, which the SSA22 protocluster is expected to become, are thought to form as the result of successive major mergers in the cluster core. A group of massive galaxies has already been identified in the protocluster core by \citet{kubo2015b}, which may be the site where the future BCG is forming. A larger-than-expected portion of these galaxies are already quiescent and bulge-dominated, suggesting significant evolution and possible previous merger activity. JWST observations of this system and morphological analyses of its components could allow the detection of merger signatures, giving us a possible window into the early stages of BCG formation.

\section{Summary}
\label{sec:summary}

We have pursued multiple avenues of morphological analysis on protocluster and field galaxies detected in new and archival \HST\ WFC3 F160W images of the SSA22 protocluster. We fit single S\'ersic models to galaxies detected in our images to extract effective sizes and \sersic\ indices, and then examined the residual images after model subtraction for evidence of tidal features. We also calculated the Gini coefficient $G$, moment of light $M_{20}$, and concentration statistic $C$ for protocluster and field galaxies detected in our images. For a third point of comparison we used a visual classification scheme modeled on \citet{hine2015} to examine the observed merger fractions among protocluster and field LBGs.

To supplement our morphological analysis, we fit the UV-to-near-IR SEDs and non-parametric SFHs of a sample of protocluster and field galaxies, in order to measure stellar masses and SFR.

Our main results and conclusions are as follows:

\begin{itemize}
	\item{Using two-sample KS tests to compare the \sersic\ fit morphologies of SSA22 protocluster 
	LBGs to a combined sample of field LBGs from SSA22 and GOODS-N, we find no significant differences
	in the protocluster and field distributions of any of the model parameters, including effective size 
	$r_e$ ($p_{KS} \ge 0.37$) and \sersic\ index $n$ ($p_{KS} \ge 0.17$).}
	\item{We find evidence of tidal features in the residual images of both SSA22 protocluster and field LBGs after
	subtracting the best-fit \sersic\ model. Based on this analysis we estimate rough merger fractions of
	0.13--0.44 for protocluster LBGs and 0.14--0.28 for field LBGs. We estimate that the largest and brightest of
	the plausible tidal features have masses $\log_{10} M_{\star}/M_{\odot} = 9.78$, suggesting that they may be
	as massive as the Small Magellanic Cloud.}
	\item{Using two-sample KS tests comparing the non-parametric morphologies of SSA22 protocluster and field galaxies,
	we find no significant differences in the protocluster and field distributions of $M_{20}$
	($p_{KS} \ge 0.30$), and $C$ ($p_{KS} \ge 0.30$). We find a marginal difference between the $G$ distributions of 
	protocluster and field \citetalias{steidel2003} LBGs ($p_{KS} = 0.04$). However, this is not supported by the KS 
	tests on any of the other measures of the galaxies' concentration (i.e., $n$ and $C$), and is not evident when
	\citetalias{micheva2017} LBGs are included in the KS test ($p_{KS} = 0.25$). We note that only one of the galaxies 
	in which we identify a plausible tidal feature is classified as a merger by the \citet{lotz2008} cuts in the 
	$G$--$M_{20}$ plane. We thus hesitate to estimate a merger fraction based on non-parametric morphological analysis.}
	\item{By performing visual merger classifications of selected F160W galaxy cutouts for a direct comparison to \citet{hine2015}
	we estimate merger fractions $0.38^{+0.37}_{-0.20}$ among \citetalias{steidel2003} protocluster LBGs and $0.41^{+0.11}_{-0.09}$
	among \citetalias{steidel2003} field LBGs. We note that visual classifications from our SSA22 images are limited by the number
	of \citetalias{steidel2003} LBGs in our images, and that the number of mergers may be undercounted due to limited depth.}
	\item{We find that the SED fits to our small sample of protocluster and field LBGs are consistent 
	with elevated star formation in the protocluster's oldest stellar population, with the mean protocluster 
	SFH being significantly elevated over the mean field SFH between lookback times 100 Myr -- 2 Gyr. The mean
	protocluster LBG is also more massive and more attenuated by a factor of 2 compared to the mean field galaxy.
	The mean protocluster LBG is thus slightly redder in terms of IR color, with $J-{\rm F444W}=1.68_{-0.43}^{+0.46}$,
	than the mean field LBG, which has 
	$J-{\rm F444W}=1.31_{-0.17}^{+0.17}$. However, young stars are still the dominant contributor to the SEDs of both
	protocluster and field LBGs.}
	\item{Based on our results, we conclude that the observed enhancement in the SSA22 protocluster AGN fraction may
	be due to the larger average stellar mass (and hence larger average SMBH mass) of galaxies in the protocluster. In addition, 
	the protocluster LBGs we have studied here appear to have formed more stellar mass earlier than their field counterparts.}
\end{itemize}

Our results are limited throughout by the small number statistics of protocluster LBGs; we are only able to identify 24 protocluster LBGs in our F160W images, and requirements on converged \sersic\ fits and available photometry mean that in practice we can only use a fraction of them in our analysis. These limits also mean that our analyses are difficult to control for mass; studies of merger fractions in particular are sensitive to mass, as merger fraction increases with stellar mass at fixed redshift. Our focus on Lyman-break selected galaxies in this work may also exclude more massive galaxies with more evolved SED shapes \citep{wang2019}, which may be involved in ongoing mergers or have morphologies that indicate past mergers. We are hopeful that the increased sensitivity of JWST will allow the construction of true mass-selected samples, which will allow deeper studies of the connections between stellar mass, AGN fraction, and overdensity.

\acknowledgements{We gratefully acknowledge support from STScI grant \textit{HST}-GO-13844.013-A (EBM, BDL, BB). DMA acknowledges support from the Science and Technology Facilities Council (ST/P000541/1; ST/T000244/1). This work has made use of the Rainbow Cosmological Surveys Database, which is operated by the Centro de Astrobiolog\'ia (CAB/INTA), partnered with the University of California Observatories at Santa Cruz (UCO/Lick, UCSC).}\\

\facilities{HST (WFC3), Subaru, Spitzer}\\

\software{\texttt{astropy} \citep{astropy2013, astropy2018}, 
\adrizzle\ \citep{fruchter2010}, 
\sextractor\ \citep{bertin1996}, 
\galfitm\ \citep{vika2013}, 
\galapagostwo\ \citep{barden2012}, 
\ligh\ \citep{eufrasio2017}, 
\texttt{P\'EGASE} \citep{fioc1997, fioc1999}}

\bibliographystyle{aasjournal}
\bibliography{SSA22_WFC3}

\appendix
\section{Reliability of Morphological Measurements}
\label{app:morph_sim}

\begin{figure}
	\centering
	\plotone{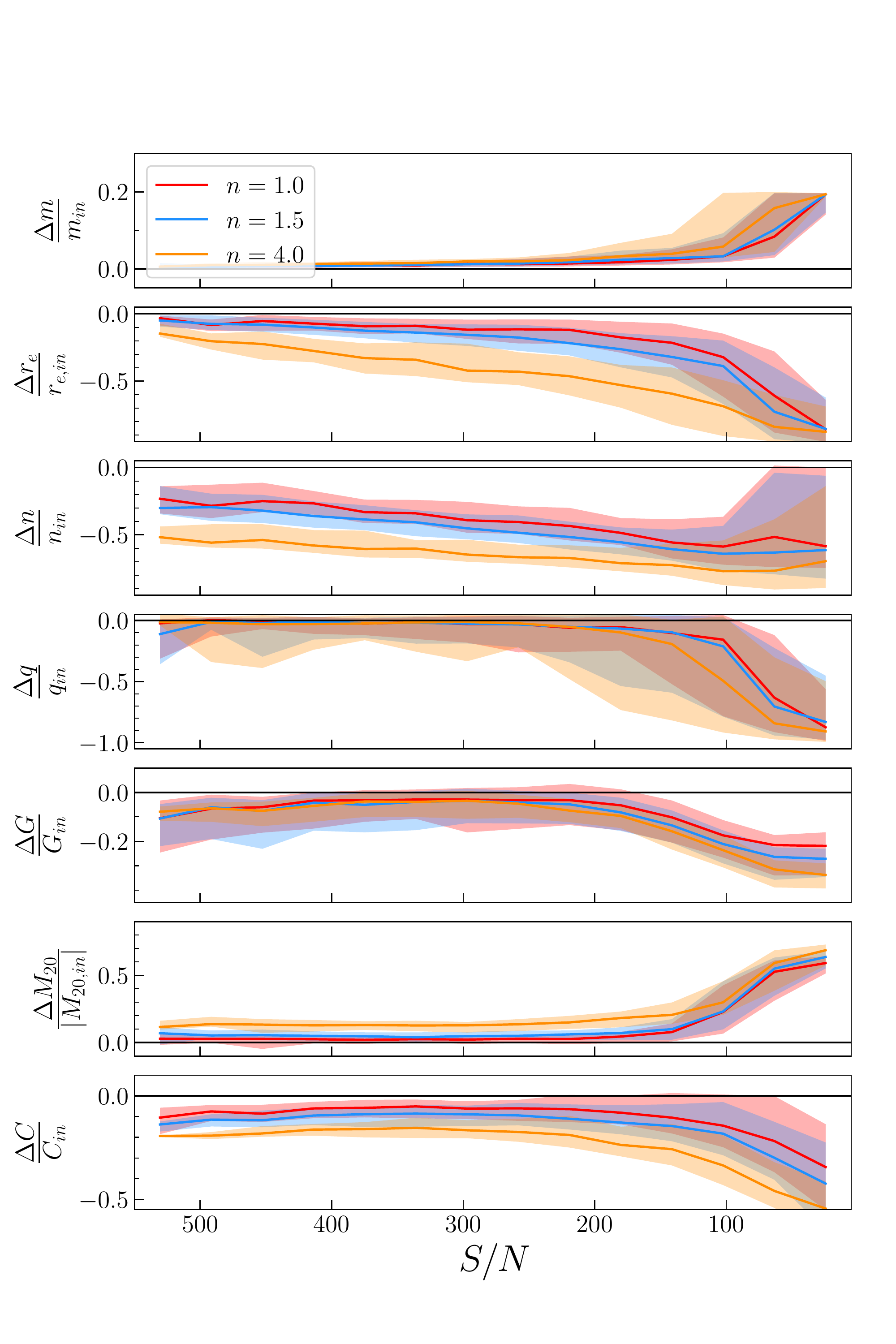}
	\caption{\label{fig:morphsim}For each morphological measurement, the median relative 
	difference between the initial values in three sets of noise-free, un-convolved
	S\'ersic models and the extracted values after PSF convolution and the addition of 
	noise is shown as a function of the final signal-to-noise ratio in a 
	$1''$ diameter aperture. The shaded regions show the 16th to 84th percentile interval.}
\end{figure}

We assessed the reliability of our model-fitting and non-parametric analyses with a Monte Carlo technique. We created three sets of \galfitm\ \sersic\ models with indices $n = \{1.0, 1.5, 4.0\}$. The other parameters of the model were allowed to vary randomly over their observed ranges. We convolved these synthetic galaxies with our PSF model, added them to blank sky frames from our images, and re-measured their morphologies with \galfitm\ and our non-parametric analysis procedure as described in \autoref{sec:npmorphologies}. We show the relative differences between the recovered morphologies and morphologies as measured from un-convolved, noise-free models in \autoref{fig:morphsim}.

We find that the reliability of the \galfitm-recovered \sersic\ radius $r_e$ and axis ratio $q$ are strongly dependent on the signal-to-noise ratio, with \galfitm\ consistently underestimating the effective radii of low-$S/N$ galaxies by as much as $75\%$ and consistently converging on unrealistically small axis ratios for low-$S/N$ galaxies. Galaxies with bulge-like $n=4$ profiles are more strongly affected in both cases. There are large errors in the recovered value of the \sersic\ index $n$ for all initial values of $n$ and all values of $S/N$. The recovered value is significantly smaller than the true value in all cases, indicating that the over-representation of galaxies with $n \le 1.0$ in our results may be due to underestimation of the ``true" $n$ for these galaxies. However, we note also that the consistent underestimation of $n$ and $r_e$ is expected from the nature of these simulations; convolution with the PSF spreads out the light profile of the model, making it flatter, and the observed effective radius of the model galaxies is naturally decreased by the addition of noise. Since the effects of PSF blurring and Poisson noise cannot be removed from the images in any practical situation, the ``true'' values of $n$ and $r_e$ are not necessarily recoverable, and our measurements serve only as a description of the data.

The non-parametric morphological measures are generally more stable with $S/N$. $M_{20}$ is the most strongly affected, with low-$S/N$ galaxies measured to have significantly larger $M_{20}$; that is, the models are observed to be clumpier after convolution with the PSF and the addition of noise.

With the exception of $n$, $r_e$, and $C$ (which also depends on measurements of the effective size of the galaxy and is thus biased low even for high $S/N$), the median relative errors in the morphological measurements are $<10\%$ within the $2\sigma$ range for $S/N \gtrsim 100$ and input \sersic\ indices 1.0 and 1.5.

As the majority of galaxies in our catalog have \sersic\ indices $<4$, we take these results to show that our morphological analyses are reliable for our \citetalias{steidel2003} sample (with one exception, our \citetalias{steidel2003} protocluster LBGs have $S/N \gtrsim 100$, and all of them have $n < 2$).  In general, we expect the \sersic\ fit results in the catalog to be reliable for the other galaxies in our sample (provided they meet our other criteria for acceptable fits) for $S/N \gtrsim 100$.

\section{SED Fit Results}
\label{app:SED}

\begin{figure*}
	\centering
	\includegraphics[scale=0.90]{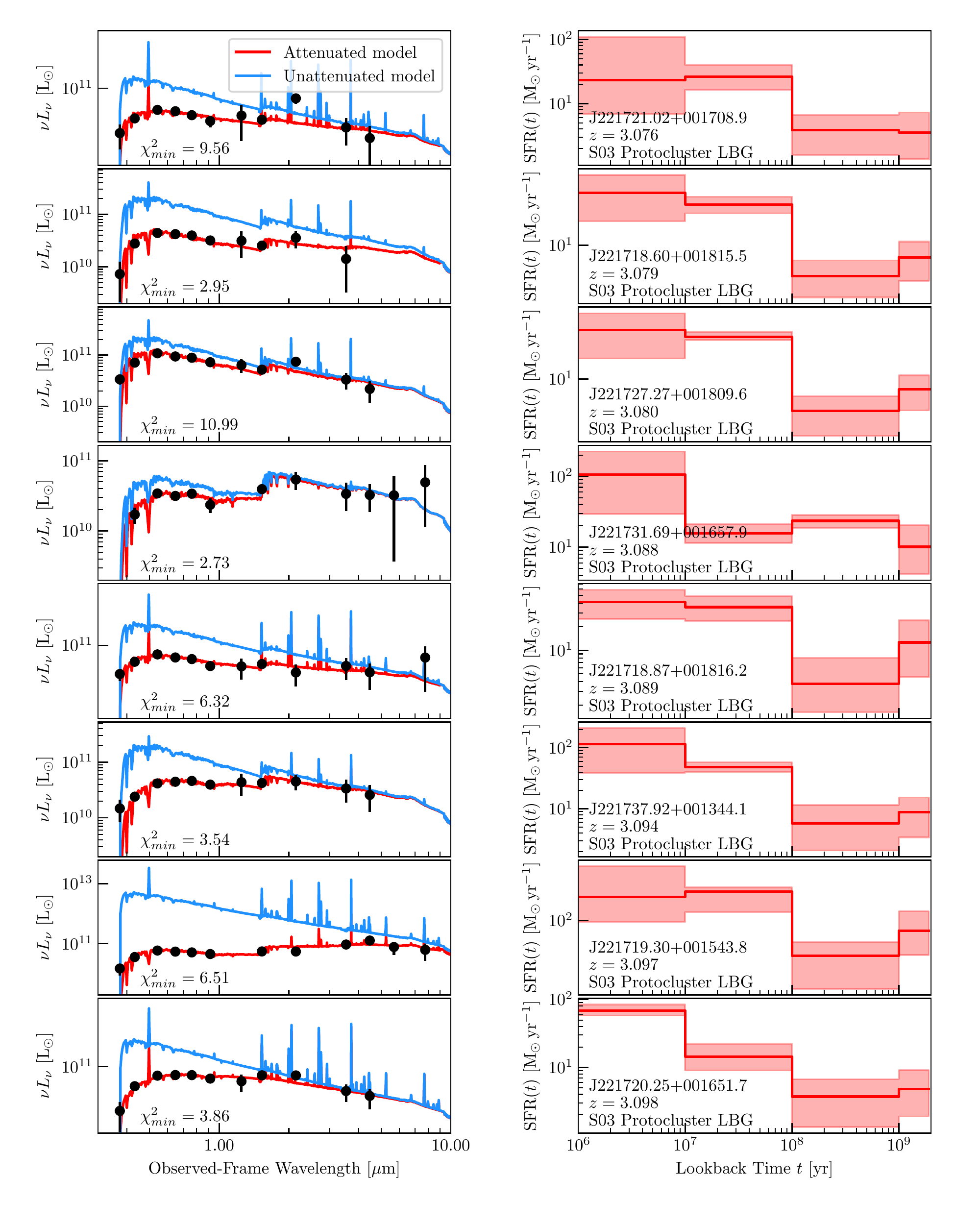}
	\caption{\label{fig:appB_S03_proto}\textit{Left:} We show the best-fit SED model for each of the 8 galaxies in our sample of \citet{steidel2003} protocluster LBGs that have SED fits. \textit{Right:} We show the median SFH for the same galaxies. For clarity we have truncated the last bin of the SFH at 2 Gyr. The shaded regions indicate the 16th to 84th percentile interval.}
\end{figure*}

\begin{figure*}
	\centering
	\includegraphics[scale=0.90]{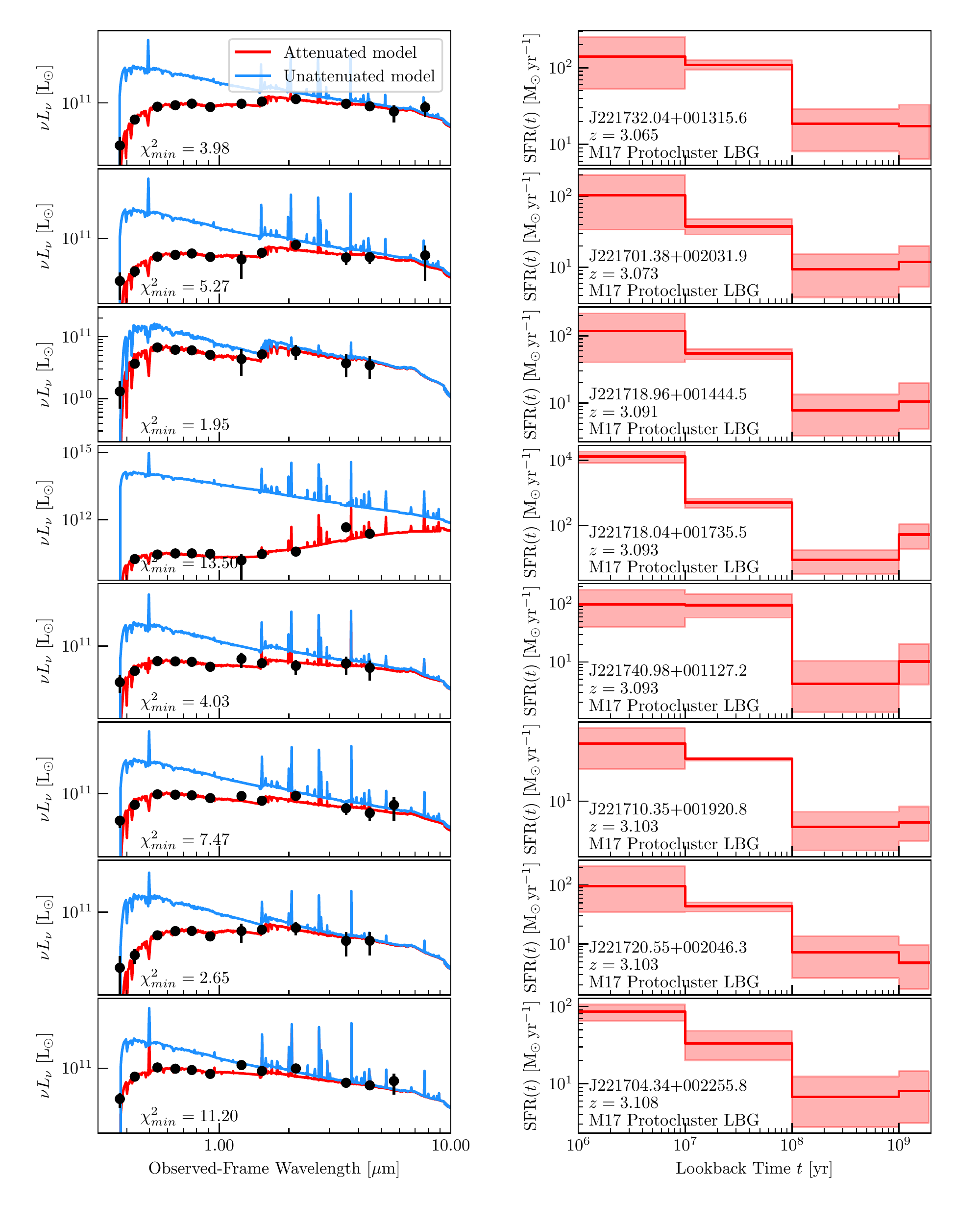}
	\caption{\label{fig:appB_M17_proto}Same as \autoref{fig:appB_S03_proto}, showing the 8 \citetalias{micheva2017} protocluster LBGs with SED fits.}
\end{figure*}

\begin{figure*}
	\centering
	\includegraphics[scale=0.90]{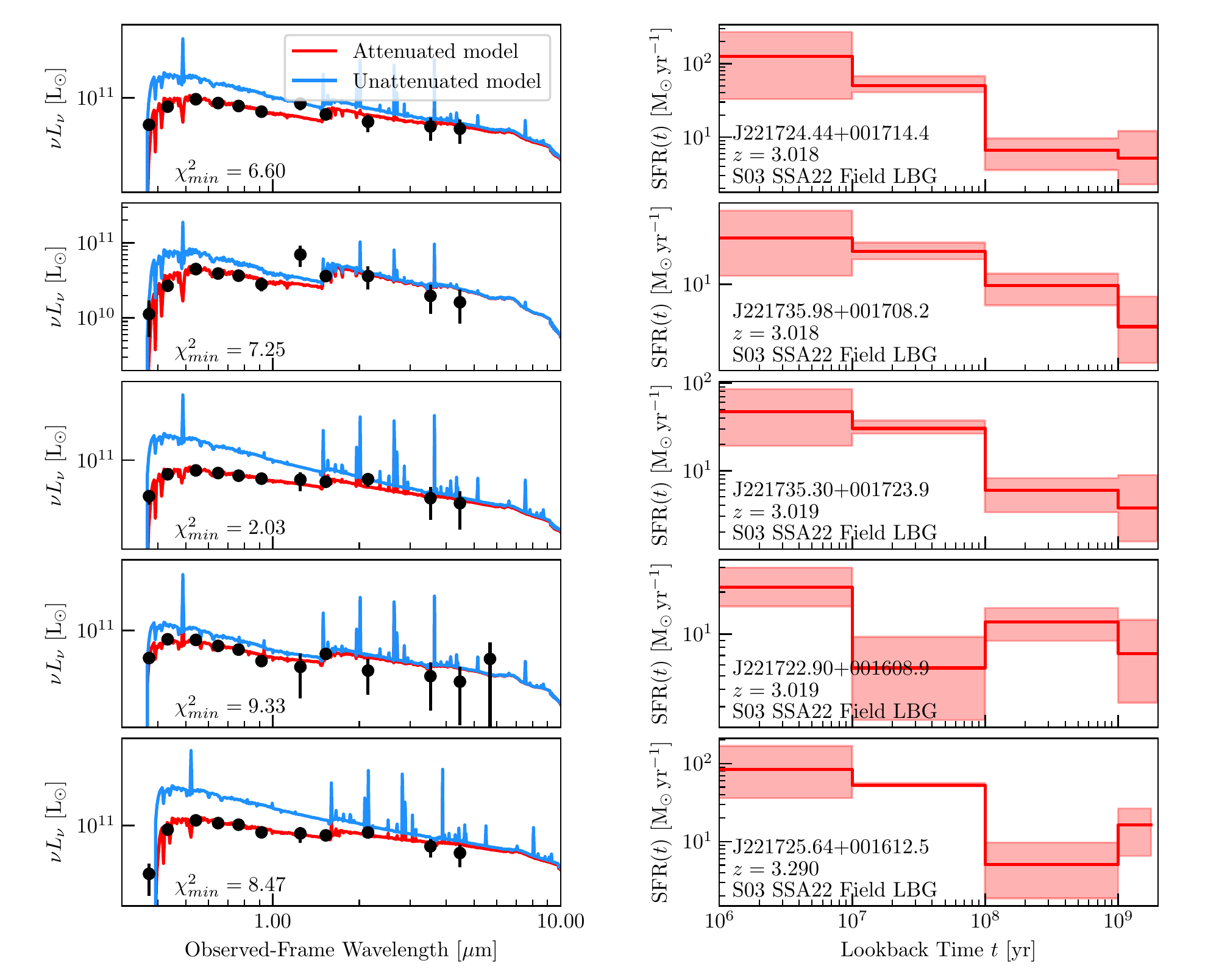}
	\caption{\label{fig:appB_S03_SSA22_field}Same as \autoref{fig:appB_S03_proto}, showing the 5 \citetalias{steidel2003} SSA22 field LBGs with SED fits.}
	\includegraphics[scale=0.90]{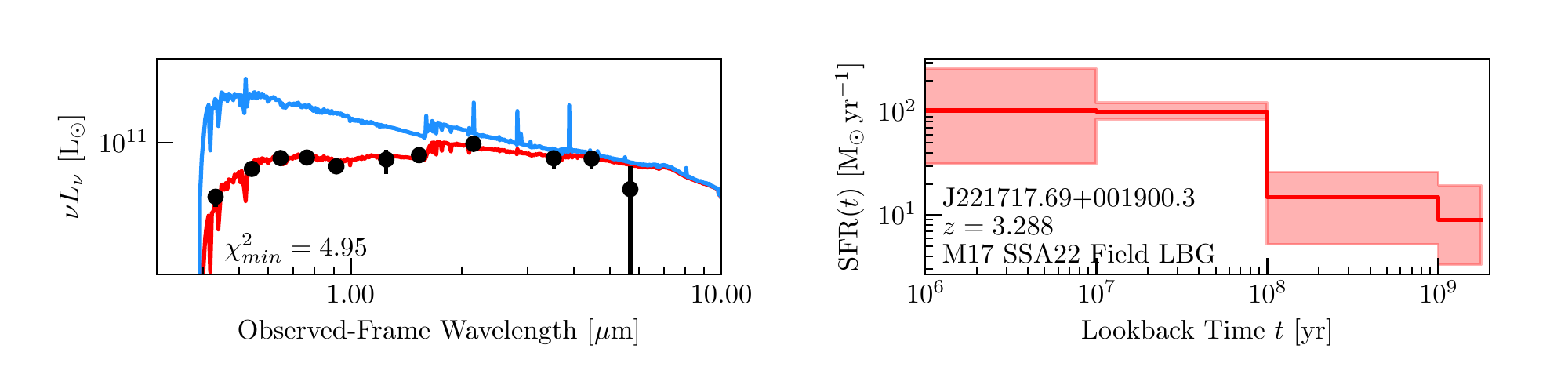}
	\caption{\label{fig:appB_M17_SSA22_field}Same as \autoref{fig:appB_S03_proto}, showing the single \citetalias{micheva2017} SSA22 field LBG with an SED fit.}
\end{figure*}

\begin{figure*}
	\centering
	\figurenum{18}
	\includegraphics[scale=0.90]{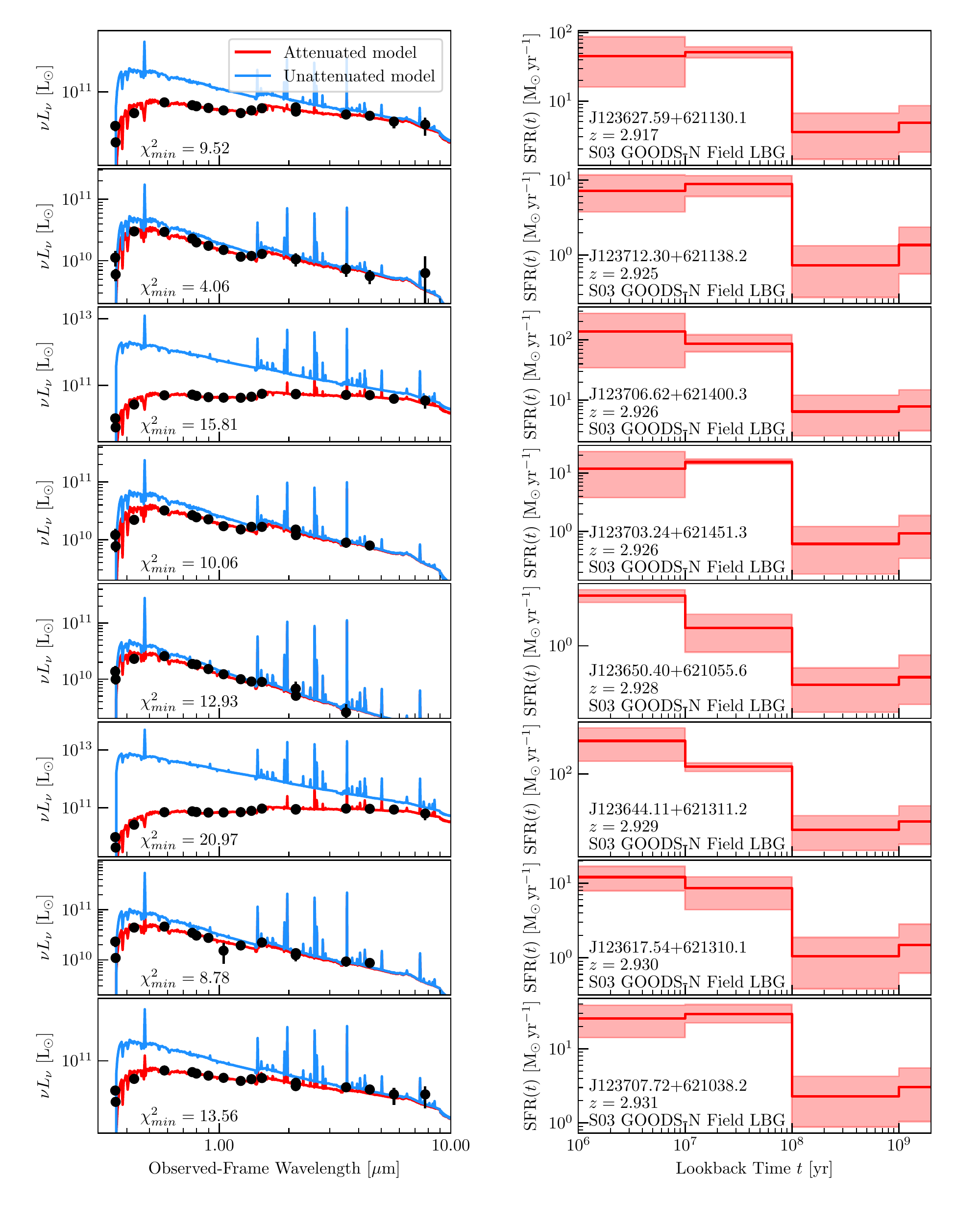}
	\caption{\label{fig:appB_S03_GOODSN_field_1to7}Same as \autoref{fig:appB_S03_proto}, showing the first 8 \citetalias{steidel2003} GOODS-N field LBGs listed in \autoref{tab:GOODSNphys}. We show the SED and SFH fits for the remaining GOODS-N LBGs in the online version of this article.}
\end{figure*}

We show the SED fit results with $Z = 0.655 Z_{\odot}$ for our SSA22 LBG samples in \autoref{tab:SSA22phys}, \autoref{fig:appB_S03_proto}, \autoref{fig:appB_M17_proto}, \autoref{fig:appB_S03_SSA22_field}, and \autoref{fig:appB_M17_SSA22_field}. We also show the SED fit results with $Z = 0.655 Z_{\odot}$ for our sample of GOODS-N field LBGs in \autoref{tab:GOODSNphys} and example fits in \autoref{fig:appB_S03_GOODSN_field_1to7}. For the full set of GOODS-N SED plots, please refer to the online version of this article.

\input{table_example_SED_fits_SSA22.tex}
\input{table_example_SED_fits_GOODSN.tex}

\begin{figure*}
	\centering
	\figurenum{18}
	\includegraphics[scale=0.90]{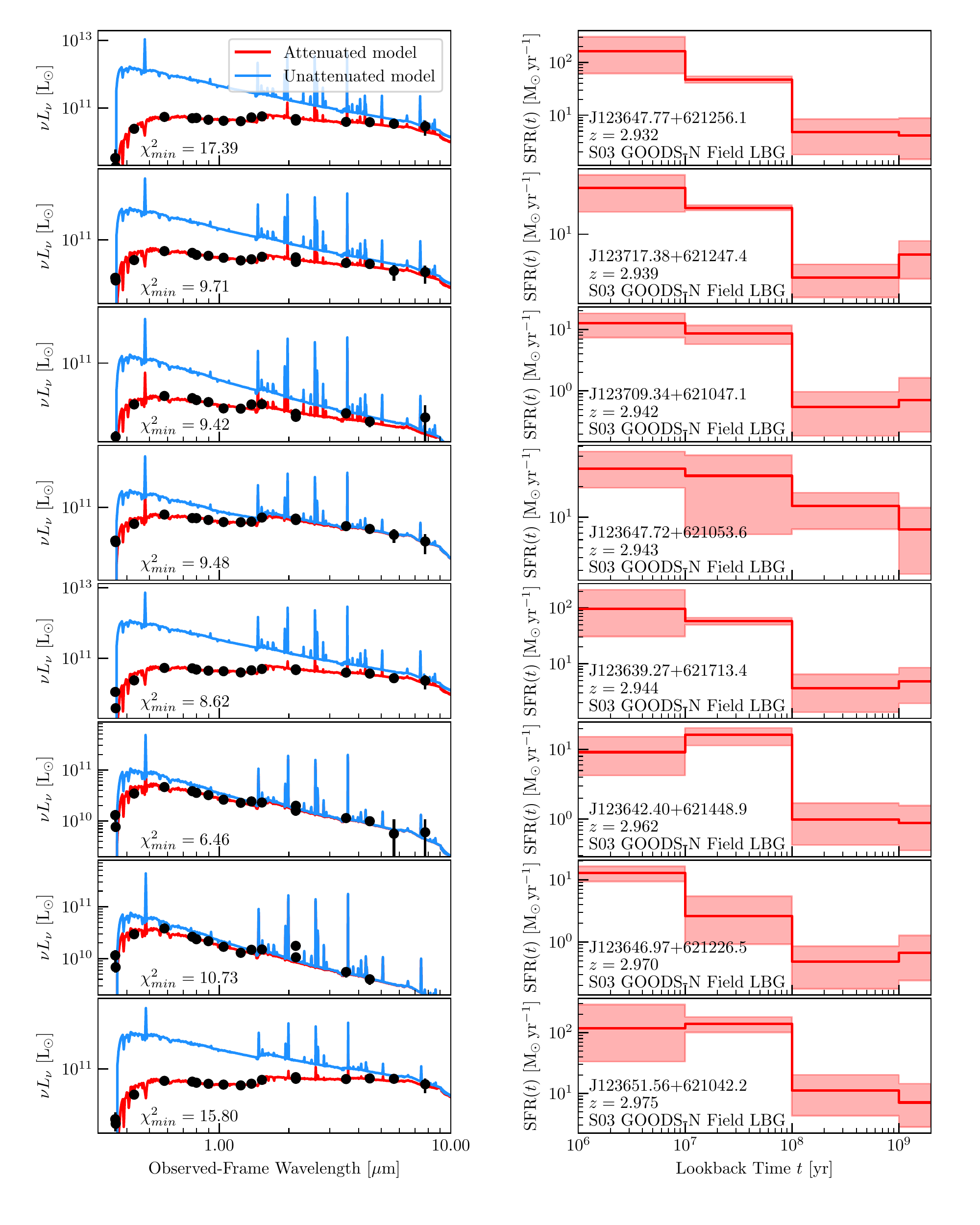}
	\caption{\textit{Continues.}}
\end{figure*}

\begin{figure*}
	\centering
	\figurenum{18}
	\includegraphics[scale=0.90]{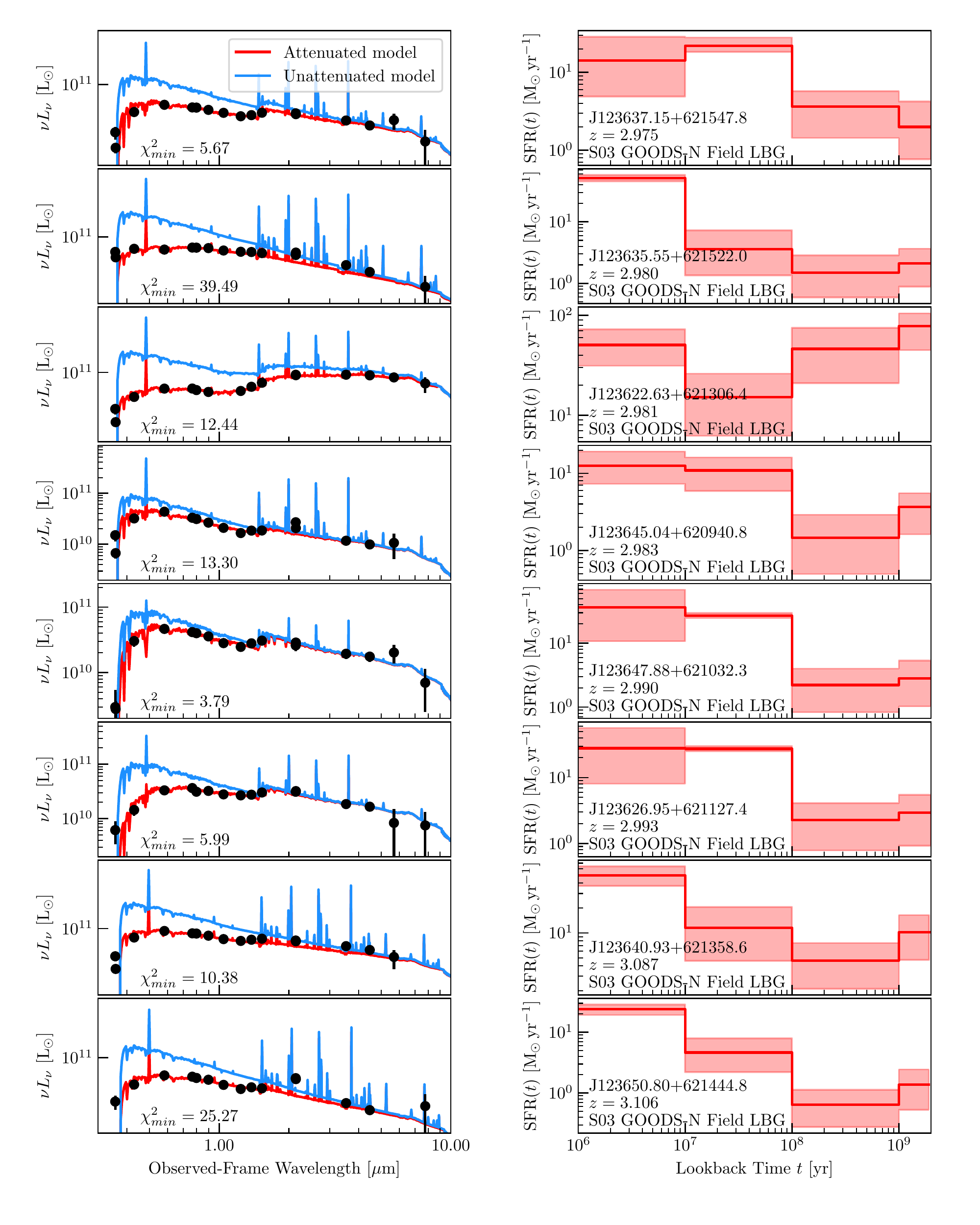}
	\caption{\textit{Continues.}}
\end{figure*}

\begin{figure*}
	\centering
	\figurenum{18}
	\includegraphics[scale=0.90]{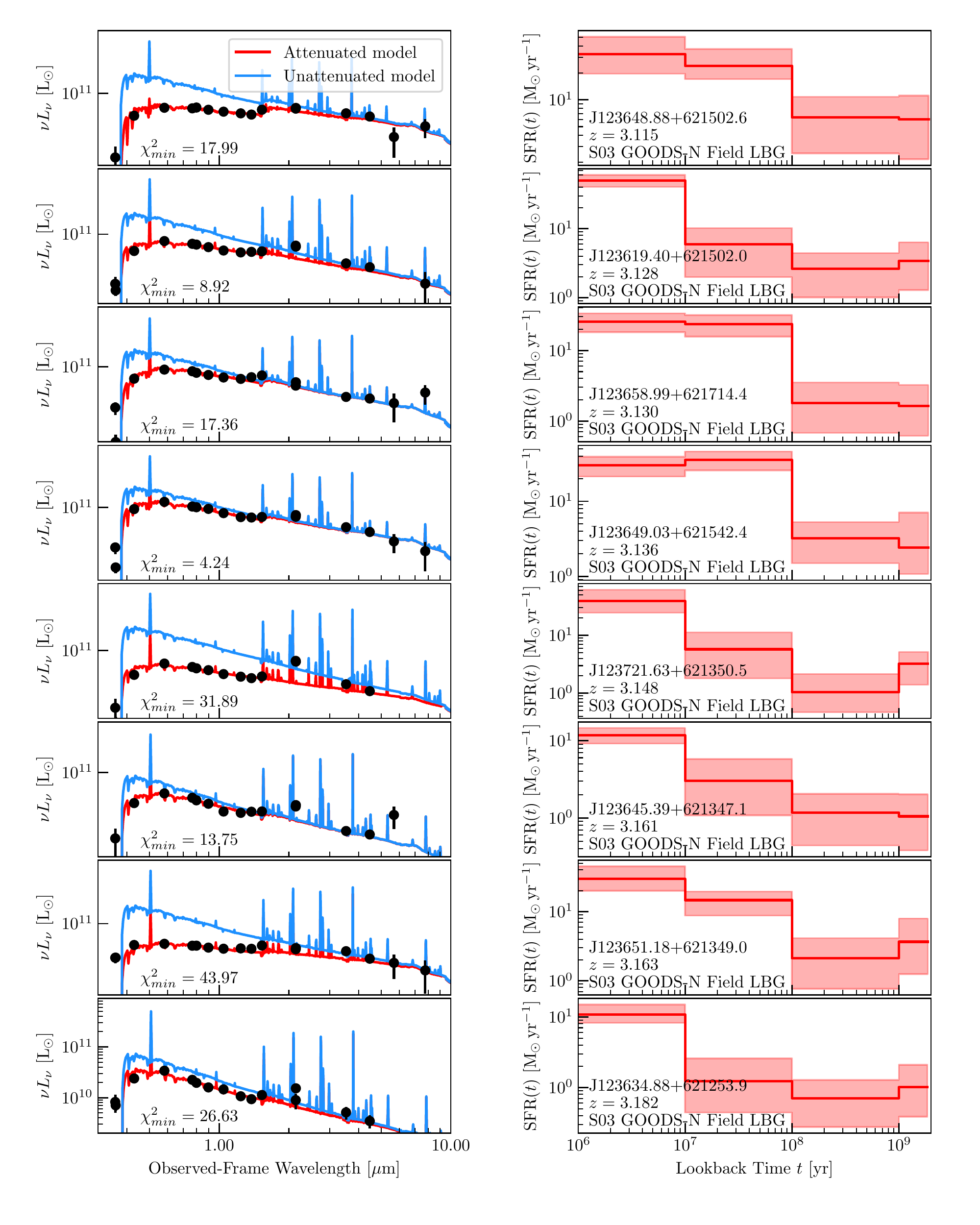}
	\caption{\textit{Continues.}}
\end{figure*}

\begin{figure*}
	\centering
	\figurenum{18}
	\includegraphics[scale=0.90]{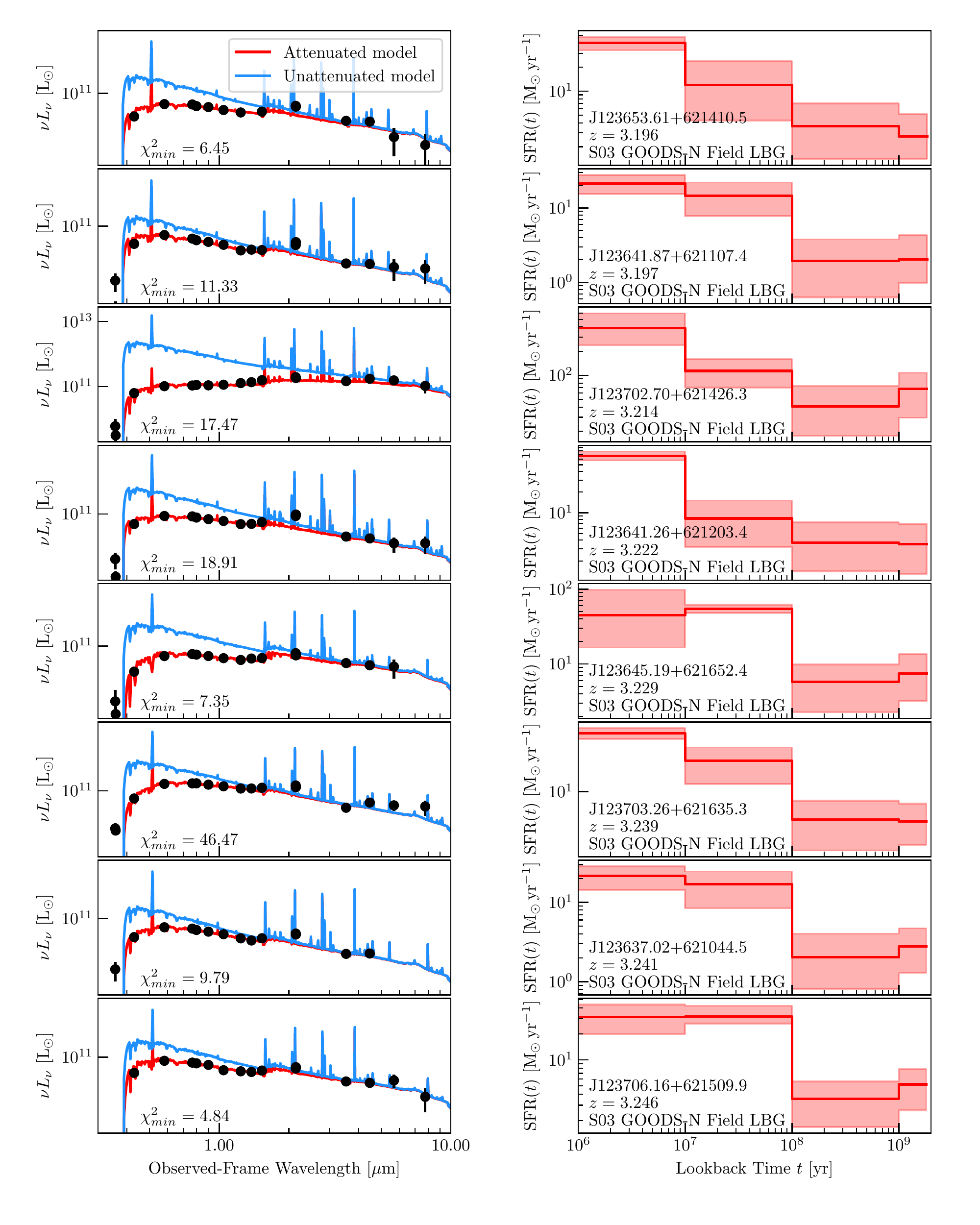}
	\caption{\textit{Continues.}}
\end{figure*}

\end{document}

%% file: table1.tex
\begin{deluxetable*}{l c c c c c r}[tb]
	\tablecolumns{7}
	\tablecaption{\label{table:obs}Summary of \HST\ WFC3 F160W survey fields.}
	\tablehead{\colhead{ID} & \colhead{R.A.} & \colhead{Dec.} & \colhead{Pos. Angle} &
	\colhead{Exp. Time} & \colhead{Proposal No.} & \colhead{PI}\\
	\colhead{} & \colhead{(deg)} & \colhead{(deg)} & \colhead{(deg)} & \colhead{(s)} &
	\colhead{} & \colhead{}}
	\startdata
	\cutinhead{New Observations\tablenotemark{a}}
	TARGET1  & 334.29 & 0.3046 & -47.95  & 5223.50 & 13844 & Lehmer \\ 
	TARGET2	 & 334.32 & 0.3001 & 121.10  & 5223.50 & 13844 & Lehmer \\
	TARGET3	 & 334.33 & 0.3371 & 111.91  & 5223.50 & 13844 & Lehmer \\
	TARGET4	 & 334.39 & 0.2734 & -67.27  & 5223.50 & 13844 & Lehmer \\
	TARGET5	 & 334.40 & 0.2185 & -71.17  & 5223.50 & 13844 & Lehmer \\
	TARGET6	 & 334.49 & 0.2507 & -72.31  & 5223.50 & 13844 & Lehmer \\
	\cutinhead{Archival Observations\tablenotemark{b}}
	SSA22AC6M4	   & 334.42 & 0.1908 & 111.75 & 2611.75 & 11735 & Mannucci \\
	SSA22AC30  	   & 334.33 & 0.2624 & 111.77 & 2611.75 & 11735 & Mannucci \\
	SSA22AM16 	   & 334.38 & 0.2194 &  68.60 & 2611.75 & 11735 & Mannucci \\
	SSA22AM38C48   & 334.33 & 0.3109 & 126.09 & 2611.75 & 11735 & Mannucci \\
	\tableline
	SSA-22A-IR   & 334.34 & 0.2888 & 115.00  &	2611.75 &   11636 &   Siana \\
	SSA-22A-IR2  & 334.35 & 0.2846 &  85.00  &	2611.75 &   11636 &   Siana \\
	\tableline
	SSA22.4.IR & 334.21 &	 0.3260 &   108.00 &	2611.75 &   14747 &   Robertson \\
	SSA22.5.IR & 334.24 &	 0.3473 &	-74.60 & 	2611.75 &   14747 &   Robertson \\
	SSA22.6.IR & 334.28 &	 0.3636 &	111.00 &	2611.75 &   14747 &  Robertson \\
	SSA22.7.IR & 334.30 &	 0.3157 &	106.00 &	2611.75 &   14747 &   Robertson\\
	\enddata
	\tablenotetext{a}{Shown in red in \autoref{fig:obs}.}
	\tablenotetext{b}{Shown in blue in \autoref{fig:obs}.}
\end{deluxetable*}

%% file: table2.tex
\begin{longrotatetable}
\begin{deluxetable*}{l c c c c c c c c c c c c c c c r h}
\tabletypesize{\footnotesize}
\movetabledown=1in
\tablecaption{\label{table:morphcatalog}Catalog excerpt for protocluster LBGs 
with acceptable fits as defined in \autoref{sec:models}, showing the 
\galfitm-extracted morphological parameters.}
\tablehead{\colhead{ID} & \colhead{$S/N$\tablenotemark{a}} & \colhead{$m_{\rm F160W}$\tablenotemark{b}} & 
\colhead{$\delta m_{\rm F160W}$} & \colhead{$r_e$\tablenotemark{c}} &\colhead{$\delta r_e$} &
\colhead{$n$\tablenotemark{d}} & \colhead{$\delta n$} & \colhead{$q$\tablenotemark{e}} & \colhead{$\delta q$} & \colhead{PA\tablenotemark{f}} & \colhead{$\delta$PA} & 
\colhead{$z$\tablenotemark{g}} & \colhead{$z$ Src.\tablenotemark{h}} & \colhead{$\Sigma_{LAE}$\tablenotemark{i}} & \colhead{Vis. Class.\tablenotemark{j}} & \colhead{S03 Name\tablenotemark{k}} & \nocolhead{Note}\\
\colhead{} & \colhead{} & \colhead{} & \colhead{} & \colhead{(arcsec)} &
\colhead{(arcsec)} & \colhead{} & \colhead{} & \colhead{} & \colhead{} & \colhead{(deg)} & \colhead{(deg)} & 
\colhead{} & \colhead{} & \colhead{$\rm(arcmin^{-2})$} & \colhead{} & \colhead{} & \nocolhead{}}
\startdata
J221710.35+001920.8 & 223.5 & 23.50 & 0.07 & 0.09 & 0.03 & 1.15 & 1.41 & 0.45 & 0.22 & 47.76 & 17.56 & 3.103 & M17 & 0.64 & \nodata & \nodata & \nodata \\
J221704.34+002255.8 & 188.2 & 24.80 & 0.32 & 0.18 & 0.11 & 1.28 & 3.04 & 0.72 & 0.52 & -42.55 & 98.42 & 3.108 & M17 & 0.77 & \nodata & \nodata & \nodata \\
J221732.04+001315.6 & 183.6 & 22.93 & 0.06 & 0.32 & 0.03 & 0.65 & 0.17 & 0.78 & 0.06 & 30.23 & 14.57 & 3.065 & M17 & 0.67 & C2 & \nodata & \nodata \\
J221737.92+001344.1 & 166.5 & 24.61 & 0.19 & 0.19 & 0.07 & 0.59 & 0.95 & 0.73 & 0.30 & -68.91 & 49.33 & 3.094 & S03a & 1.26 & M2 & SSA22a-MD14 & \nodata \\
J221731.69+001657.9 & 163.4 & 23.94 & 0.05 & 0.28 & 0.03 & 0.47 & 0.19 & 0.58 & 0.06 & -56.33 & 7.60 & 3.088 & S03a & 0.91 & C2 & SSA22a-M28 & \nodata \\
J221720.25+001651.7 & 157.0 & 23.32 & 0.06 & 0.38 & 0.03 & 0.82 & 0.18 & 0.73 & 0.06 & -78.20 & 9.94 & 3.098 & S03a & 0.84 & C2 & SSA22a-C35 & \nodata \\
J221718.87+001816.2 & 156.8 & 24.06 & 0.06 & 0.09 & 0.02 & 1.78 & 1.16 & 0.77 & 0.20 & 12.01 & 32.32 & 3.089 & S03e & 1.20 & M2 & SSA22a-D17 & \nodata \\
J221718.96+001444.5 & 135.1 & 23.78 & 0.10 & 0.19 & 0.03 & 0.90 & 0.64 & 0.53 & 0.16 & 21.98 & 13.89 & 3.091 & M17 & 0.54 & \nodata & \nodata & \nodata \\
J221701.38+002031.9 & 133.6 & 24.33 & 0.13 & 0.32 & 0.08 & 0.99 & 0.65 & 0.27 & 0.12 & -14.07 & 9.27 & 3.073 & M17 & 0.73 & \nodata & \nodata & \nodata \\
J221720.20+001731.6 & 123.1 & 24.61 & 0.07 & 0.19 & 0.02 & 1.42 & 0.73 & 0.22 & 0.17 & -16.46 & 6.21 & 3.065 & M17 & 1.07 & \nodata & SSA22a-C47 & \nodata \\
J221718.04+001735.5 & 120.8 & 24.43 & 0.09 & 0.27 & 0.05 & 1.78 & 1.18 & 0.30 & 0.12 & 31.35 & 7.55 & 3.093 & M17 & 0.86 & \nodata & \nodata & \nodata \\
J221731.51+001631.0 & 118.2 & 24.79 & 0.09 & 0.26 & 0.05 & 0.63 & 0.38 & 0.50 & 0.13 & 56.67 & 11.93 & 3.098 & M17 & 0.88 & \nodata & SSA22a-M25 & \nodata \\
J221740.98+001127.2 & 108.7 & 24.28 & 0.10 & 0.21 & 0.04 & 0.95 & 0.66 & 0.32 & 0.16 & 69.28 & 8.14 & 3.093 & M17 & 0.59 & C2 & \nodata & \nodata \\
J221719.30+001543.8 & 106.4 & 23.56 & 0.07 & 0.58 & 0.04 & 0.51 & 0.15 & 0.44 & 0.04 & -75.46 & 3.67 & 3.097 & S03a & 0.64 & M3 & SSA22a-C30 & \nodata \\
J221736.90+001712.8 & 98.1 & 25.08 & 0.08 & 0.29 & 0.04 & 1.17 & 0.51 & 0.24 & 0.09 & 0.22 & 5.07 & 3.099 & M17 & 0.95 & \nodata & SSA22a-M31 & \nodata \\
J221721.02+001708.9 & 79.0 & 25.13 & 0.05 & 0.26 & 0.00 & 0.26 & 0.06 & 0.41 & 0.01 & 68.08 & 0.15 & 3.076 & S03e & 1.03 & C2 & SSA22a-C39 & \nodata \\
\enddata
\tablenotetext{a}{Signal-to-noise in a $1''$ diameter aperture.}
\tablenotetext{b}{Integrated S\'ersic model magnitude.}
\tablenotetext{c}{S\'ersic model effective radius.}
\tablenotetext{d}{S\'ersic model index.}
\tablenotetext{e}{S\'ersic model axis ratio $b/a$.}
\tablenotetext{f}{S\'ersic model position angle.}
\tablenotetext{g}{Redshift from literature; see \autoref{sec:catalog_generation}.}
\tablenotetext{h}{Redshift source: S03e=\citet{steidel2003} \lya\ emission; S03a=\citet{steidel2003} absorption-line; M17=Taken from \citet{micheva2017} catalogs.}
\tablenotetext{i}{Local LAE surface density; see \autoref{sec:physcorr}.}
\tablenotetext{j}{Consensus visual classification; see \autoref{sec:visual-classification}.}
\tablenotetext{k}{\citetalias{steidel2003} catalog designation.}
\end{deluxetable*}
\end{longrotatetable}

%% file: table3.tex
\begin{deluxetable*}{l c c h c c c c c c c c c c h}
\tablecaption{\label{table:photcatalog}Catalog excerpt for protocluster LBGs 
with acceptable fits as defined in \autoref{sec:models}, showing additional photometry.}
\tablehead{\colhead{ID} & \colhead{$u^*$} & \colhead{$B$} & \nocolhead{$NB497$} &
\colhead{$V$} & \colhead{$R$} & \colhead{$i'$} & \colhead{$z'$} & \colhead{$J$} &
\colhead{$K$} & \colhead{$3.6 \mu m$} & \colhead{$4.5 \mu m$} & 
\colhead{$5.8 \mu m$} & \colhead{$8.0 \mu m$} & \nocolhead{$\log_{10} L_x$\tablenotemark{a}}}
\startdata
J221710.35+001920.8 & 26.57 & 25.31 & \nodata & 24.31 & 24.11 & 23.94 & 23.91 & 23.40 & 22.78 & 23.07 & 23.13 & 22.32 & 23.05 & \nodata \\
J221704.34+002255.8 & 26.77 & 25.13 & \nodata & 24.23 & 24.08 & 23.96 & 23.99 & 23.04 & 22.65 & 23.04 & 22.95 & 22.40 & 22.85 & \nodata \\
J221732.04+001315.6 & 27.61 & 25.66 & \nodata & 24.49 & 24.16 & 23.85 & 23.85 & 23.26 & 22.34 & 22.11 & 22.02 & 22.12 & 21.49 & \nodata \\
J221737.92+001344.1 & 26.82 & 26.10 & \nodata & 25.21 & 24.91 & 24.66 & 24.61 & 24.13 & 23.49 & 23.25 & 23.29 & \nodata & 22.53 & \nodata \\
J221731.69+001657.9 & 27.98 & 26.46 & \nodata & 25.42 & 25.28 & 24.99 & 25.16 & 25.34 & 23.29 & 23.25 & 23.03 & 22.78 & 21.98 & \nodata \\
J221720.25+001651.7 & 27.57 & 25.80 & \nodata & 24.83 & 24.56 & 24.36 & 24.36 & 24.16 & 23.18 & 23.63 & 23.69 & 24.76 & 23.27 & \nodata \\
J221718.87+001816.2 & 26.41 & 25.50 & \nodata & 24.78 & 24.73 & 24.62 & 24.80 & 24.45 & 24.20 & 23.27 & 23.38 & 22.84 & 21.92 & \nodata \\
J221718.96+001444.5 & 26.95 & 25.64 & \nodata & 24.69 & 24.55 & 24.37 & 24.33 & 24.13 & 23.22 & 23.15 & 22.97 & 23.78 & 22.86 & \nodata \\
J221701.38+002031.9 & 27.50 & 26.68 & \nodata & 25.43 & 25.08 & 24.78 & 24.78 & 24.58 & 23.01 & 23.30 & 23.01 & 23.29 & 22.29 & \nodata \\
J221720.20+001731.6 & \nodata & \nodata & \nodata & \nodata & \nodata & \nodata & \nodata & \nodata & \nodata & \nodata & \nodata & \nodata & \nodata & \nodata \\
J221718.04+001735.5 & 28.79 & 26.45 & \nodata & 25.64 & 25.30 & 25.08 & 24.91 & 25.27 & 23.68 & 20.42 & 20.88 & 22.87 & 22.42 & \nodata \\
J221731.51+001631.0 & \nodata & \nodata & \nodata & \nodata & \nodata & \nodata & \nodata & \nodata & \nodata & \nodata & \nodata & \nodata & \nodata & \nodata \\
J221740.98+001127.2 & 26.86 & 26.01 & \nodata & 25.14 & 24.93 & 24.75 & 24.81 & 23.98 & 23.78 & 23.09 & 23.09 & \nodata & \nodata & \nodata \\
J221719.30+001543.8 & 26.80 & 25.66 & \nodata & 24.82 & 24.68 & 24.54 & 24.45 & 25.43 & 23.27 & 22.14 & 21.55 & 21.82 & 21.72 & \nodata \\
J221736.90+001712.8 & \nodata & \nodata & \nodata & \nodata & \nodata & \nodata & \nodata & \nodata & \nodata & \nodata & \nodata & \nodata & \nodata & \nodata \\
J221721.02+001708.9 & 27.20 & 26.20 & \nodata & 25.43 & 25.28 & 25.29 & 25.39 & 24.72 & 23.17 & 24.22 & 24.54 & \nodata & \nodata & \nodata \\
\enddata
\tablecomments{In each column we show the AB magnitude of the galaxy in the given band, as measured by \citet{kubo2013}. We label the IRAC bands (channel 1--4) by their reference wavelemgth ($3.6\ \micron - 8.0\ \micron$).}
\end{deluxetable*}

%% file: table4.tex
\begin{deluxetable*}{l c c c c c c c c}
\tablecolumns{9}
\tablecaption{\label{table:ksall}One- and two-dimensional KS test statistics and
               probabilities for derived properties and correlations of
               protocluster and field LBGs.}
\tablehead{\colhead{} & \multicolumn{4}{c}{S03 LBGs\tablenotemark{a}} & \multicolumn{4}{c}{S03 \& M17 LBGs\tablenotemark{b}} \\ \cmidrule(lr){2-5} \cmidrule(lr){6-9} \colhead{Protocluster/Field Comparison} & \colhead{$N_{\rm proto}$\tablenotemark{c}} & \colhead{$N_{\rm field}$\tablenotemark{d}} &  \colhead{$D_{KS}$\tablenotemark{e}} & \colhead{$p_{KS}$\tablenotemark{f}} & \colhead{$N_{\rm proto}$} & \colhead{$N_{\rm field}$} & \colhead{$D_{KS}$} & \colhead{$p_{KS}$}}
\startdata
$m$ 					& 9 & 41 & 0.43 & 0.09 & 16 & 42 & 0.28 & 0.28 \\
$\log r_e$ 				& 9 & 41 & 0.30 & 0.41 & 16 & 42 & 0.26 & 0.37 \\
$\log n$ 				& 9 & 41 & 0.36 & 0.23 & 16 & 42 & 0.31 & 0.17 \\
$q$ 				    & 9 & 41 & 0.25 & 0.66 & 16 & 42 & 0.27 & 0.31 \\
\midrule
$G$ 					& 13 & 13 & 0.54 & 0.04 & 24 & 15 & 0.32 & 0.25 \\
$M_{20}$ 				& 13 & 13 & 0.38 & 0.30 & 24 & 15 & 0.28 & 0.41 \\
$C$ 					& 13 & 13 & 0.38 & 0.30 & 24 & 15 & 0.23 & 0.61 \\
\midrule
$M_{\star}$ 		    & 8 & 45 & 0.58 & 0.01 & 15 & 46 & 0.58 & 0.00 \\
SFR   					& 8 & 45 & 0.49 & 0.05 & 15 & 46 & 0.54 & 0.00 \\
sSFR  					& 8 & 45 & 0.24 & 0.75 & 15 & 46 & 0.22 & 0.57 \\
Mass-weighted age   	& 8 & 45 & 0.23 & 0.80 & 15 & 46 & 0.19 & 0.75 \\
\cutinhead{Joint Distribution Tests}
$m - \log r_e$ 			& 9 & 41 & 0.28 & 0.59 & 14 & 42 & 0.26 & 0.42 \\
$m - \log n$ 			& 9 & 41 & 0.40 & 0.20 & 14 & 42 & 0.28 & 0.32 \\
$m - q$					& 9 & 41 & 0.43 & 0.12 & 14 & 42 & 0.26 & 0.40 \\
$\log r_e - \log n$ 	& 9 & 41 & 0.32 & 0.44 & 14 & 42 & 0.26 & 0.40 \\
$\log r_e - q$			& 9 & 41 & 0.29 & 0.55 & 14 & 42 & 0.25 & 0.45 \\
$\log n - q$ 			& 9 & 41 & 0.36 & 0.30 & 14 & 42 & 0.29 & 0.29 \\
\midrule
$G - M_{20}$ 			& 13 & 13 & 0.54 & 0.05 & 24 & 15 & 0.40 & 0.10 \\
$C - M_{20}$ 			& 13 & 13 & 0.46 & 0.13 & 24 & 15 & 0.32 & 0.28 \\
$G - C$ 				& 13 & 13 & 0.54 & 0.05 & 24 & 15 & 0.40 & 0.10 \\
\enddata
\tablecomments{For tests on parametric morphological properties and physical properties we include galaxies from our comparison sample of GOODS-N field LBGs. For the physical properties we show only the results for the SED fits with $Z = 0.655 Z_{\odot}$; the results of the tests are not significantly different for the fits with $Z = Z_{\odot}$. We exclude the \citetalias{micheva2017} LBG J221718.04+001735.5 from the KS tests on the SED-fitting derived parameters due to likely contamination of its near-IR photometry by a nearby point source.}
\tablenotetext{a}{I.e., comparing protocluster and field LBGs from the \citetalias{steidel2003} catalog only.}
\tablenotetext{b}{I.e., comparing protocluster and field LBGs from both the \citetalias{steidel2003} and \citetalias{micheva2017} catalogs.}
\tablenotetext{c}{Number of protocluster LBGs in comparison.}
\tablenotetext{d}{Number of field LBGs in comparison.}
\tablenotetext{e}{Two sample KS test statistic.}
\tablenotetext{f}{Two sample KS test $p$-value.}
\end{deluxetable*}

%% file: table5_gehrels.tex
\begin{deluxetable*}{l c c c c c c c c r}
\tablecaption{\label{tab:visual-classification}Number of galaxies in each category and calculated merger fraction for each 
of our visual classification samples. The categories M$1-4$ and C$1-2$ are defined in \autoref{sec:visual-classification}. The uncertainty on the merger fraction is calculated
using Poisson statistics.}
\tablehead{\colhead{} & \multicolumn{6}{c}{Number in Category} & \colhead{} & \colhead{} & \colhead{} \\ \cmidrule(lr){2-7} \colhead{Sample} & \colhead{M1} & \colhead{M2} & \colhead{M3} & \colhead{M4} & \colhead{C1} & \colhead{C2} & \colhead{Mergers} & \colhead{Isolated} & \colhead{Merger Fraction}}
\startdata
Protocluster & 0 & 2 & 1 & 0 & 1 & 4 & 3 & 5 & $0.38^{+0.37}_{-0.20}$ \\
X-ray AGN & 1 & 2 & 0 & 0 & 2 & 1 & 3 & 3 & $0.50^{+0.49}_{-0.27}$ \\
Total Field & 9 & 6 & 4 & 1 & 17 & 12 & 20 & 29 & $0.41^{+0.11}_{-0.09}$ \\
GOODS-N Field & 8 & 6 & 4 & 1 & 11 & 8 & 19 & 19 & $0.50^{+0.14}_{-0.11}$ \\
SSA22 Field & 1 & 0 & 0 & 0 & 6 & 4 & 1 & 10 & $0.09^{+0.21}_{-0.08}$ \\
\enddata
\end{deluxetable*}

%% file: table6.tex
\begin{deluxetable*}{l c c}
	\tabletypesize{\footnotesize}
	\tablecaption{\label{table:sedfilter}SED fitting filters for each sample.}
	\tablehead{\colhead{Sample} & \colhead{Observatory/Instrument} & \colhead{Filter(s)}}
	\startdata
	SSA22\tablenotemark{a}		& SUBARU/SUPRIMECAM				& $u^*~B~V~R~i'~z'$ \\
	\nodata		& SUBARU/MOIRCS					& $J~K_s$ \\
	\nodata		& \textit{HST}/WFC3				& F160W  \\
	\nodata		& \textit{Spitzer}/IRAC			& 3.6 \micron\ 4.5 \micron\ 5.8 \micron\ 8.0 \micron\ \\
	\tableline
	GOODS-N\tablenotemark{b} 	& KPNO/4m/MOSAIC				            & $U$ \\
	\nodata	 	& LBT/LBC						            & $U$ \\
	\nodata		& \textit{HST}/ACS/WFC						& F435W F606W F775W F814W F850LP \\
	\nodata		& \textit{HST}/WFC3    					    & F105W F125W F140W F160W \\
	\nodata	 	& CFHT/WIRCam						        & $K_s$ \\
	\nodata		& SUBARU/MOIRCS					            & $K_s$ \\
	\nodata		& \textit{Spitzer}/IRAC			            & 3.6 \micron\ 4.5 \micron\ 5.8 \micron\ 8.0 \micron\ \\
	\nodata     & \textit{Spitzer}/MIPS			            & 24 \micron\ 70 \micron\ \\
	\enddata
	\tablenotetext{a}{With the exception of F160W, the SSA22 photometry was measured by \citet{kubo2013}. F160W photometry was extracted from our images using the same procedures as \citet{kubo2013}.}
	\tablenotetext{b}{See \citet{barro2019} for description of the procedures used to extract the GOODS-N photometry.}
\end{deluxetable*}

%% file: table7.tex
\begin{deluxetable}{lcc}[htb]
	\tabletypesize{\small}
	\tablecaption{\label{tab:sfr_enhancement}SFR enhancement as a function of time for both assumed metallicities.}
	\tablecolumns{3}
	\tablehead{\colhead{} & \multicolumn{2}{c}{${\rm SFR/SFR_{Field}}$} \\ \cmidrule(lr){2-3} 
	\colhead{Epoch} & \colhead{$Z = 0.655 Z_\odot$} & \colhead{$Z = Z_\odot$}}
	\startdata
	$0 - 10$ Myr 	& $1.84_{-0.65}^{+1.45}$	& $2.02_{-0.70}^{+0.82}$ \\
	$10 - 100$ Myr 	& $2.36_{-0.63}^{+0.46}$	& $1.64_{-0.26}^{+0.29}$ \\
	$0.1 - 1$ Gyr	& $1.57_{-0.48}^{+0.59}$	& $1.73_{-0.47}^{+0.60}$ \\
	$1 - 2$ Gyr  	& $2.17_{-0.81}^{+1.03}$	& $1.91_{-0.53}^{+0.56}$ \\
	\enddata
	\tablecomments{Uncertainties are reported for the 16\% to 84\% confidence interval.}
\end{deluxetable}

%% file: table_example_SED_fits_SSA22.tex
\begin{deluxetable*}{l c c c c c c c r}
\tablecolumns{9}\tablecaption{\label{tab:SSA22phys}We show the SED-fit derived physical properties for our samples of SSA22 LBGs.}
\tablehead{\colhead{ID} & \colhead{$z$\tablenotemark{a}} & \colhead{$\chi^2_{min}$\tablenotemark{b}} & \colhead{$P_{null}(\chi^2_{min})$\tablenotemark{c}} & \colhead{$M_*$\tablenotemark{d}} & \colhead{SFR\tablenotemark{e}} & \colhead{sSFR\tablenotemark{f}} & \colhead{Age\tablenotemark{g}} & \colhead{Ref.\tablenotemark{h}} \\
\colhead{} & \colhead{} & \colhead{} & \colhead{} & \colhead{($10^9\ {\rm M_\odot}$)} & \colhead{($\rm M_\odot\ yr^{-1}$)} & \colhead{($10^9\ {\rm yr^{-1}}$)} & \colhead{($10^8\ {\rm yr}$)} & \colhead{}}
\startdata
\cutinhead{SSA22 Protocluster}
J221721.02+001708.9 & 3.076 & 9.56 & 0.05 & 7.10 & 29.01 & 4.22 & 6.42 & S03 \\
J221718.60+001815.5 & 3.079 & 2.95 & 0.57 & 10.30 & 60.10 & 6.11 & 5.86 & S03 \\
J221727.27+001809.6 & 3.080 & 10.99 & 0.05 & 11.06 & 68.97 & 6.57 & 5.74 & S03 \\
J221731.69+001657.9 & 3.088 & 2.73 & 0.84 & 21.88 & 25.92 & 1.14 & 7.14 & S03 \\
J221718.87+001816.2 & 3.089 & 6.32 & 0.39 & 12.42 & 36.50 & 2.94 & 9.24 & S03 \\
J221737.92+001344.1 & 3.094 & 3.54 & 0.62 & 13.26 & 55.25 & 4.20 & 6.91 & S03 \\
J221719.30+001543.8 & 3.097 & 6.51 & 0.37 & 76.70 & 340.37 & 3.97 & 8.02 & S03 \\
J221720.25+001651.7 & 3.098 & 3.86 & 0.70 & 6.69 & 20.40 & 3.01 & 7.55 & S03 \\
J221732.04+001315.6 & 3.065 & 3.98 & 0.41 & 30.42 & 114.46 & 3.72 & 6.78 & M17 \\
J221701.38+002031.9 & 3.073 & 5.27 & 0.51 & 16.02 & 45.55 & 2.85 & 7.91 & M17 \\
J221718.96+001444.5 & 3.091 & 1.95 & 0.92 & 15.80 & 61.85 & 3.95 & 7.11 & M17 \\
J221718.04+001735.5 & 3.093 & 13.50 & 0.04 & 197.17 & 1707.56 & 9.05 & 2.42 & M17 \\
J221740.98+001127.2 & 3.093 & 4.03 & 0.40 & 17.55 & 99.49 & 6.15 & 6.14 & M17 \\
J221710.35+001920.8 & 3.103 & 7.47 & 0.28 & 10.27 & 72.38 & 7.43 & 4.33 & M17 \\
J221720.55+002046.3 & 3.103 & 2.65 & 0.85 & 11.74 & 49.95 & 4.41 & 5.62 & M17 \\
J221704.34+002255.8 & 3.108 & 11.20 & 0.08 & 12.03 & 38.65 & 3.19 & 7.24 & M17 \\
\cutinhead{SSA22 Field}
J221724.44+001714.4 & 3.018 & 6.60 & 0.36 & 12.76 & 58.88 & 4.75 & 5.65 & S03 \\
J221735.98+001708.2 & 3.018 & 7.25 & 0.30 & 10.14 & 27.81 & 2.69 & 5.94 & S03 \\
J221735.30+001723.9 & 3.019 & 2.03 & 0.92 & 8.62 & 32.55 & 3.93 & 6.25 & S03 \\
J221722.90+001608.9 & 3.019 & 9.33 & 0.16 & 12.15 & 7.27 & 0.60 & 8.39 & S03 \\
J221725.64+001612.5 & 3.290 & 8.47 & 0.21 & 15.33 & 57.08 & 3.73 & 8.01 & S03 \\
J221717.69+001900.3 & 3.288 & 4.95 & 0.55 & 22.39 & 102.38 & 4.68 & 5.23 & M17\\
\enddata
\tablenotetext{a}{Redshift from literature; see \autoref{sec:catalog_generation}.}
\tablenotetext{b}{Minimum $\chi^2$ of of SED fitting chain.}
\tablenotetext{c}{Probability of \textit{accepting} the hypothesis that the data were generated by the best-fit model.}
\tablenotetext{d}{Stellar mass produced by median SFH.}
\tablenotetext{e}{Star formation rate averaged over the last 100 Myr of the median SFH.}
\tablenotetext{f}{Specific star formation rate averaged over the last 100 Myr of the median SFH.}
\tablenotetext{g}{Mass-weighted age of the median SFH.}
\tablenotetext{h}{Catalog reference. S03=\citet{steidel2003}, M17=\citet{micheva2017}}
\end{deluxetable*}

%% file: table_example_SED_fits_GOODSN.tex
\begin{deluxetable*}{l c c c c c c r}
\tablecolumns{8}\tablecaption{\label{tab:GOODSNphys}We show the SED-fit derived physical properties for our sample of GOODS-N field LBGs from the \citet{steidel2003} catalog.}
\tablehead{\colhead{ID} & \colhead{$z$\tablenotemark{a}} & \colhead{$\chi^2_{min}$\tablenotemark{b}} & \colhead{$P_{null}(\chi^2_{min})$\tablenotemark{c}} & \colhead{$M_*$\tablenotemark{d}} & \colhead{SFR\tablenotemark{e}} & \colhead{sSFR\tablenotemark{f}} & \colhead{Age\tablenotemark{g}} \\
\colhead{} & \colhead{} & \colhead{} & \colhead{} & \colhead{($10^9\ {\rm M_\odot}$)} & \colhead{($\rm M_\odot\ yr^{-1}$)} & \colhead{($10^9\ {\rm yr^{-1}}$)} & \colhead{($10^8\ {\rm yr}$)}}
\startdata
J123627.59+621130.1 & 2.917 & 9.52 & 0.48 & 9.66 & 51.92 & 5.45 & 6.29 \\
J123712.30+621138.2 & 2.925 & 4.06 & 0.91 & 2.02 & 8.67 & 4.23 & 7.71 \\
J123706.62+621400.3 & 2.926 & 15.81 & 0.20 & 17.26 & 94.71 & 5.53 & 5.85 \\
J123703.24+621451.3 & 2.926 & 10.06 & 0.26 & 2.26 & 15.24 & 6.84 & 5.17 \\
J123650.40+621055.6 & 2.928 & 12.93 & 0.07 & 0.56 & 2.59 & 4.55 & 6.42 \\
J123644.11+621311.2 & 2.929 & 20.97 & 0.03 & 27.66 & 170.82 & 6.30 & 5.30 \\
J123617.54+621310.1 & 2.930 & 8.78 & 0.27 & 2.33 & 9.06 & 3.81 & 7.72 \\
J123707.72+621038.2 & 2.931 & 13.56 & 0.19 & 5.90 & 28.90 & 4.85 & 6.51 \\
J123647.77+621256.1 & 2.932 & 17.39 & 0.10 & 11.02 & 60.34 & 5.56 & 5.24 \\
J123717.38+621247.4 & 2.939 & 9.71 & 0.47 & 6.60 & 38.16 & 5.83 & 6.55 \\
J123709.34+621047.1 & 2.942 & 9.42 & 0.31 & 1.51 & 9.04 & 5.83 & 5.76 \\
J123647.72+621053.6 & 2.943 & 9.48 & 0.49 & 14.09 & 26.07 & 1.80 & 7.78 \\
J123639.27+621713.4 & 2.944 & 8.62 & 0.57 & 10.50 & 63.71 & 6.17 & 5.57 \\
J123642.40+621448.9 & 2.962 & 6.46 & 0.78 & 2.37 & 15.47 & 6.70 & 4.97 \\
J123646.97+621226.5 & 2.970 & 10.73 & 0.22 & 1.07 & 3.68 & 3.54 & 7.35 \\
J123651.56+621042.2 & 2.975 & 15.80 & 0.15 & 23.33 & 141.81 & 6.14 & 4.69 \\
J123637.15+621547.8 & 2.975 & 5.67 & 0.84 & 5.36 & 21.20 & 4.05 & 5.95 \\
J123635.55+621522.0 & 2.980 & 39.49 & 0.00 & 3.07 & 8.41 & 2.83 & 7.98 \\
J123622.63+621306.4 & 2.981 & 12.44 & 0.26 & 73.39 & 19.02 & 0.26 & 11.20 \\
J123645.04+620940.8 & 2.983 & 13.30 & 0.15 & 4.00 & 11.07 & 2.78 & 9.34 \\
J123647.88+621032.3 & 2.990 & 3.79 & 0.96 & 5.43 & 28.01 & 5.21 & 6.24 \\
J123626.95+621127.4 & 2.993 & 5.99 & 0.74 & 5.51 & 27.89 & 5.06 & 6.28 \\
J123640.93+621358.6 & 3.087 & 10.38 & 0.32 & 9.80 & 15.63 & 1.55 & 9.90 \\
J123650.80+621444.8 & 3.106 & 25.27 & 0.00 & 1.71 & 6.55 & 3.97 & 7.56 \\
J123648.88+621502.6 & 3.115 & 17.99 & 0.08 & 9.57 & 24.85 & 2.53 & 7.21 \\
J123619.40+621502.0 & 3.128 & 8.92 & 0.45 & 4.40 & 10.39 & 2.36 & 8.02 \\
J123658.99+621714.4 & 3.130 & 17.36 & 0.07 & 4.00 & 23.86 & 5.92 & 4.99 \\
J123649.03+621542.4 & 3.136 & 4.24 & 0.94 & 6.45 & 34.75 & 5.41 & 5.09 \\
J123721.63+621350.5 & 3.148 & 31.89 & 0.00 & 3.27 & 9.51 & 3.10 & 8.49 \\
J123645.39+621347.1 & 3.161 & 13.75 & 0.09 & 1.66 & 3.96 & 2.41 & 7.32 \\
J123651.18+621349.0 & 3.163 & 43.97 & 0.00 & 4.85 & 16.13 & 3.43 & 7.52 \\
J123634.88+621253.9 & 3.182 & 26.63 & 0.00 & 1.21 & 2.26 & 1.91 & 8.33 \\
J123653.61+621410.5 & 3.196 & 6.45 & 0.49 & 5.20 & 14.86 & 3.04 & 6.40 \\
J123641.87+621107.4 & 3.197 & 11.33 & 0.33 & 3.71 & 15.34 & 4.09 & 6.22 \\
J123702.70+621426.3 & 3.214 & 17.47 & 0.06 & 71.93 & 140.47 & 1.95 & 8.63 \\
J123641.26+621203.4 & 3.222 & 18.91 & 0.04 & 5.66 & 14.62 & 2.61 & 6.90 \\
J123645.19+621652.4 & 3.229 & 7.35 & 0.60 & 11.77 & 54.49 & 4.64 & 6.17 \\
J123703.26+621635.3 & 3.239 & 46.47 & 0.00 & 7.19 & 34.98 & 4.91 & 5.41 \\
J123637.02+621044.5 & 3.241 & 9.79 & 0.20 & 4.22 & 17.54 & 4.24 & 6.53 \\
J123706.16+621509.9 & 3.246 & 4.84 & 0.77 & 7.19 & 32.01 & 4.40 & 6.55 \\
\enddata
\tablenotetext{a}{Redshift from \citet{steidel2003}.}
\tablenotetext{b}{Minimum $\chi^2$ of of SED fitting chain.}
\tablenotetext{c}{Probability of \textit{accepting} the hypothesis that the data were generated by the best-fit model.}
\tablenotetext{d}{Stellar mass produced by median SFH.}
\tablenotetext{e}{Star formation rate averaged over the last 100 Myr of the median SFH.}
\tablenotetext{f}{Specific star formation rate averaged over the last 100 Myr of the median SFH.}
\tablenotetext{g}{Mass-weighted age of the median SFH.}
\end{deluxetable*}